\documentclass[ijoc,nonblindrev]{informs3} 

\OneAndAHalfSpacedXI

\usepackage{hyperref}
\usepackage{subfig}
\usepackage[ruled,vlined]{algorithm2e}

\hypersetup{
    colorlinks=true,
    linkcolor=red,
    citecolor = blue,
    filecolor=magenta,      
    urlcolor=cyan,
}
\usepackage{natbib}
 \bibpunct[, ]{(}{)}{,}{a}{}{,}%
 \def\bibfont{\small}%

\DeclareMathAlphabet\mathrsfso      {U}{rsfso}{m}{n}
\usepackage{multirow}
\newcommand{\Exp}{{\rm E {}}}
\newcommand{\prob}{{\rm P {}}}
\newcommand{\RNum}[1]{\uppercase\expandafter{\romannumeral #1\relax}}

\newcommand{\revision}[1]{{\color{black}#1}}
\newcommand{\minorrevision}[1]{{\color{black}#1}}

\newcommand{\norm}[1]{\left\lVert#1\right\rVert}

\TheoremsNumberedThrough     

\EquationsNumberedThrough    

\begin{document}


\RUNAUTHOR{Abouee-Mehrizi, Mirjalili, and Sarhangian}


\TITLE{Platelet Inventory Management with Approximate Dynamic Programming}
\RUNTITLE{Platelet Inventory Management with ADP}

\ARTICLEAUTHORS{%
\AUTHOR{Hossein Abouee-Mehrizi}
\AFF{Department of Management Sciences, University of Waterloo, Waterloo, Ontario, N2L 3G1 Canada, \EMAIL{haboueem@uwaterloo.ca} \URL{}}
\AUTHOR{Mahdi Mirjalili, Vahid Sarhangian}
\AFF{Department of Mechanical and Industrial Engineering, University of Toronto, Toronto, Ontario, M5S 3G8 Canada, \EMAIL{mhdmjli@mie.utoronto.ca, sarhangian@mie.utoronto.ca} \URL{}}
} 







\ABSTRACT{
We study a stochastic perishable inventory control problem with endogenous (decision-dependent) uncertainty in shelf-life of units. Our primary motivation is determining ordering policies for blood platelets. Hospitals typically order their required platelets from a central supplier and as such the shelf-life of units at the time of delivery can be subject to significant variability. Determining optimal ordering quantities is a challenging task due to the short maximum shelf-life of platelets (3-5 days after testing) and high uncertainty in daily demand. We formulate the problem as an infinite-horizon discounted Markov Decision Process (MDP). The model captures salient features observed in our data from a network of Canadian hospitals and allows for fixed ordering costs. We show that with uncertainty in shelf-life, the value function of the MDP is non-convex and key structural properties valid under deterministic shelf-life no longer hold. Hence, we propose an Approximate Dynamic Programming (ADP) algorithm to find approximate policies. We approximate the value function using a linear combination of basis functions and tune the parameters using a simulation-based policy iteration algorithm. We evaluate the performance of the proposed policy using extensive numerical experiments in parameter regimes relevant to the platelet inventory management problem. We further leverage the ADP algorithm to evaluate the impact of ignoring shelf-life uncertainty. Finally, we evaluate the out-of-sample performance of the ADP algorithm in a case study using real data and compare it to the historical hospital performance and other benchmarks. After tuning the parameters, the ADP policy can be computed online in a few seconds and results in more than 50\% lower expiration and shortage rates compared to the historical rates. In addition, it performs better or as well as \minorrevision{other benchmarks including} an exact policy that ignores uncertainty in shelf-life and becomes hard to compute for larger instances of the problem. 
}%


\maketitle

%
\vspace*{-0.5cm}
\section{Introduction}
Every year, more than 2.5 million blood platelet units are distributed in the US and Canada \citep{free2023continued}. The majority of these units are collected from donors and distributed to hospitals through central suppliers, e.g., the Red Cross in the US and Canadian Blood Services (CBS) in Canada. Large hospitals typically require daily deliveries, whereas smaller hospitals have less frequent deliveries. Due to the high uncertainty in demand and the short shelf-life of platelet units (3-5 days after testing), determining order quantities and frequencies that strike a balance between availability of units in the hospital and wastage is a challenging task.  Motivated by these challenges, a large body of inventory management literature has focused on developing optimal or close-to-optimal ordering policies for platelets. A common assumption in this body of literature, and more broadly the literature on perishable inventory management, is that the shelf-life of units is deterministic. In practice, when ordering from a central supplier, the shelf-life of deliveries could be subject to significant uncertainty. For example, our data from a network of Canadian hospitals in Hamilton, Ontario (See Section \ref{sec:casestudy} for details) shows significant uncertainty in the remaining shelf-life of delivered units. In addition, as a result of the ordering practices of the supplier, the age of delivered units can depend on the size of the orders. This introduces endogeneity in shelf-life uncertainty, because the uncertainty is impacted by the ordering decisions.  

Finding optimal or approximately optimal ordering policies under shelf-life uncertainty is known to be difficult (see, e.g., \citealt{nahmias2011perishable}). The few studies in the literature that account for uncertainty in shelf-life assume that orders expire in the same sequence in which they enter inventory (see, e.g., \citealt{nahmias1977ordering}, \citealt{chen2014coordinating}, and \citealt{chao2015approximation}).  In practice, this assumption does not hold in general. As a result, key structural properties that hold when shelf-life is deterministic, e.g., convexity of the cost function in the ordering decision, may no longer hold, making the computation of an optimal policy or high-quality heuristics more difficult. Due to these complexities, the effect of uncertainty on the optimal policy and sub-optimality of policies that ignore it are also not well-understood. 

In this paper, we propose an Approximate Dynamic Programming (ADP) approach for platelet inventory management under endogenous shelf-life uncertainty. We consider a periodic-review inventory control problem where both the demand and shelf-life of orders are subject to uncertainty. Demand is non-stationary but independent across the periods of the horizon. Unmet demand is assumed to be lost, i.e., satisfied through emergency orders. The shelf-life of the deliveries is uncertain and potentially endogenous to the order size.  Demand is satisfied according to the Oldest-Unit, First-Out (OUFO) allocation policy and lead time of deliveries is zero. We consider four cost components: a linear expiration, lost demand (shortage), holding cost, and a fixed ordering cost. The fixed ordering cost is typically ignored in the literature (\citealt{nahmias2011perishable}) but is relevant for smaller hospitals that require less frequent deliveries. 

We formulate the problem as a discounted infinite-horizon Markov Decision Process (MDP).  Using small instances of the problem that can be solved exactly, we examine the structure of the value function and the optimal policy. We show that in general the value function is non-convex (even in the absence of a fixed cost) and the optimal policy does not satisfy the typical structural properties valid under deterministic shelf-life. We thus propose a simulation-based ADP approach to obtain approximately optimal policies. We parameterize the value function as a linear combination of a small number of basis functions and use a simulation-based policy iteration algorithm to tune the parameters. We evaluate the performance of the ADP by comparing it to the exact solution for small instances of the problem, and a Myopic policy (that minimizes the expected one-period cost and ignores future costs), as well as an information relaxation lower-bound for larger instances. We leverage the ADP policy to examine the impact of ignoring uncertainty of the shelf-life and its dependence on the order size in various system and cost parameter regimes. Finally, we examine the performance of the proposed approach through a case study using data from a Canadian hospital. Our main results and contributions can be summarized as follows.

\begin{itemize}
    \item \emph{Approximate Policy Iteration algorithm and the choice of basis functions:} We identify a small set of basis-functions that provide an accurate approximation of the value function under endogenous uncertainty in the remaining shelf-life. We examine the performance of basis function candidates using the optimal value function for small problem instances. We find that a quadratic approximation structure together with the value function of the corresponding non-perishable problem performs well in most scenarios relevant to platelet inventory management where shortage is costlier than expiration. When the unit cost of expiration is large and dominates the other three inventory costs, including interaction terms can reduce error significantly. In our numerical study, including cubic terms can slightly reduce the optimality gap to below 10\%, while the interaction term can reduce the optimality gap to below 5\%. Among all cases, and using the quadratic basis functions, optimality gap is on average less than 2\%, and 80\% less than that of the Myopic policy. 
    \item \emph{Impact of ignoring uncertainty in shelf-life}:
     We examine the impact of ignoring shelf-life uncertainty using the exact optimal policy when maximum shelf-life is three days and our ADP policy when maximum shelf-life is \minorrevision{longer}. We do so by quantifying the sub-optimality of policies assuming deterministic and exogenous shelf-life. Ignoring shelf-life uncertainty could lead to significant sub-optimality in relevant parameter regimes where the chance of receiving older units with remaining shelf-life of one is high and the fixed ordering and unit wastage costs are relatively large. The sub-optimality gap is reduced when considering an appropriate exogenous uncertainty associated with each endogenous case, but the remaining gap attributed to the impact of ignoring endogeneity could still be large (up to 32\%) when the dependence on order-size is strong enough and fixed ordering cost is large.  
     
    \item \emph{Case study:}
    Our proposed ADP allows for online calculation of optimal policies in a few seconds and only requires time-based (e.g., day of the week) real-time information. As such, it can be easily implemented in practice. Using platelet inventory data from a Canadian hospital, we compute the counterfactual out-of-sample performance of the ADP against that of the hospital in 2017 \minorrevision{as well as three other benchmarks: an exact solution under deterministic shelf-life; an $(s,S)$ policy where the parameters are chosen to optimize the empirical in-sample cost; and a greedy policy that uses the information-relaxation lower-bound as an approximation of the value function. The ADP method achieves significant reductions with respect to all performance measures compared to the historical performance with 35\%, 5\%, 33\%, 57\%, and 66\% reduction in fraction of days with non-zero order size, total order size over the year, daily average holding, wastage, and shortage rates, respectively. It also outperforms or performs as well as the benchmarks while scaling well for larger instances. }
\end{itemize}

The rest of the paper is organized as follows. Section \ref{sec:lit} provides a review of the related literature. In Section \ref{sec:mdp}, we present the DP model to formulate our periodic-review perishable inventory problem with endogenous shelf-life uncertainty. \minorrevision{In Section \ref{sub:structure} we discuss and numerically illustrate the structure of the value function and optimal policy.} In Section \ref{sec:adp}, we propose our simulation-based ADP algorithm. In Section \ref{sec:numerics}, we examine the performance of the ADP approach and measure the value of accounting for shelf-life uncertainty and its dependence on order size. In Section \ref{sec:casestudy}, we present the results of our case study using real data. Finally, we conclude the paper in Section \ref{sec:conclusion}.

\section{Literature Review} \label{sec:lit}
In this section, we briefly review three relevant streams of literature, i.e., inventory control of perishable products, \revision{blood products}, and ADP.
\subsection{Related Literature on Perishable Inventory Control}\label{sec:perishlit}
Our work relates to the large body of literature on ordering policies for perishable products, see \cite{nahmias2011perishable} for a comprehensive review of single-product periodic-review perishable inventory problems. The main focus of this stream is on characterizing the optimal ordering policy under certain assumptions, i.e., shelf-life of orders is deterministic, lead time and fixed ordering costs are zero, demand is independent and identically distributed (iid), and allocation policy is First-In-First-Out (FIFO). 

DP is the main approach to characterize the structure of the optimal policy, used first in \cite{nahmias1973optimal} for a two-period shelf-life problem and later in \cite{fries1975optimal} and \cite{nahmias1975optimal} for the general shelf-life problems with lost and backlogged demand, respectively. These studies demonstrate that optimal order quantities depend on the age distribution of units available in the inventory and hence computing the optimal policy using exact DP is hard due to the large state-space. They further find that a unit increase in the level of fresher inventory decreases the optimal order quantity at a rate less than one, but more quickly compared to a unit increase in the level of older inventory.

Given the complexities of finding optimal policies, other studies have mainly focused on developing heuristic solutions that are easy to implement. Examples of proposed heuristics are base-stock or TIS (Total-Inventory-to-S) policies that bring up the total inventory level to a fixed value (\citealt{nahmias1976myopic}, \citealt{nandakumar1993near}, \citealt{cooper2001pathwise}, \citealt{bu2023asymptotic}), inspired from the literature of non-perishable inventory systems, and 
NIS (New-Inventory-to-S) policies that bring up the inventory level of newer items to a fixed value (\citealt{brodheim1975evaluation}, \citealt{deniz2010managing}, \citealt{civelek2015blood}), inspired from the monotone sensitivity of the optimal ordering policy. Other examples include \cite{rajendran2020hybrid} who propose the $(s, S, \alpha, Q)$ policy which places an order when the total inventory is below the reorder level $s$ with the order size depending on $\alpha$ parameter and develop a stochastic mixed-integer program to find the optimal parameters. These heuristics typically do no have performance guarantees. One exception is approximation algorithms that extend the marginal cost accounting scheme \citep{levi2007approximation} developed for nonperishable problems to perishable problems, e.g., \cite{chao2015approximation} and \cite{zhang2023truncated}. Recently, \cite{bu2023asymptotic} establish the asymptotic optimality of base-stock policies when lifetime, demand population size, unit penalty cost, or unit outdating cost becomes large. \cite{farrington2023going} propose GPU-accelerated approaches to extend the use of value iteration to larger instances of perishable inventory problems.

\subsection{Related Literature on Inventory Management of Blood Products}\label{sec:bloodlit}

\revision{Finding good policies for ordering perishable blood products (Red-cells, Platelets, Plasma, etc.) has been the primary motivation of many studies in the perishable inventory management literature; see, e.g., \cite{karaesmen2011managing} and \cite{haijema2017blood}. A more recent body of works aims to leverage existing granular data to determine ordering quantities; see \cite{li2023blood} for a recent review. A common approach in this literature is to first predict the demand using predictive models from the statistical machine learning literature and then optimize the ordering decisions to hedge against the prediction errors (see, e.g., \citealt{li2021decision}, \citealt{motamedi2022}, \citealt{li2022demand}). Another approach is to combine the prediction and optimization steps by replacing the loss function of the prediction model with the objective function of the decision-making problem, as done in the single-period Newsvendor setting in \cite{ban2019big}. \cite{guan2017big} formulate base-stock levels as a linear function of exogenous features correlated with demand and propose a convex optimization model to determine ordering quantities that minimize the empirical expiration rate subject to a minimum inventory on hand. \cite{abouee2022data} (see also, \citealt{mirjalili2022data}) consider minimizing a weighted \revision{sum} of expiration and shortage costs and propose a robust optimization model that accounts for shelf-life uncertainty. A challenge with these approaches is the potential need for a large number of \revision{covariates} that may not be available in real-time (e.g., number of patients with abnormal platelets in the hospital). \cite{abouee2022data} shows that even a small number of covariates that are available online (e.g., day of the week, month of the year, and average demand in previous days) can provide high-quality ordering decisions when the predictions directly consider the down-stream inventory costs. However, including fixed ordering costs in this framework is not straightforward as it renders the optimization problem nonconvex. With zero fixed costs, the optimal policy tends to order on a daily basis. This is reasonable for large hospitals placing daily orders, but is less practical for smaller hospitals that order less frequently. The ADP approach proposed here relies on a high-fidelity yet still parsimonious stochastic model whose parameters are estimated offline, and allows for fixed ordering costs. }

\cite{nahmias1977ordering} is the first study on perishable inventory that considers both demand and shelf-life are random with a restrictive assumption that successive orders expire in the same sequence that they enter inventory (i.e., noncrossover lifetimes). This condition is sufficient to show that all of the properties of the optimal policy, including the convexity of the cost function in ordering decision, are carried over from the deterministic shelf-life problem considered in \cite{nahmias1975optimal}. More recently, \cite{chen2014coordinating} and \cite{chao2015approximation} use the noncrossover lifetimes assumption to extend their approach to problems with random shelf-life. To the best of our knowledge, no other studies have investigated the perishable inventory problem with shelf-life uncertainty as we study in this paper. Moreover and in contrast to most studies in the literature, we consider the fixed ordering cost such as costs of delivery per order. \cite{nahmias1978fixed} is the first study on perishable inventory that considers the fixed ordering cost together with the deterministic shelf-life assumption. In general, as we illustrate, the problem with nonzero fixed ordering cost and shelf-life uncertainty is nonconvex and hard to solve. See Section \ref{sub:structure} for details.

\subsection{Related Literature on ADP}

\revision{To address the large state-space of our perishable inventory control problem, we propose an ADP algorithm; see \cite{powell2007approximate} and \cite{Bert05}. Our approach is based on approximation in the value space. It relies on a high-fidelity stochastic model of the inventory system and leverages simulation with variance reduction. The main idea is to parameterize the value function as a linear function of basis functions and then tune the parameters. Nevertheless, there is no universal method for selecting the basis functions. As such, a primary contribution of our work is to propose and evaluate the performance of a class of basis functions in our specific setting.}  

The ADP methods in value space can be classified into two main categories: simulation-based and linear programming algorithms. The latter is based on formulating the optimality equations associated with an infinite-horizon DP model as a linear program where the value function is represented using an appropriate approximation structure (\citealt{de2003linear}). This approach is used in various applications including perishable inventory problems (\citealt{sun2014quadratic}), healthcare scheduling (\citealt{saure2012dynamic}, \citealt{gocgun2014dynamic}, \citealt{saure2020dynamic}), and revenue management (\citealt{ke2019approximate}), among others. The linear programming algorithms are mostly restricted to affine approximation structures, but can still provide good results with no need for iterative learning. Simulation-based algorithms consist of a broader range of methods that can be classified into model-free, e.g., Q-learning (\citealt{watkins1992q}), and model-based methods (e.g., \citealt{saure2015simulation}). The latter is based on approximation of the value function of an arbitrary policy which can be embedded within a policy iteration algorithm, while the former is based on direct approximation of the optimal value function. Our work is related to model-based approaches that use either direct or indirect methods for policy evaluation (\citealt{Bert05}). Direct methods use simulation to generate value function estimates for different system states and fit a predefined approximation function to these estimates by minimizing a normed criterion of error, see, e.g., \cite{chen2021managing} for application to perishable inventory, \cite{maxwell2010approximate, maxwell2013tuning} for ambulance redeployment, and \cite{astaraky2015simulation} for healthcare scheduling problems. Indirect methods, such as temporal difference learning (\citealt{sutton1988learning}), update the parameters of value function approximation using iterative methods that do not require the entire simulation trajectory, e.g., \cite{dai2019inpatient}. In this work, we use a simulation-based approximate policy iteration approach together with the direct method for policy evaluation.

ADP has been applied to perishable inventory problems in the literature, but without shelf-life uncertainty. For instance, \cite{chen2021managing} considers managing perishable inventory with multiple demand classes each requiring different levels of freshness and associated with different lost-demand costs. They assume zero lead time and fixed ordering cost together with deterministic shelf-life to characterize structural properties of the value function and leverage them to design their adaptive approximation approach. \cite{sun2014quadratic} use approximation in value space for a perishable product whose demand is either satisfied based on the FIFO allocation policy or backlogged. They consider a positive lead time, zero fixed ordering cost, and deterministic shelf-life.

We demonstrate the performance of our proposed ADP through extensive numerical experiments. For cases where the exact solution is not known, we compare the performance of the ADP with a lower-bound obtained using the information relaxation approach established in \cite{brown2010information} for finite-horizon and later in \cite{brown2017information} for infinite-horizon discounted problems. Specifically, we partially relax the nonanticipativity constraint with respect to shelf-life uncertainty and solve simplified DPs with multiple shelf-life samples and take the sample average of corresponding value functions as the lower-bound on cost. See \cite{brown2022information} for a review of the information relaxation approach.

\section{MDP Formulation} \label{sec:mdp}
We formulate the problem of finding an optimal ordering policy for a perishable product with maximum shelf-life of $m$ periods as an $\alpha$-discounted infinite-horizon discrete-time MDP. In each period, the decision maker decides how many units to order from a supplier. We assume the supplier satisfies all orders with zero lead time but the shelf-life of supplied units is subject to uncertainty and could depend on the order size.

Denote the remaining shelf-life of received units by an $m$-dimensional random vector $\textbf{Y} = (Y_{m}, \dots, Y_{1}) \in \mathbb{Z}^{m}_+$, where $Y_{k}$ is the number of units received with remaining shelf-life of $k \in \{1, \dots, m\}$ periods. Let $ z= \sum_{k=1}^{m}Y_{k}$ denote the order size. We assume that given an order size of $z$, each received unit has remaining shelf-life $k$ with probability $p_{k}(z)$, independent of other units. Note that the probabilities can depend on the order size. Given these assumptions, $\textbf{Y}$ has a \revision{multinomial} distribution with $z$ trials, $m$ events, and probability mass function:
\begin{align}\label{eq:pmf}
    P (Y_{m} = y_{m}, \, \dots, Y_{1}=y_{1}) = \frac{z!}{y_{m}!\,y_{m-1}!\,\dots\,y_{1}!}p_m(z)^{y_{m}}p_{m-1}(z)^{y_{m-1}}\dots p_1(z)^{y_{1}}.
\end{align}
In our numerical experiments, we assume that $p_{k}(z)$, $k\in \{1, \dots, m\}$ has a \revision{multinomial} logit form, that is, the log-odd ratio of receiving units with each remaining shelf-life $k>1$ versus $k=1$ is an affine function of the order size $z$. The functions $p_{k}(z)$ can then be estimated using a \revision{multinomial} \revision{logistic} regression. We illustrate this using data in Section \ref{subsub:shelf-life}.

Denote by $D_t$ the demand in period $t$. We assume that demand follows a \emph{periodic} distribution with period $\gamma$, that is, $D_t \overset{d}{=} D_{t+\gamma}$ where $\overset{d}{=}$ denotes equality in distribution. We denote by $\tau \in \{0, \dots, \gamma-1\}$ the time-index within the demand period. For instance, periodic demand with $\gamma=7$ captures temporal variation in demand across days of the week with $\tau \in \{0, \dots, 6\}$ corresponding to each day of the week (Monday to Sunday). We assume demand is independent across different periods. 

\subsection{State Variable and Decisions} 
Denote by $(\tau, \textbf{X}) \equiv (\tau, X_{m-1}, \dots, X_{1})$ the state of the MDP. $(\tau, \textbf{X})$ is an $m$-dimensional vector with the first element 
denoting the time-index within the demand period (e.g., day of the week) and included to obtain a time-homogeneous MDP. The last $m-1$ elements (which we refer to as the inventory state) keep track of the number of units in inventory with different ages at the beginning of each period. The state-space is then the set  $ \{0, \dots, \gamma-1\} \times \mathcal{X} $ where $\mathcal{X}\subset \mathbb{Z}^{m-1}_+$.  We denote a realization of the inventory state at a given time period with $\textbf{x} = (x_{m-1}, x_{m-2}, \dots, x_{1})$, where $x_{i}$ is the number of units with the remaining shelf-life of $i \in \{1, \dots, m-1\}$ in the inventory.

After observing the state at the beginning of each time period, the decision maker can place \minorrevision{an order of size $z \in \mathcal{Y}=\{0, \dots, \bar{z}\}$ where $\bar{z}<\infty$ is the maximum order size.} An admissible ordering policy is a (measurable) map $\mu: \{0, \dots, \gamma-1\} \times \mathcal{X} \xrightarrow{} \mathcal{Y}$, determining the number of orders placed in each time period. \revision{Note that the bounded order size together with maximum shelf-life of $m<\infty$ implies that the state-space is also bounded.}

\subsection{Dynamics}
\revision{The sequence of events in each period is as follows. At the beginning of each period, the decision maker observes the state of the system and places an order. The order is then delivered and the inventory state is updated. Next, the demand for the period is realized and satisfied using the available inventory.} We assume that demand is satisfied with the oldest units available in the inventory, i.e., the  allocation policy is OUFO. Further, we assume unsatisfied demand is lost, and units with remaining shelf-life of one at the end of each period are expired. 

Let the current state be $(\tau, \textbf{x})$. The transition probabilities under a policy $\mu$ are induced by the following equations that determine the state at the beginning of the next period $(\tau', \textbf{x}')$:
    \begin{align}
            &\tau' = (\tau+1) \mbox{ mod } \gamma,\label{eq:dynamics_1}\\
            &X'^\mu_{j} = \bigg(x_{j+1} + Y_{j+1}^\mu - \Big(D_\tau - \sum_{i=1}^j (x_{i} + Y_{i}^\mu)\Big)^+\bigg)^+, \quad \quad \forall j \in \{1,\dots, m-2\},\\  
            &X'^\mu_{m-1} = \bigg(Y_{m}^\mu - \Big(D_\tau - \sum_{i=1}^{m-1} (x_{i}+Y_{i}^\mu)\Big)^+\bigg)^+, \label{eq:dynamics_2}
    \end{align}
where $a^+ = \max(a,0)$ and $\textbf{Y}^\mu = (Y_m^\mu, \dots, Y_1^\mu)$ denotes the remaining shelf-life of units (independent of the time-period) under policy $\mu$. (Note that we use the superscript $\mu$ to indicate dependence on the ordering policy.) To simplify notation, we define $s(\tau, \textbf{x}, Y_m^\mu, \dots, Y_1^\mu, D_{t})$ to be the function that takes $(\tau,\textbf{x})$ and returns the inventory state at the beginning of the next period according to \eqref{eq:dynamics_1}-\eqref{eq:dynamics_2}, and for each realization of random variables $\textbf{Y}^\mu$ and $D_{\tau}$, i.e., $\textbf{X}'^\mu=s(\tau, \textbf{x}, Y_m^\mu, \dots, Y_1^\mu, D_{\tau})$. 

\subsection{Cost Structure and Optimality Equations}
We consider four cost components: a fixed ordering cost \revision{$\kappa$}, and linear holding, shortage, and wastage costs with unit costs denoted by $h$, $l$, and $\theta$, respectively. 
The expected infinite-horizon discounted cost associated with any ordering policy $\mu$ starting from state $(\tau,\textbf{x})$ is,
\begin{align}
    v^{\mu}(\tau, \textbf{x}) = \Exp \Bigg[\sum_{t=0}^{\infty} \alpha^t C\big(\textbf{X}^\mu_t, \, \mu (\tau_t,\, \textbf{X}^\mu_t)\big) \mid (\tau_0, \textbf{X}^\mu_0) = (\tau,\textbf{x}) \Bigg],\label{eq:expcs}
\end{align}
where $0\leq \alpha < 1$ is the discount-factor and, 
\begin{align}
C\big(\textbf{X}^\mu, \, \mu (\tau,\, \textbf{X}^\mu)\big) &= \kappa \: \textbf{1}_{\big\{\mu (\tau,\, \textbf{X}^\mu)>0\big\}} +h \: \Big(\sum_{i=1}^{m-1}X^\mu_{i} + \mu (\tau,\, \textbf{X}^\mu_t) -D_{\tau}\Big)^+ \nonumber\\
&+l \: \Big(D_{\tau}-\sum_{i=1}^{m-1}X^\mu_{i} - \mu (\tau,\, \textbf{X}^\mu)\Big)^+ +\theta \: (X^\mu_{1}+Y^\mu_{1}-D_{\tau})^+. \label{eq:cost}
\end{align}
\minorrevision{The cost in period $t$ is comprised of a fixed ordering cost associated with any (non-zero) order quantity (incurred at the beginning of the period and after placing the order), the linear holding cost of units remaining in the inventory, and the shortage and wastage costs incurred in that period (all incurred at the end of the period and after realizing the demand). }

\revision{The value function $v(\tau,\textbf{x})$ is the \emph{minimum} expected cost in \eqref{eq:expcs} among all admissible policies. For $ (\tau,\textbf{x}) \in \{0,\dots,\gamma-1\} \times \mathcal{X}$ define the DP mappings, 
\begin{align}
     (\mathcal{T}_{\mu} v)(\tau,\textbf{x}) & \equiv \Exp \bigg[C\big(\textbf{x}, \, \mu(\tau,\textbf{x}) \big) + \alpha v\big(s(\tau, \textbf{x}, Y^\mu_m, \dots, Y^\mu_1, D_{\tau})\big)\bigg],\\
    (\mathcal{T}v)(\tau,\textbf{x}) & \equiv \min_{z\in \mathcal{Y}} \Exp \bigg[C\big(\textbf{x}, \, z\big) + \alpha v\big(s(\tau, \textbf{x}, Y^z_m, \dots, Y^z_1, D_{\tau})\big)\bigg]. \label{eq:mincost}
\end{align}
An optimal policy then satisfies the Bellman equation, 
\begin{align}
    (\mathcal{T}v)(\tau,\textbf{x}) = v(\tau,\textbf{x}).\label{eq:opteq}
\end{align}
Given that $0\leq \alpha < 1$; the state-space is countable; and that the cost function in \eqref{eq:cost} is bounded, the Bellman equation admits a unique solution (Theorem 6.2.5. in \citealt{puterman2014markov}).}

\section{Structure of the Optimal Policy and Value Function} \label{sub:structure}
In this section, we examine the structure of the optimal policy and value function. More specifically, we show through numerical examples that key structural properties known to hold under the deterministic shelf-life assumption, do not hold in general when the shelf-life is random. In the following, we first review these properties in more details. 

Under the assumptions of the deterministic shelf-life and zero fixed ordering cost, two key structural properties hold: First, the expected cost function is convex in the ordering decision, and hence the value function (after minimization) is also convex (see, e.g., Theorem 2.2. in \citealt{nahmias2011perishable}). Second, the optimal ordering policy satisfies the following bounded sensitivity (see, e.g., Theorem 2.2. in \citealt{nahmias2011perishable} or Theorem 3 in \citealt{li2014multimodularity}): 
\begin{align}
    -1 \leq \Delta_{x_{m-1}}\mu^*(\tau, \textbf{x}) \leq \Delta_{x_{m-2}}\mu^*(\tau, \textbf{x}) \leq \dots \leq \Delta_{x_{1}}\mu^*(\tau, \textbf{x}) \leq 0, \quad \forall \tau, \textbf{x}, \label{ineq:StructLit}
\end{align}
\revision{where $\Delta_{x_{i}} \mu^*(\tau, \textbf{x})$ denotes the first difference of $\mu^*$ with respect to $x_i$, defined as $(\mu^*(\tau, \textbf{x} +\delta\textbf{e}_i) -\mu^*(\tau, \textbf{x}))/\delta$ where $\textbf{e}_i$ is a vector with a value of one in its $i$th component and zero elsewhere, and $\delta$ is a small positive value. When $\mu^* $ is differentiable, then $\Delta_{x_{i}} \mu^*(\tau, \textbf{x}) \equiv \partial \mu^* / \partial x_i$.}
The condition implies that the optimal ordering policy is a decreasing function of the available inventory of different ages. That is, a unit increase in the on-hand units of newer inventory (with longer remaining shelf-life) decreases the optimal ordering quantity more rapidly than a unit increase in the on-hand units of older inventory with at most one unit. \cite{li2014multimodularity} use multimodularity to show the bounded sensitivity \eqref{ineq:StructLit} of the optimal ordering policy for a system with clearance sale (Multimodularity implies increasing differences and convexity; see Lemma 1 in \citealt{li2014multimodularity}). \revision{As we illustrate numerically next, (1) the expected cost-to-go function minimized in the right-hand-side of \eqref{eq:mincost} is generally nonconvex in the ordering decision; (2) the value function after minimization remains nonconvex (and hence not multimodular); and (3) the optimal policy $\mu^*$ no longer satisfies the bounded sensitivity property in \eqref{ineq:StructLit}.}

\begin{figure}
    \centering
    \includegraphics[angle=-1.1, width=\textwidth]{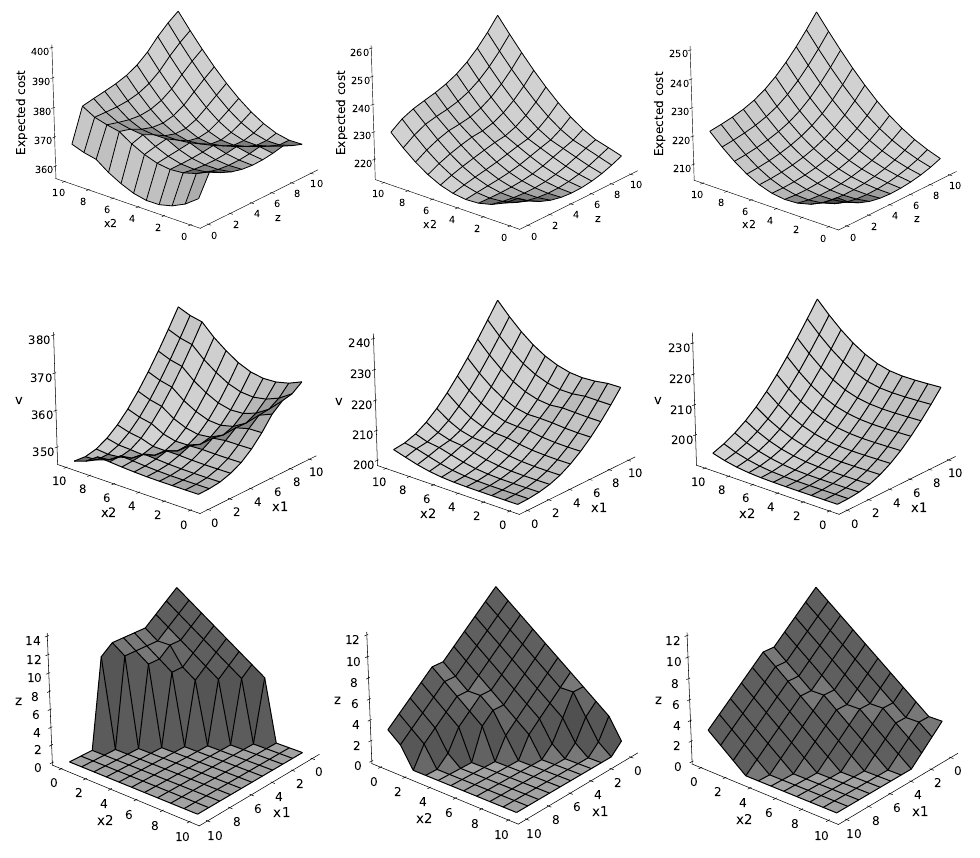}
    \caption{Comparing the expected cost function, value function, and optimal policy obtained under the fixed ordering cost of $\kappa=10$ and endogenous shelf-life uncertainty (left column) with those obtained under zero fixed ordering cost and endogenous (middle column) or deterministic shelf-life (right column) assuming $h=1, \, l =20, \, \theta = 5$ and the demand and other parameters are the same across columns.}%
    \label{fig:structure}%
\end{figure}

We consider a small size problem with $m=3$ which can be solved exactly using the value iteration algorithm. We specify $p_k(z)$ for $k\in\{1,2,3\}$, using the \revision{multinomial} logit model (see Section \ref{subsub:shelf-life} for details):
\begin{align}
    p_3(z) &= \frac{\exp(0.5+0.8z)}{1+\exp(1+0.4z)+\exp(0.5+0.8z)}, \\
    p_2(z) &=\frac{\exp(1+0.4z)}{1+\exp(1+0.4z)+\exp(0.5+0.8z)}, \\ 
    p_1(z) &= 1 - p_3(z) - p_2(z).
\end{align}
We fix $h=1, \, l = 20, \, \theta =5,$ and consider two cases of $\kappa\in\{0,10\}$ for the fixed ordering cost. The periodic demand follows truncated \revision{negative binomial} distributions fitted to real demand data (see Section \ref{subsub:demand}). Each row of Figure \ref{fig:structure} presents respectively the expected cost function for $x_1 = 8$ and varying $x_2$ and $z$; the value function after minimization; and the optimal policy starting at $\tau =0$. 

For the general problem with the \emph{nonzero} fixed ordering cost and endogenous shelf-life uncertainty (left column), we observe that the expected cost and value function are nonconvex and the sensitivity of the resulting optimal ordering policy to available inventory does not satisfy \eqref{ineq:StructLit}. For example, when $x_2=0$, a unit increase in $x_1$ from 7 to 8 decreases the optimal ordering quantity by 10 units to zero. Intuitively, with the nonzero fixed ordering cost, it can be optimal to place larger orders less frequently and keep more units in the inventory to avoid ordering in every period. We note that the value functions are not $K$-convex \citep{scarf1960optimality} either, which is the case for a non-perishable inventory problem with positive fixed ordering cost.

For the problem with \emph{zero} fixed ordering cost and endogenous shelf-life uncertainty (middle column), we observe the expected cost and value function are non-convex and the optimal policy does not satisfy \eqref{ineq:StructLit}. For example, when $x_2=6$, a unit increase in $x_1$ from 1 to 2 decreases the optimal ordering quantity from 5 to 4, while when $x_1 = 1$, a unit increase in $x_2$ from 6 to 7 does not change the optimal ordering quantity of 5 units. Thus, the optimal ordering quantity is not necessarily more sensitive to the fresher inventory. Intuitively, with endogenous shelf-life uncertainty, it is optimal to place larger (smaller) orders per period if the chance of receiving newer inventory increases (decreases) in the size of orders.   

To better understand the source of non-convexity after including the endogenous shelf-life uncertainty, we next consider a single-period problem with continuous demand and probability density (cumulative distribution) function $g(\cdot)$ ($G(\cdot)$). For simplicity, we fix the fraction of orders received with different remaining ages to the mean values in the \revision{multinomial} distribution, i.e., assume that $p_k(z)z$ orders have remaining shelf-life $k\in\{1,2,3\}$. We have,
\begin{align}
    v^z(x_2, x_1) = \Exp \Big[h\big(x_2 +x_1 + z -D\big)^+ +l\big(D-x_2 -x_1 - z\big)^+ +\theta\big(x_1+p_1(z)z-D\big)^+\Big],
\end{align}
which can be rewritten as,
\begin{align}
    v^z(x_2, x_1) = h \int_0^{x_2 +x_1+z} \big(x_2 +x_1 + z -u\big) g(u)du &+l \int_{x_2 +x_1+z}^\infty \big(u-x_2 -x_1 - z\big)g(u)du \\ &+\theta \int_0^{x_1+p_1(z)z}\big(x_1+p_1(z)z-u\big)g(u)du. \nonumber
\end{align}
Using the Leibniz integral rule, the first and second \minorrevision{derivatives} with respect to $z$ are,
\begin{align}
    \frac{d v^z(x_2,x_1)}{d z} &= (h+l) G(x_2+x_1+z) - l + \theta \Big(\big(p'_1(z)z + p_1(z)\big) G(x_1+p_1(z)z)\Big), \\ \label{eq:deriv}
    \frac{d^2 v^z(x_2, x_1)}{dz^2} &= (h+l) g(x_2 +x_1+z) + \theta \big(p''_1(z)z + 2p'_1(z)) G(x_1+p_1(z)z)\big) \\ \nonumber
    &\hspace{3.2cm}+ \theta \big(p'_1(z)z + p_1(z))^2 g(x_1+p_1(z)z)\big).
\end{align}
The first and last term in \eqref{eq:deriv} are positive but there is no guarantee for the middle term and hence the second derivative to remain positive. Therefore, the expected cost is not in general jointly convex in the ordering decision and state variables and hence the value function  after minimization is also not in general convex. We note that with exogenous probabilities the second term is zero and hence one can show convexity for this specific example. However, it is not in general straightforward to prove convexity with random shelf-lives (see, \citealt{nahmias1977ordering}).

\section{Simulation-Based ADP} \label{sec:adp}
\revision{In the absence of structural properties, the large state-space of our problem makes using exact approaches for $m>3$ computationally intractable. For instance, for a problem with $m=5$, $\gamma=7$, and assuming maximum $20$ units for each $x_i, \, i \in \{1, \dots, 4\}$, we need to compute and store $v(\tau, \textbf{x})$ for $7\times21^4 = 1,361,367$ states in \emph{each iteration} of the value iteration algorithm.} 

To address the computational difficulties when dealing with a large state-space, we approximate the value function for each $\tau \in\{0, \dots, \gamma-1\}$ with a linear combination of basis functions. That is,
\begin{equation}
    \tilde{v}(\tau, \textbf{x}) = \sum_{i=1}^k \beta_{\tau,i}\phi_i(\tau, \textbf{x}), \quad \forall \textbf{x} \in \mathcal{X}, \label{approx}
\end{equation} 
where $\boldsymbol{\phi}(\tau, \textbf{x})=(\phi_1(\tau, \textbf{x}), \dots, \phi_k(\tau, \textbf{x}))$ is a vector of $k$ basis functions, and $\boldsymbol{\beta}_\tau = (\beta_{\tau,1}, \dots, \beta_{\tau,k})$ is the associated coefficient vector. We assume that the coefficient vector can vary but the structure and number of basis functions are the same in each time index $\tau$ within the demand period. 

\minorrevision{With an arbitrary accurate approximation $\tilde{v}$ for the true value function $v$, one can get arbitrary close to the optimal policy through greedy minimization of the right-hand-side of \eqref{eq:opteq} after replacing $v$ with $\tilde{v}$.} To formalize, denote the maximum norm by $\norm{\cdot}$ and let $\tilde{\mu}$ denote the greedy policy obtained based on $\tilde{v}$. We have $\norm{v^{\tilde{\mu}} -v}\leq 2 \alpha \norm{\tilde{v}-v} / (1-\alpha)$ (\citealt{BertNeuro}, Proposition 6.1.). Therefore, owing to the bounded state-space, there exists an $\epsilon_0>0$ such that if $\norm{\tilde{v}-v}<\epsilon_0$ then $\tilde{\mu}$ is the optimal policy. The key is therefore to select an appropriate class of basis functions and tune the corresponding parameters such that $\tilde{v} \approx v$. Unfortunately, there is no universal approach for selecting the basis functions and the appropriate choice highly depends on the specific problem. In the following section, we use the numerical observations made in Section \ref{sub:structure} to motivate a class of basis function candidates for our problem.

\subsection{Basis Functions} \label{basisfun}
For each $\tau \in\{0, \dots, \gamma-1\}$, we propose the following decomposition of the value function into $k$ terms:
\begin{equation}
    \tilde{v}(\tau, \textbf{x}) = \beta_{\tau,1} + \beta_{\tau,2} v_1(\tau, \sum_{i=1}^{m-1}x_i) + v_2(\tau, \textbf{x}), \quad \forall \textbf{x} \in \mathcal{X}. \label{BF_approx}
\end{equation}
\revision{The first basis function is assumed to be $\phi_1(\tau, \textbf{x}) = 1$. The second basis function $\phi_2(\tau, \textbf{x}) = v_1(\tau, \sum_{i=1}^{m-1}x_i)$ is the value function of the corresponding nonperishable inventory problem (i.e., assuming $\theta=0$) included to approximately capture the effect of fixed costs. \minorrevision{(Note that obtaining $v_1$ requires solving a one-dimensional MDP for each $\tau$}.) The last $k-2$ terms  $v_2(\tau, \textbf{x}) = \sum_{i=3}^k \beta_{\tau,i}\phi_i(\tau, \textbf{x})$ are included to capture the nonconvex structure of the value function due to shelf-life uncertainty and approximate the expiration cost not included in $v_1$. Starting with linear functions and gradually increasing the order, we consider the following four polynomial candidates for $v_2$}: 
\begin{enumerate}
\item Choice 1:  $x_j , \quad j=1, \dots, m-1,$

\item Choice 2:  $x_j,  \, x_j^2, \quad j=1, \dots, m-1,$

\item Choice 3:  $x_j, \, x_j^2, \, x_j^3, \quad  j=1, \dots, m-1,$ 

\item Choice 4:  $x_j, \, x_j^2,  \, x_jx_{j'}, \quad j,j' = 1, \dots, m-1, \,  j\ne j.$
\end{enumerate}
Although including higher order basis functions could reduce the function approximation error, it requires tuning more parameters and hence adds to the computational complexity of the algorithm. Therefore, we prefer not to include them unless they significantly improve the performance. In Section \ref{perfbasisfun}, we examine the performance of the above choices using small problem instances for which we can exactly solve the problem. In Section \ref{app:basisfunctions} of the Online Appendix, we provide an analytical justification for excluding higher-order basis functions for a special case of the problem.

\subsection{Approximate Policy Iteration Algorithm} \label{subsec:api}
The steps of the approximate policy iteration algorithm are similar to the exact algorithm except that we use simulation to implement the policy improvement and policy evaluation steps (see, e.g., \citealt{Bert05}). Specifically, we start with coefficient vectors $\boldsymbol{\beta}^0_\tau = 0$ for all $\tau \in\{0, \dots, \gamma-1\}$, and iteratively update the coefficient vectors according to the following steps \revision{(see Online Appendix \ref{app:algorithm} for the complete algorithm)}:
\begin{enumerate}
    \item \emph{Policy improvement}: we start from a random initial period and inventory state and make ordering decisions \revision{in iteration $n$} using,
    \begin{align}
    \mu^n(\tau, \textbf{x}) & = \argmin_{z \in \mathcal{Y}} \Exp \bigg[C\big(\textbf{x}, \, z\big) + \alpha \sum_{i=1}^k\beta^{n-1}_{\tau+1,i}\phi_{i}\big(s(\tau, \textbf{x}, Y_m^z, \dots, Y_1^z, D_\tau)\big)\bigg], \label{greedy}
    \end{align} 
    and by following a simulated sample path of shelf-life and demand generated for a sufficiently long horizon. For example, we generate samples for 100 periods (see Section \ref{subsec:perf} for more details) in our numerical experiments. This allows us to obtain the ordering policy $\mu^n$ for each state $\textbf{x} \in \mathcal{X}$ \emph{visited} in each $\tau \in\{0, \dots, \gamma-1\}$ during the simulation. As shelf-life samples depend on the order size, we generate samples for all possible order sizes after solving \eqref{greedy} by exploiting Common Random Numbers (CRN) (see, e.g., \citealt{glasserman1992some}).
    We repeat $Q$ replications of the above procedure to create a set of inventory states visited in each $\tau$ under policy $\mu^n$, denoted by  $\big \{\textbf{x}_{\tau,\psi}^n:\psi=1,\dots,\Psi_\tau(Q)\big\}$, where $\Psi_\tau(Q)$ is the total number of states visited with $Q$ replications and demand index $\tau$.
    \item \emph{Policy evaluation}:  
    we denote by $c_{\tau,\psi}^n$ the Monte Carlo estimate of the expected long-run discounted cost incurred by starting from state $\textbf{x}_{\tau,\psi}^n$ in period $\tau$ and following policy $\mu^n$ for the long but finite horizon. We then solve the following least squares problems and update the coefficient vectors:
    \begin{align}
            &\boldsymbol{\beta}_\tau^{*} = \argmin_{\boldsymbol{\beta}_\tau} \sum_{\psi=1}^{\Psi_\tau(Q)}\Big[c_{\tau,\psi}^n-\boldsymbol{\beta}_\tau \cdot \boldsymbol{\phi}(\tau, \textbf{x}_{\tau,\psi}^n) \Big]^2 , \quad \quad \forall \tau \in \{0,\dots,\gamma-1\}, \label{ls} \\
            &\boldsymbol{\beta}_\tau^{n} = (1-\lambda_n) \times \boldsymbol{\beta}_\tau^{n-1} + \lambda_n \times \boldsymbol{\beta}_\tau^{*}, \quad \quad \quad \forall \tau \in \{0,\dots,\gamma-1\}, \label{upd}
    \end{align}
    where we use $\lambda_n = 1/n$ to smooth the coefficient vectors in the $n$th iteration.
    
    \item We increase $n$ by one and go to Step 1 until $n$ reaches a predetermined total number of iterations before the algorithm terminates.
\end{enumerate}

A few comments are in order. \revision{First, in contrast to the exact policy iteration algorithm which converges to the optimal policy, there is no guarantee that the approximate policy iteration algorithm converges. Instead, the expected cost of the approximate policy oscillates in a neighborhood of the value function. To elaborate, let $\{\tilde{v}^{n}\}$ denote the sequence of approximate value functions obtained by the algorithm, and let $v^{\mu_n}$ denote the expected cost under the policy obtained in iteration $n$ of the algorithm. Assume that $\norm{\tilde{v}^n (\tau,\textbf{x}) - v^{\mu_n}(\tau,\textbf{x})} \leq \epsilon$ and $\norm{(\mathcal{T} \tilde{v}^n) (\tau,\textbf{x}) - (\mathcal{T}_{\mu^{n+1}} \tilde{v}^n)(\tau,\textbf{x})} \leq \delta$ for $n=0,1,\ldots$. Then, $\epsilon$ is a worst-case bound on the error in policy evaluation, including both the error due to simulation and function approximation, and $\delta$ is a bound on the error incurred in updating the policy. We have (\citealt{BertNeuro}, Prop. 6.2):
\begin{equation}\label{eq:bound}
    \limsup_{n \to \infty} \: \max_{(\tau,\textbf{x})} |v^{\mu_n}(\tau,\textbf{x})-v(\tau,\textbf{x})| \leq \frac{\delta+2\alpha \epsilon}{(1-\alpha)^2}.
\end{equation}
Therefore, with sufficiently large number of iterations the value function will be in a neighborhood proportional to a constant multiple of $\epsilon$ and $\delta$. Empirically, it has been observed that the algorithm reaches this neighborhood in a few iterations and we also observe this for our problem in the numerical experiments of Section \ref{sec:numerics}. Further, while \eqref{eq:bound} is a worst-case bound and cannot be computed without knowing $\epsilon$ and $\delta$, it is constructive for choosing the parameters of the algorithm. Specifically, we can control the simulation error component (through the number of replications) in $\epsilon$ and bring it close to zero. Similarly, by running sufficient simulation replications and leveraging CRN we can make $\delta$ effectively zero, such that the error is limited to that of function approximation.} 

Second, the minimization in $\eqref{greedy}$ involves computing an expectation with respect to a multi-dimentional random variable, which becomes computationally difficult for realistic instances of the problem. For such cases, we use Monte Carlo simulation to find the ordering decision for a given state using the approximate value function. In particular, we estimate the expected cost for each possible order size and choose the value that minimizes the sample average estimate. This approach is subject to sampling error, but by utilizing CRN we reduce the variance of the estimates when comparing policies with different order sizes and increase the probability of correctly selecting the best ordering quantity \citep{nelson2021foundations}. To implement CRN, we note that different order sizes require a different number of pseudorandom numbers. For example, with shelf-life $m$ and \minorrevision{maximum order size $\bar{z}$, we generate $\bar{z}$ independent samples of the uniform distribution in each replication and use the first $z$ samples to construct a multinomial sample $(y_1^z, \dots, y_m^z)$ for each $z\in \{1, \dots, \bar{z}\}$ using the inversion method.}

\revision{Finally, the policy improvement step (Step 1) can be completed in a few seconds for realistic instances of the problem (see Online Appendix \ref{app:algorithm} for details). Hence, after tuning the coefficients based on the observed states during the \emph{offline} phase of the algorithm, ordering decisions can be computed \emph{online} from any starting state. The online calculation eliminates the need to compute and store the policy for the entire state-space. It is also practically appealing, because ordering decisions need to be made at most on a daily basis.} 

\subsection{Information Relaxation Lower-Bound} \label{subsec:lb}
In this section, we propose a partial information relaxation approach to obtain a lower-bound on the value function of our infinite-horizon problem (\citealt{brown2017information}). The lower-bound allows us to quantify the performance of the ADP algorithm in cases where computing the exact solution is not feasible. \minorrevision{In addition, we use it to construct another benchmark for the ADP approach in the case study of Section \ref{sec:casestudy}.} In Section \ref{subsubsec:perfm3}, we examine the tightness of the lower-bound for small instances of the problem with $m=3$. We use the lower-bound to evaluate the ADP performance in the absence of the optimal policy for cases with $m=5$ (see Section \ref{sec:adperfm5}). For larger problems with $m=8$, computing the lower-bound also becomes computationally expensive. Hence, we measure the ADP performance compared to the Myopic solution \minorrevision{(see Section \ref{sec:adperfm8} of the Online Appendix)}.

The lower-bound is obtained by partially relaxing the nonanticipative requirement of the ordering policies. More specifically, consider an alternative problem where the decision maker has knowledge of the uncertainty in the remaining shelf-life of ordered units, while the demand is still uncertain. Denote by $\boldsymbol{\omega} \equiv \{\omega_t\}_{t\geq 0}$ a sequence of the realizations of iid random variables that determine the uncertainty in the remaining shelf-life over the infinite horizon. Each realization consists of $M$ uniform random variables used to generate remaining shelf-life samples for each $z \in \mathcal{Y}$ according to the \revision{multinomial} distribution in each period. Since in practice we cannot generate an infinitely long sequence, \cite{brown2017information} propose to use an equivalent formulation of the original DP in which there is no discounting but the length of the horizon is random and follows a \revision{geometric} distribution with success probability $1-\alpha$. In this formulation, we can equivalently express the expected cost associated with any ordering policy $\mu$ as
\begin{align}
    v^{\mu}(\tau, \textbf{x}) = \Exp \Bigg[\sum_{t=0}^{T} C\big(\textbf{X}^\mu_t, \, \mu (\tau_t,\, \textbf{X}^\mu_t)\big) \mid (\tau_0, \textbf{X}^\mu_0) = (\tau,\textbf{x}) \Bigg],
\end{align}
where $T$ is Geometric with parameter $1-\alpha$. The transition probabilities are also modified such that with probability $1-\alpha$ the next state can be an absorbing state. 
We can now generate $\boldsymbol{\omega}$ given a sample of the length of the horizon. We define by $\hat{v}(\tau,\textbf{x},\boldsymbol{\omega})$ the value function of the relaxed problem where the demand is random but the length of the horizon and remaining shelf-life for all possible order sizes in all periods is known. 
It then follows (\citealt{brown2017information}) that the value function of the original problem satisfies
$
    v(\tau,\textbf{x}) \geq \Exp [\hat{v}(\tau, \textbf{x}, \boldsymbol{\omega})], 
$
where the expected value is with respect to all possible realizations of the length of the horizon and the corresponding  $\boldsymbol{\omega}$, and can be estimated via Monte Carlo simulation. 

\section{Numerical Study} \label{sec:numerics}
In this section, we conduct an extensive numerical study to first determine appropriate choices of basis functions and then evaluate the performance of the ADP approach. 
In addition, we leverage the ADP in the absence of the optimal policy to examine the impact of ignoring shelf-life uncertainty and its dependence on order size and identify parameter regimes where ignoring them leads to costlier ordering decisions.  \minorrevision{All code and data for the experiments can be
found in the accompanying GitHub repository (\citealt{PLTADP}).}

\subsection{Experimental Setup} 
We use platelet inventory data from HGH (see Section \ref{sec:casestudy} for details) to inform the selection of models for demand and shelf-life uncertainty in our experiments. \revision{See Appendix \ref{ap:data} for evidence of the validity of the parametric choices.}

\subsubsection{Demand.} \label{subsub:demand} We assume that the demand is periodic with $\gamma=7$ and follows a truncated \revision{negative binomial} distribution, i.e.,
\begin{align}\label{eq:NBinom}
    &\prob(D_\tau = x) = \frac{\Gamma(x+n_\tau)}{\Gamma(n_\tau)x!} q_\tau^{n_\tau} (1-q_\tau)^x, \quad \forall x \in\{0, 1 , \dots, M-1\}, \, \tau \in \{0, \dots, 6\}, \\ \nonumber
    &\prob(D_\tau = M) = 1 - \sum_{i=0}^{M-1} \prob(D_\tau = i), 
\end{align}
where $x$ represents the number of failures in a sequence of Bernoulli trials before a target number of successes, denoted by $n_\tau$, is achieved and $M$ is the truncation value.  The probability of success in each trial can be written as $q_\tau=n_\tau/(n_\tau+\delta_\tau)$, where $\delta_\tau$ is the expected value of the fitted distribution for period $\tau$. We use our data to estimate the parameters as reported in Table \ref{tab:demparam} and set $M=20$ based on the maximum historical demand observed in the data. In Online Appendix \ref{app:sensdem}, we present the means and variances after truncation and conduct sensitivity analysis for demand.

\begin{table}
\centering
\caption{Estimated parameters of the \revision{negative binomial} distribution for different days of the week.}
\label{tab:demparam}
\begin{tabular}{|c|ccccccc|}
\hline
 $\tau$ & 0 & 1 & 2 & 3 & 4 & 5 & 6 \\
 \hline
 $n_\tau$ & 3.5 & 11.0 & 7.2 & 11.1 & 5.9 & 5.5 & 2.2 \\
 \hline
 $\delta_\tau$ & 5.7 & 6.9 & 6.5 & 6.2 & 5.8 & 3.3 & 3.4 \\
 \hline 
\end{tabular}
\end{table}

\subsubsection{Shelf-Life.} \label{subsub:shelf-life} As discussed in Section \ref{sec:mdp}, we assume the following \revision{multinomial} logit form, 
\begin{equation}\label{eq:MLogit}
    \log\bigg(\frac{p_k(z)}{p_1(z)}\bigg) = c_0^k + c_1^k z , \quad \forall k\in\{2,3, \dots, m\},
\end{equation}
which requires us to determine $2\times (m-1)$ parameters to characterize an endogenous shelf-life uncertainty. The values of $c_0^k$ characterize a base exogenous distribution (when $z=0$) and the values of $c_1^k$ change the base exogenous distribution for every positive order size. 

\begin{figure}%
    \centering
    \captionsetup[subfigure]{labelformat=empty}
    \subfloat[(A) Order size of $z=0$.]{\includegraphics[width=0.5\textwidth]{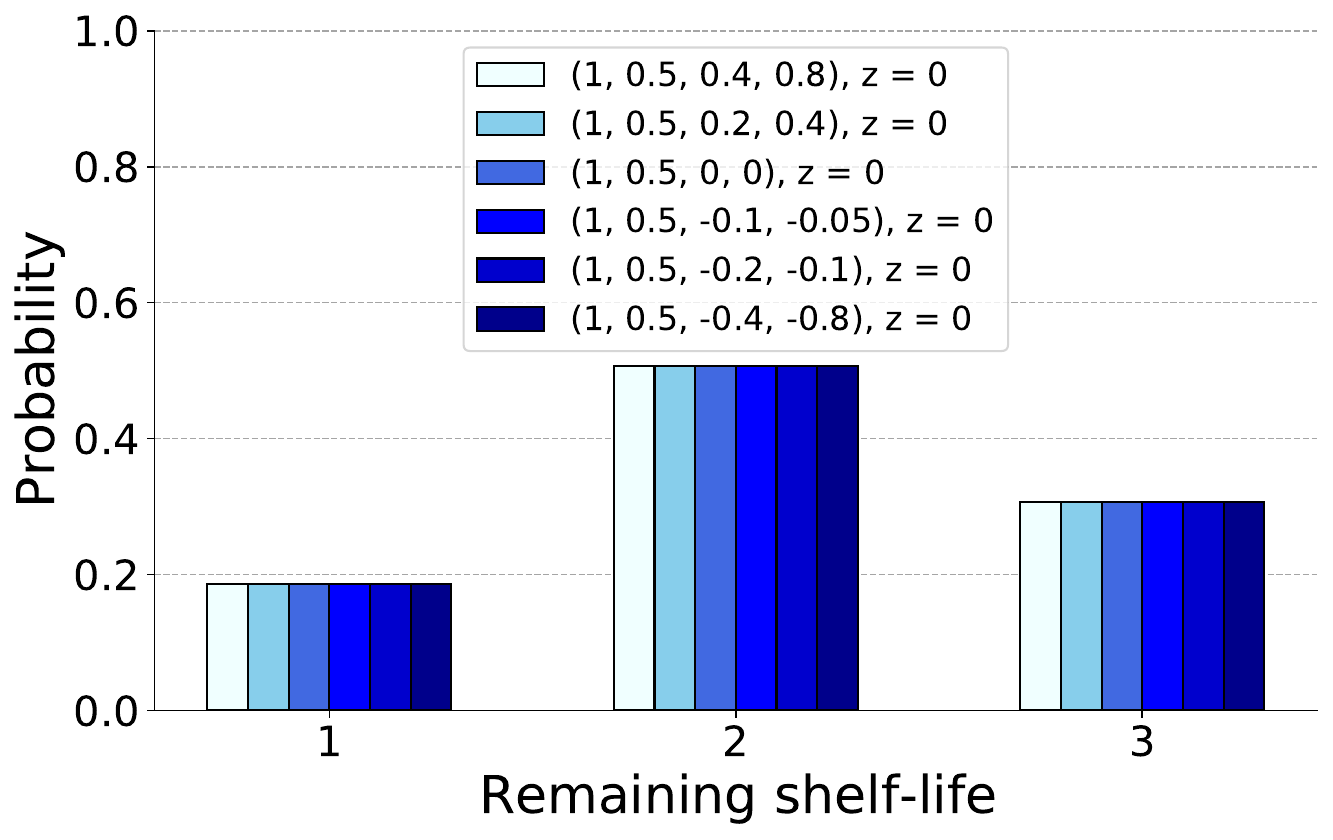}}%
    \subfloat[(B) Order size of $z=5$.]{\includegraphics[width=0.48\textwidth]{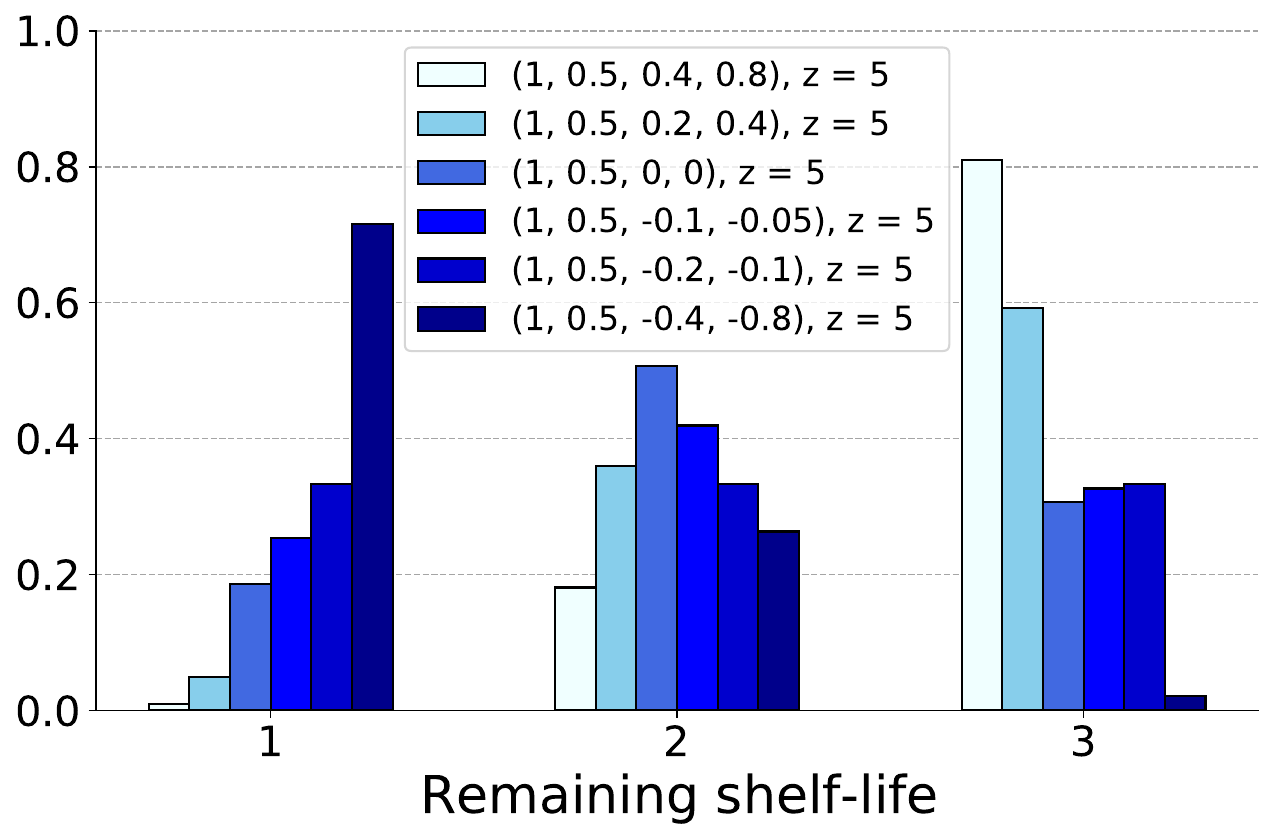} }\\%
    \subfloat[(D) Order size of $z=15$.]{\includegraphics[width=0.5\textwidth]{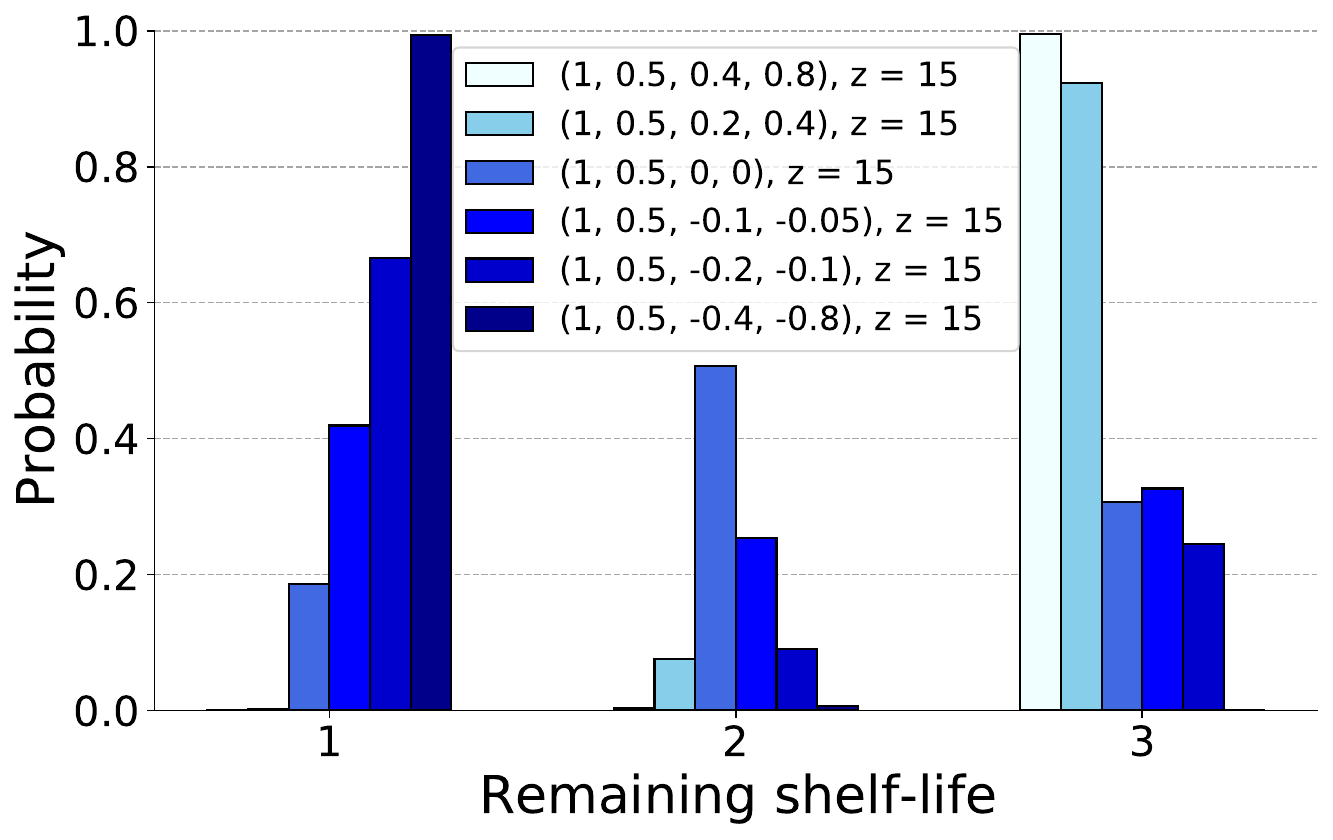} }
    \subfloat[(C) Order size of $z=10$.]{\includegraphics[width=0.48\textwidth]{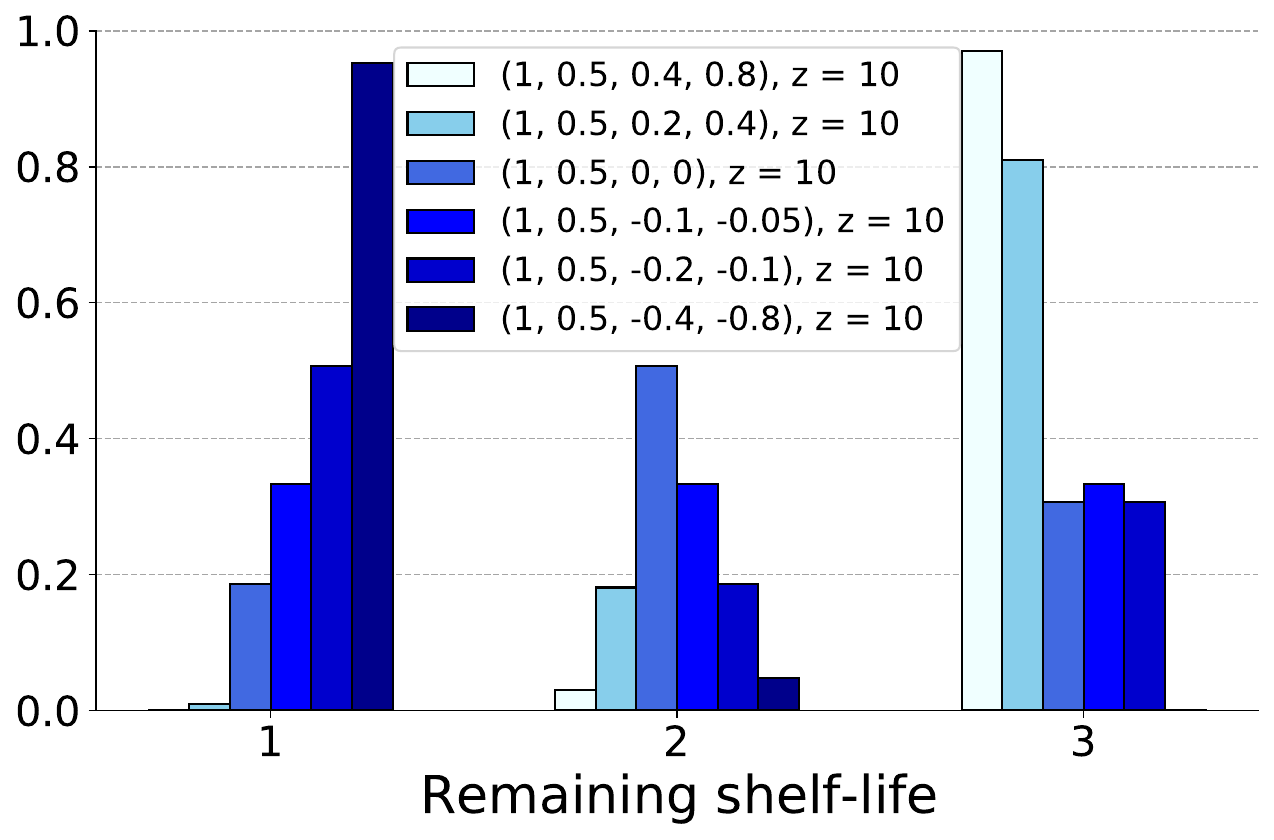} }
    \caption{Effect of order size coefficients in the \revision{multinomial} \revision{logistic} model on the remaining shelf-life distribution for  order sizes of 0, 5, 10, and 15. The larger absolute magnitude of positive (negative) coefficients decreases (increases) the probability of receiving units with remaining shelf-life of one day more rapidly.}%
    \label{fig:shelf-life}%
\end{figure}

For problem instances with $m=3$, we assume $c_0^2 = 1$ and $c_0^3 = 0.5$ resulting in the base exogenous distribution: $p_3 = 0.3, \, p_2 = 0.5, \, p_1 = 0.2$, and consider different cases for $(c_1^2, c_1^3)\in \{(0.4, 0.8), (0.2, 0.4), (0, 0), (-0.1, -0.05), (-0.2, -0.1), (-0.4, -0.8) \}$. Based on \eqref{eq:MLogit}, for cases with positive (negative) coefficients, a unit increase in order size is associated with $c_1^2$ and $c_1^3$ increase (decrease) in the log-odd ratio of receiving units with remaining shelf-life of two and three versus one, respectively. As a result, the probability of receiving units with remaining shelf-life of one decreases (increases) in order size until it converges to zero (one) for large enough order sizes. Also, the speed of convergence increases in the absolute values of the coefficients. Figure \ref{fig:shelf-life} illustrates the impact of parameters, i.e., $(c_0^2, \, c_0^3, \, c_1^2, \, c_1^3)$, on the remaining shelf-life distribution for different order sizes $z \in \{0, 5, 10, 15\}$.

For problem instances with $m=5$, we use platelet inventory data at HGH to fit a \revision{multinomial} logistic regression and use it as the baseline, that is $c_0^2 = 1.9, \, c_0^3=c_0^4=3.1, \, c_0^5=2.5$ for the intercepts and $c_1^2 = c_1^4 = -0.03, \, c_1^3=-0.06, \, c_1^5=-0.09$ for the coefficients of the order size \revision{(see Appendix \ref{ap:data})}. The negative coefficients of the order size imply that the probability of receiving units with remaining shelf-life of one increases in order size. Intuitively, the supplier prefers to satisfy larger orders with older units as larger orders indicate a greater utilization of inventory and hence a lower likelihood of wastage at the hospital. However, the small absolute values of coefficients indicate slow increases in the probability of receiving units with remaining shelf-life of one. For instance, we have $p_1(1) = 0.016$, $p_1(20) = 0.039$, $p_1(60) = 0.155$, and $p_1(100) = 0.401$.   

We further consider an exogenous case by fitting another \revision{multinomial} logistic regression model which only includes intercepts, leading to estimates $c_0^2 = c_0^5 = 1.6, \, c_0^3= 2.6, \, c_0^4=2.8$. In addition, we consider three alternative scenarios for the endogenous uncertainty by changing the magnitude of the fitted order size coefficients in the baseline endogenous scenario (see Table \ref{tab:scen_m58} for a summary).

\begin{table}
\caption{Different parameter settings of the \revision{multinomial} logit model for larger values of $m$.}
\label{tab:scen_m58}
\centering
    \resizebox{\textwidth}{!}{
    \begin{tabular}{llccccccccccccccc}
    \hline
    \multirow{2}{*}{$m$} & \multirow{2}{*}{Scenario} & 
    \multicolumn{7}{c}{Intercepts} & & \multicolumn{7}{c}{Order size coefficients} \\
    \cline{3-9} \cline{11-17}
     &  &  $c_0^2$ & $c_0^3$ & $c_0^4$ & $c_0^5$ & $c_0^6$ & $c_0^7$ & $c_0^8$ & &  $c_1^2$ & $c_1^3$ & $c_1^4$ & $c_1^5$ & $c_1^6$ & $c_1^7$ & $c_1^8$\\
     \hline
     \multirow{5}{*}{5} & Exogenous & 1.6 & 2.6 & 2.8 & 1.6 & & &  & &  &  &  &  &  &  &  \\ 
     & Endogenous & 1.9 & 3.1 & 3.1 & 2.5 & &  & &  & -0.03 & -0.06  & -0.03 & -0.09 \\
    & Sensitivity 1 & 1.9 & 3.1 & 3.1 & 2.5& &  & &  & -0.03 & -0.06  & -0.08 & -0.09\\
    &Sensitivity 2 & 1.9 & 3.1 & 3.1 & 2.5  & &  & &  & -0.05  & -0.1 & -0.15 & -0.2 \\
    & Sensitivity 3 & 1.9 & 3.1 & 3.1 & 2.5   & &  & &  & -0.1 & -0.2 & -0.3 & -0.4\\
    \hline
     \multirow{5}{*}{8} & Exogenous  & 0.8 & 1.4 & 1.9 & 2.3 & 1.7 & 1.2 & 0.8 & & 0.0 & 0.0 & 0.0 & 0.0 & 0.0 & 0.0 & 0.0 \\
     & Endogenous &  0.8 & 1.4 & 1.9 & 2.3 & 1.7 & 1.2 & 0.8 & & -0.03 & -0.04 & -0.05 & -0.06 & -0.07 & -0.08 & -0.09 \\
    & Sensitivity 1 &  0.8 & 1.4 & 1.9 & 2.3 & 1.7 & 1.2 & 0.8 & & -0.06 & -0.08 & -0.1 & -0.12 & -0.14 & -0.16 & -0.18 \\
    &Sensitivity 2 &  0.8 & 1.4 & 1.9 & 2.3 & 1.7 & 1.2 & 0.8 & & -0.12 & -0.16 & -0.2 & -0.24 & -0.28 & -0.32 & -0.36  \\
    & Sensitivity 3 &  0.8 & 1.4 & 1.9 & 2.3 & 1.7 & 1.2 & 0.8 & & -0.24 & -0.32 & -0.4 & -0.48 & -0.56 & -0.64 & -0.72 \\
    \hline
    \end{tabular}}
\end{table}

\subsubsection{Discount Factor and Cost Parameters.} We use a discount factor of $\alpha =0.95$, which is close to one and hence suitable for long-term planning often assumed in the literature on perishable inventory systems (see, e.g., \citealt{chao2015approximation}). Intuitively, due to the perishability, the future costs of orders occur within the maximum $m$ next periods and hence the results are less sensitive to the choice of the discount factor. \revision{For the cost parameters, we normalize the unit cost of holding $h$ to 1 and set the unit cost of shortage $l$ at 20. We then consider one small (relative to shortage cost) and one large fixed ordering cost $\kappa \in\{10, 100\}$ as well as three different ratios of shortage to wastage costs, i.e., $l/\theta \in \{4, 1, 1/4\}$, leading in total to six different cost combinations. We note that in the case of platelet inventory management, shortage is considered to be costlier than expiration, i.e., $l/\theta \geq 1$ (e.g., \citealt{zhou2011inventory,civelek2015blood}).}   

\subsection{Comparing Basis Functions} \label{perfbasisfun}

\begin{figure}
    \centering
    \includegraphics[width=1.0 \textwidth]{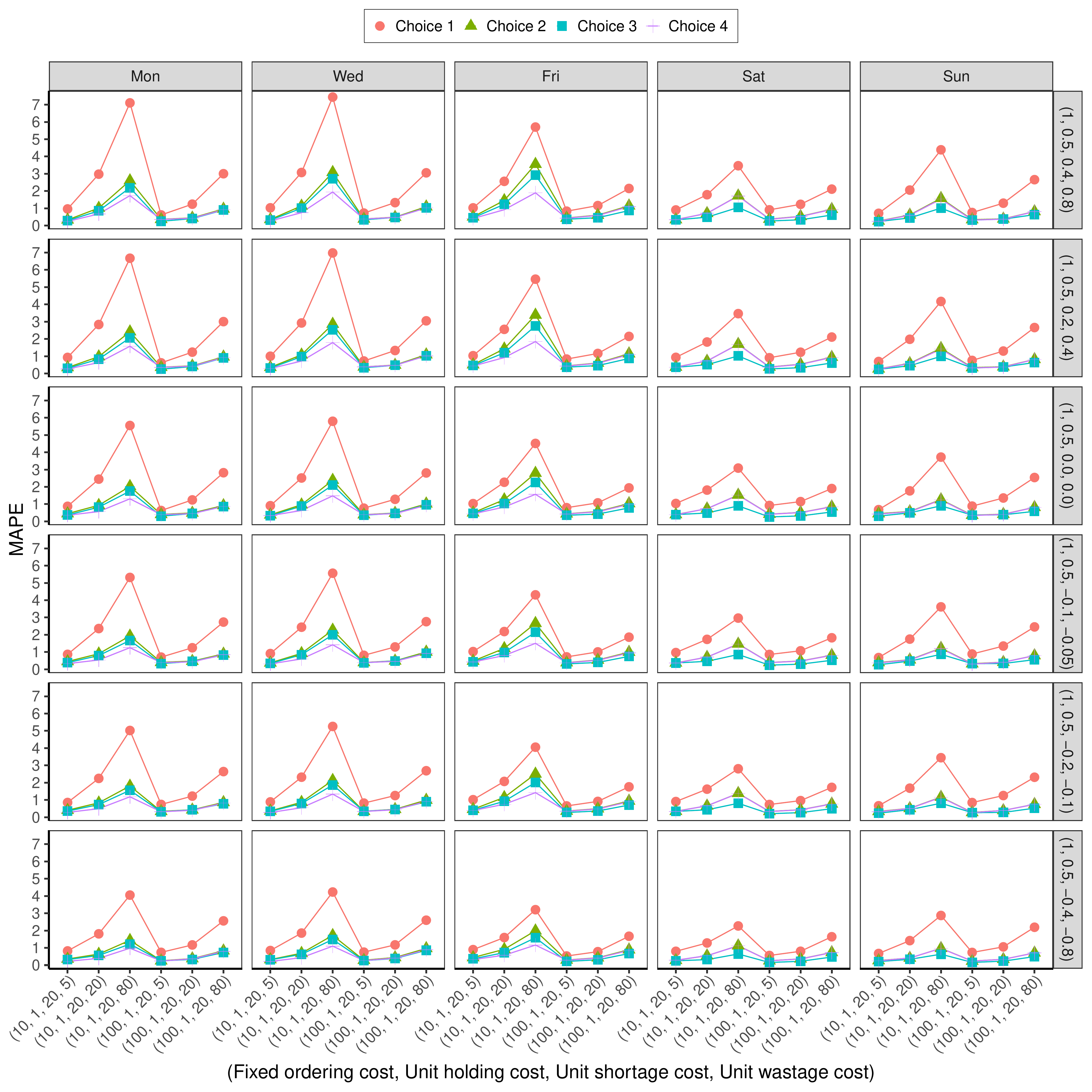}
    \caption{Performance of candidate basis functions with respect to their MAPE calculated using the optimal value function.}
    \label{fig:BF_LQC}
\end{figure}

We use small instances of the problem with $m=3$ (for which the exact value function can be calculated) to examine the choice of basis functions. 
In Figure \ref{fig:BF_LQC}, we examine the performance of each choice of basis function candidates as proposed in Section \ref{basisfun} with respect to \revision{mean absolute percentage error (MAPE)} of the resulting approximation.
Using Choice 1 of basis functions, we observe a large \revision{MAPE} when the unit cost of wastage increases. Including the quadratic terms in the approximation (Choice 2) can significantly improve the results in majority of cases.  
The value of considering cubic or interaction terms using Choice 3 or Choice 4 of basis functions depends on the cost parameters. In particular, when the unit cost of wastage is greater than shortage, the approximation using Choice 3 and 4 provides a better fit than Choice 2, with Choice 3 performing better when the fixed ordering cost is relatively large. We further examine and demonstrate these observations in Section \ref{subsubsec:perfm3} where we examine the sub-optimality of the ADP policy with different choices of basis functions.
\subsection{Performance of the ADP Approach} \label{subsec:perf}
In this section, we evaluate the performance of the ADP for different shelf-life values of $m \in \{3, 5\}$. \minorrevision{The case with $m=8$ is relegated to Online Appendix \ref{sec:adperfm8}.} For longer shelf-life values there is less expiration (unless supply is larger than demand which is not the case here) and so the problem becomes similar to one with infinite shelf-life (\citealt{sun2014quadratic}). For $m=3$ we compare the performance directly with that of the optimal policy; for $m=5$ with the lower-bound obtained from the partial information relaxation approach; and for $m=8$ with an (approximate) Myopic policy.  
We use the initial state of zero, i.e., $(\tau, \textbf{x}) = (0, 0)$ when estimating the expected total discounted cost over a finite but long horizon of 100 periods.

We initialize the ADP algorithm with $\boldsymbol{\beta}^0_\tau = 0$ for all $\tau \in\{0, \dots, \gamma-1\}$, which is equivalent to the Myopic policy that ignores future costs of orders, and use $Q=30$ replications each consisting of 100 periods. This results in at least 420 observations of the inventory state for each day of the week and satisfies a rule of thumb from the regression analysis requiring a minimum of 10 times the number of parameters, i.e., $2\times m$ for the quadratic value function approximation. 

\subsubsection{The Case with $m=3$.} \label{subsubsec:perfm3} 
\begin{figure}
    \centering
    \includegraphics[width= 1.0 \textwidth]{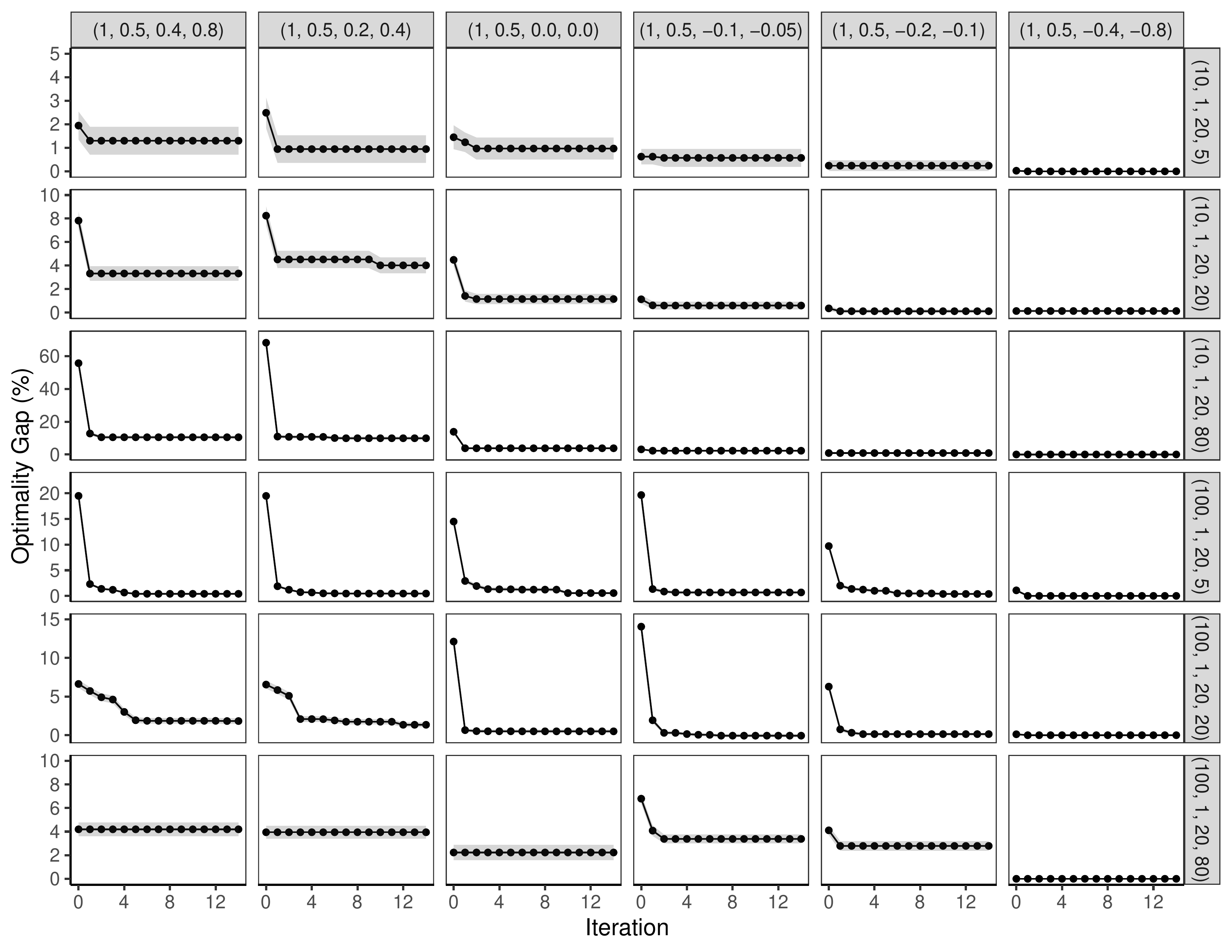}
    \caption{Relative optimality gap of the ADP policy. The black line presents the estimate of the expected optimality gap in each iteration and the gray area is the corresponding 95\% confidence interval. On average, the estimate of the expected optimality gap among all cases is 1.8\%, indicating \revision{an 80\%} reduction compared to 8.9\% of the initial Myopic solution at iteration zero.}
    \label{fig:ADPerf}
\end{figure}

We  examine the performance of approximate policies obtained using the ADP approach and Choice 2 of basis functions with respect to the resulting optimality gaps. In Figure \ref{fig:ADPerf}, we plot the estimate of the expected optimality gap together with its 95\% confidence interval in each iteration of the algorithm. On average, the estimate of the expected optimality gap among all tested scenarios is 1.8\%, which is 80\% less than the initial Myopic solution.

\begin{figure}
    \centering
    \includegraphics[width= 0.6 \textwidth]{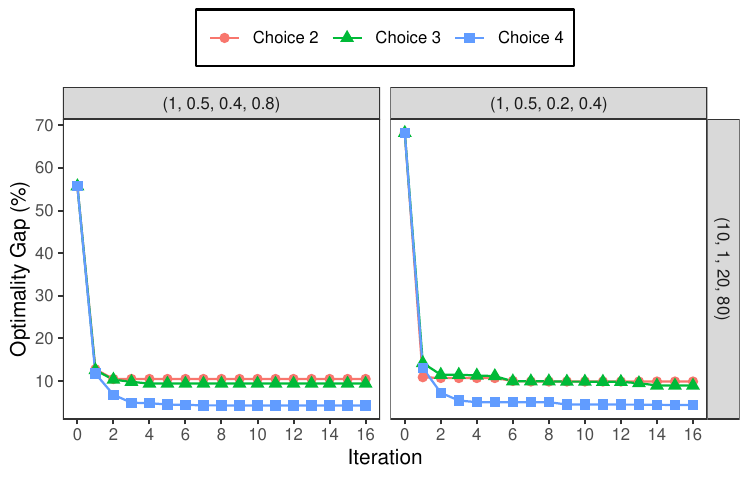}
    \caption{Performance improvement for cases with more than 5\% optimality gaps in Fig \eqref{fig:ADPerf}. The interaction term $x_1x_2$ can further improve the results compared to the cubic terms $x_1^3$ and $x_2^3$.}
    \label{fig:ADPerf-Choice234}
\end{figure}

More specifically, when the fixed ordering cost is small (i.e., $\kappa = 10$) and the coefficients of the order size in the \revision{multinomial} logistic model are negative (i.e., last three plots in the first three rows of Figure \ref{fig:ADPerf}), the initial Myopic policy performs close to optimal. Intuitively, this is because the chance of receiving units with remaining shelf-life of one is relatively high and increasing in order size. Hence, the problem is approximately similar to an independent single-period problem. However, for cases with positive coefficients of the order size, i.e., $(c_0^2, c_0^3, c_1^2, c_1^3) \in \{(1, 0.5, 0.4, 0.8), (1, 0.5, 0.2, 0.4)\}$, the estimate of the expected optimality gap for the Myopic policy respectively increases from 1.9\% and 2.5\% to 55.7\% and 68.2\% as the unit cost of expiration increases from 5 to 80. In contrast, the increase for the ADP policy is much less from 1.3\% and 0.9\% to 10.5\% and 9.9\%, respectively. Intuitively, this is because the chance of receiving fresher units with remaining shelf-life of two and three days is relatively high and increasing in order size, and hence future costs of orders become more important. In addition, we can improve the \revision{worst-case} performance of the ADP policies using Choice 3 and Choice 4 of basis functions. From Figure \ref{fig:ADPerf-Choice234}, we observe that including the interaction term in the approximation function \eqref{BF_approx} reduces the \revision{worst-case} optimality gap to 4.3\% and 4.4\%, while including the cubic terms only slightly improves the gaps to 9.4\% and 8.9\%. Therefore, Choice 4 of basis functions that includes the interaction term and requires one more parameter to tune in the algorithm, contributes more substantially to its performance and reduces the \revision{worst-case} optimality gap to below 5\%.

For the larger fixed ordering cost (i.e., $\kappa=100$), we observe that the estimate of the expected optimality gap for the Myopic policy is also considerable for negative coefficients of order size with smaller absolute values, i.e., cases where the speed of increase in the chance of receiving units with remaining shelf-life of one is slower in order size. In contrast to our previous observation, the expected optimality gap for the Myopic policy decreases as the unit cost of expiration increases. Intuitively, with a relatively larger unit cost of expiration which is still smaller than the fixed ordering cost, the Myopic policy tends to not order large amounts in each period which reduces future costs of orders. However, if the expiration cost continues to grow and eventually dominates the fixed ordering cost, the gap is expected to grow as in the first three rows of Figure \ref{fig:ADPerf}.

\begin{figure}
    \centering
    \includegraphics[width= 1.0 \textwidth]{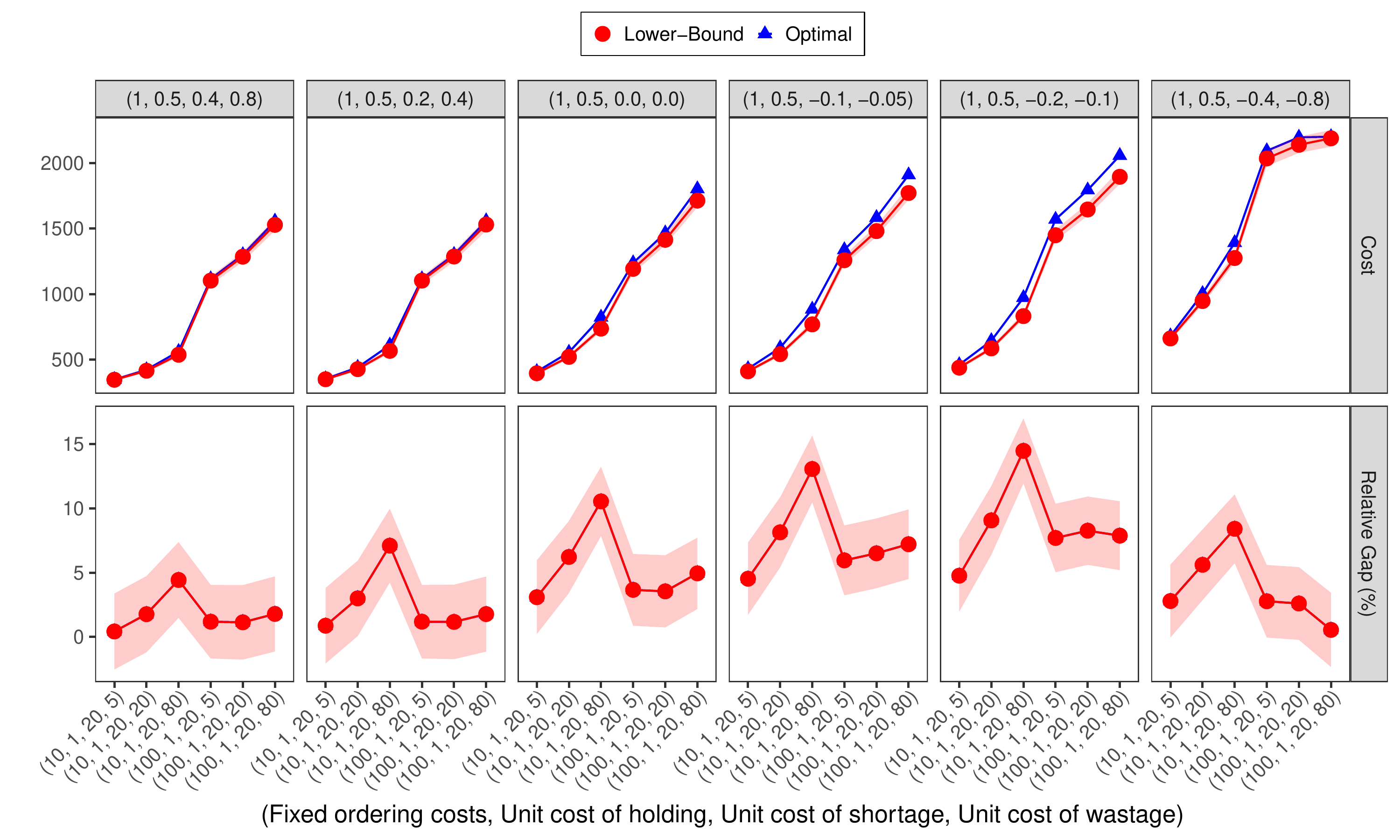}
    \caption{Quality of the lower-bound obtained using the imperfect information relaxation approach. On average, the relative gap between the lower-bound and optimal is 4.9\%.}
    \label{fig:lb}
\end{figure}

Finally, we examine the quality of the lower-bound described in Section \ref{subsec:lb} before using it as a benchmark for the case with $m=5$. We compare the sample average estimate of the lower-bound obtained using 4000 replications with the optimal value function at $(\tau, \textbf{x}) = (0, 0)$. Figure \ref{fig:lb} presents the absolute and relative gap between the lower-bound and the optimal cost for different parameter settings. When the coefficients of the order size in the multinomial logistic model are positive (i.e, the first two columns), the average and largest gap is 2.2\% and 7.1\%, but when the coefficients are negative (i.e., the last three columns), the average and largest gap is 6.7\% and 14.5\%, respectively. Further, the larger gaps occur for the smaller fixed ordering costs and higher unit costs of wastage.

\subsubsection{The Case with $m=5$.} \label{sec:adperfm5}
We next present the performance results for problem instances with a shelf-life of five days (see Table \ref{tab:scen_m58} for the shelf-life settings). 
 
\begin{figure}
    \centering
    \includegraphics[width= 0.85 \textwidth]{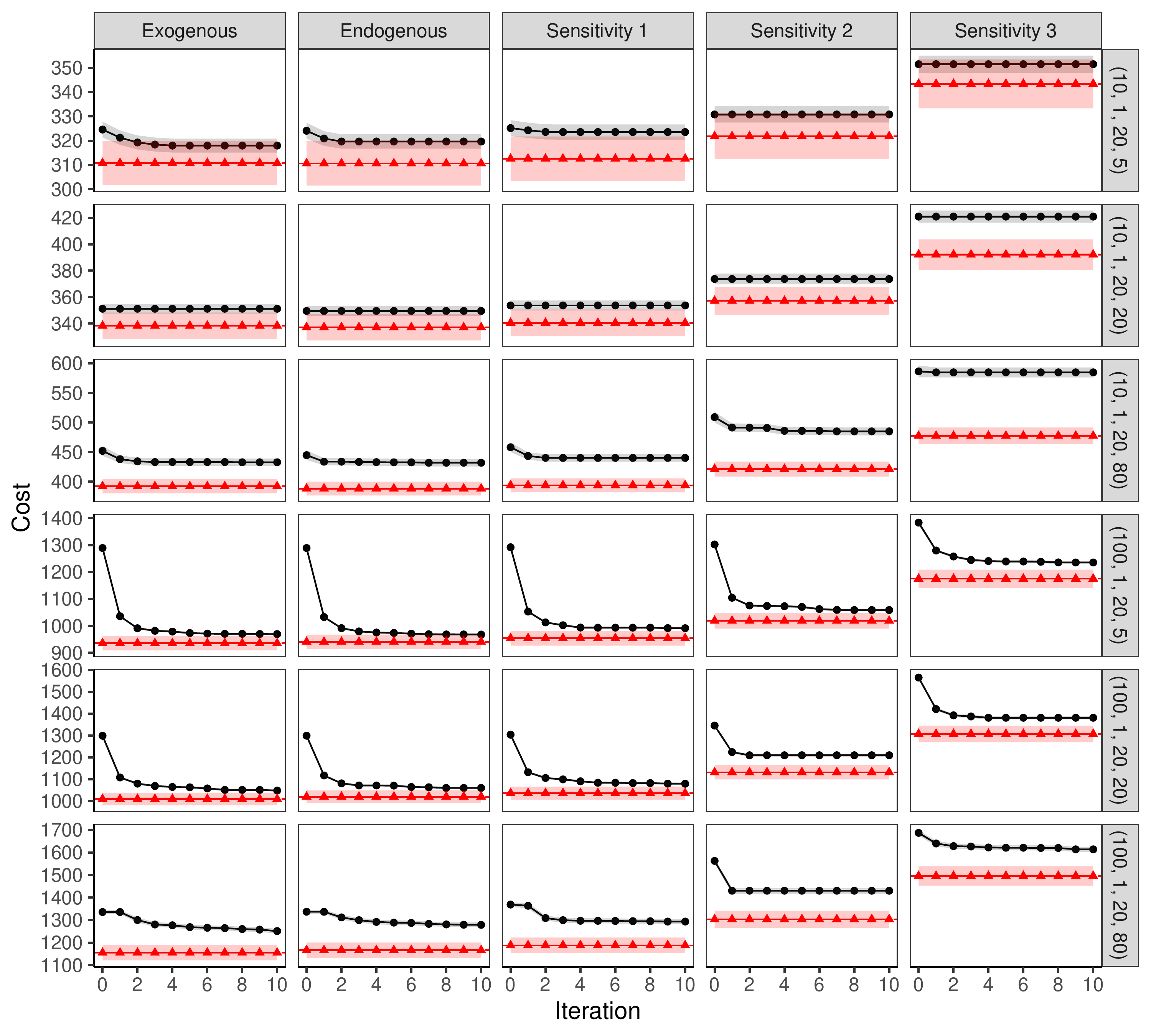}
    \caption{Performance of the ADP algorithm for $m=5$. The black line is the best estimate of the expected cost in each iteration and the red line is the estimate of the expected cost with imperfect information relaxation. The shaded areas are the 95\% confidence intervals. On average, the ADP policy can reduce the gap between the upper-bound, i.e., the Myopic policy, and the lower-bound from 16.2\% to  6.6\%.}
    \label{fig:ADPerfm5}
\end{figure}

In Figure \ref{fig:ADPerfm5}, we observe that for the fixed ordering cost of $\kappa=10$ the gap between the initial upper-bound, i.e., the Myopic policy at iteration zero, and lower-bound is on average 8.7\% (without considering the estimation errors). The ADP policy can slightly reduce the gap to 7.2\%. However, for the fixed ordering cost of $\kappa=100$, the ADP achieves significant improvements compared to the initial upper-bound and reduces the average relative gap from 23.6\% to 5.9\%. We summarize the performance of the Myopic and ADP policy for the two fixed ordering costs in Table \ref{tab:perf}, where SD is the standard deviation of the relative percentage gap compared to the lower-bound. The lower SD of the ADP compared to Myopic policy implies that ADP performs more robustly over the tested scenarios. The \revision{worst-case} gaps (e.g., 22.5\%) relate to cases with $\kappa=10$ and the unit cost of expiration $\theta = 80$. Based on the 95\% confidence intervals and our previous observation from Figure \ref{fig:lb} for similar settings, we can attribute the \revision{worst-case} gap to the quality of the lower-bound for these problem configurations.

\begin{table}
    \caption{Relative gap (\%) compared with the lower-bound.}
    \label{tab:perf}
    \centering
    \begin{tabular}{clcccc}
    \hline
    \revision{$\kappa$} & \multicolumn{1}{c}{Policy}  &  Min & Mean & Max & SD \\
    \hline
    \multirow{ 2}{*}{10}  & Myopic & 2.4 & 8.7 & 22.9 & 7.1 \\
    & ADP & 2.3 & 7.2 & 22.5 & 5.9 \\[2pt]
    \multirow{ 2}{*}{100}  & Myopic & 12.8 & 23.6 & 37.8 & 8.4\\
    & ADP & 2.9 & 5.9 & 9.7 & 2.4 \\
    \hline
    \end{tabular}
\end{table}

\subsection{Impact of Shelf-life Uncertainty}
In this section, we examine the impact of ignoring shelf-life uncertainty and its dependence on order size for cases with $m \in \{3, 5\}$ where perishability is of greater concern.
\subsubsection{The Case with $m=3$.}\label{subsub:impact3}
We obtain exact optimal policies under the assumption that the shelf-life distribution is exogenous, or deterministic and equal to three days. Intuitively, since $h<\theta$ and ordering in every period is allowed, the variability in shelf-life could significantly increase the cost by wasting more units. Hence, we focus on cases with negative coefficients for the order size covariate in the multinomial logit model, i.e., $(c_1^2, c_1^3)\in \{(-0.15, -0.05), (-0.2, -0.1), (-0.25, -0.15), (-0.3, -0.2), (-0.35, -0.25), (-0.4, -0.3) \}$. The larger absolute values of negative coefficients increase the probability of receiving units with remaining shelf-life of one  more rapidly in the order size, until it converges to the deterministic case with the maximum shelf-life of one day.  

To evaluate the impact of ignoring endogeneity, we need to consider an appropriate exogenous model of uncertainty associated with each endogenous case. To this end, we fix an ordering policy and simulate shelf-life data under that policy and the true endogenous setting. We then use the simulated data to estimate the (exogenous) probabilities of receiving units with remaining shelf-life of $k \in \{1,2,3\}$. We consider two ordering policies: the optimal policy and the Myopic policy. 

\begin{figure}
    \centering
    \includegraphics[width=\textwidth]{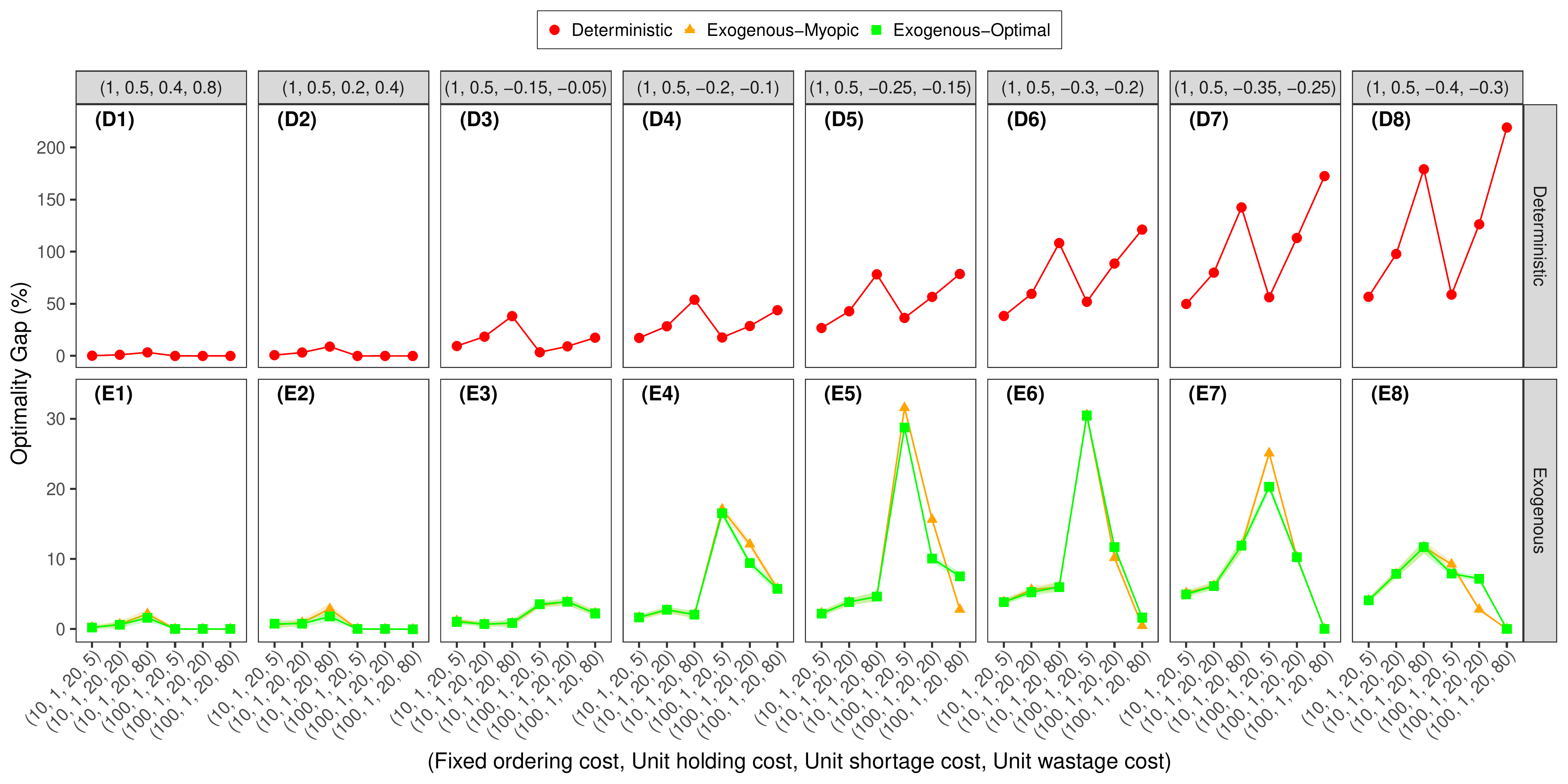}
    \caption{Impact of ignoring endogenous shelf-life uncertainty when $m=3$. On average, the estimate of the expected optimality gap among all cases is 50.9\%, 5.7\%, and 5.5\% for the deterministic, exogenous-myopic, and exogenous-optimal cases, respectively.}
    \label{fig:ImpEnd}
\end{figure}

In Figure \ref{fig:ImpEnd}, we plot the estimates of expected optimality gaps when using the policy obtained under the assumptions of deterministic and exogenous shelf-life uncertainty. The first row shows the results for deterministic shelf-life and the second row for exogenous uncertainty assumption. Each column relates to a different endogeneity setting (see Section \ref{subsub:shelf-life} for the effect of changing endogeneity parameters) and each plot illustrates the gaps for different cost settings. Overall, we investigate the impact using 48 instances of the problem.

For the policy obtained under the deterministic shelf-life, we observe that the optimality gaps increase in the absolute values of negative order size coefficients (Figure \ref{fig:ImpEnd} (D3)-(D8)) but decrease in the positive values of order size coefficients (Figure \ref{fig:ImpEnd} (D1)-(D2)). This is intuitive as in the extreme cases of (D1) and (D8), the true endogenous uncertainty converges to deterministic cases of three-day and one-day shelf-life, respectively. Further, we observe the gaps increase in the unit cost of wastage, with the increase being greater when the chance of receiving units with remaining shelf-life of one is higher, e.g., (D8). Moreover, larger fixed ordering costs can result in greater gaps, as observed from (D5) to the right, particularly when the chance of expiration and unit cost of wastage is high. Intuitively, with a larger fixed ordering cost, the policy assuming deterministic shelf-life tends to order larger amounts less frequently. However, ignoring endogeneity leads to higher optimality gaps as with larger orders the chance of receiving units with remaining shelf-life of one is higher as well. In summary, the estimate of the expected optimality gap among all cases varies between 0.0\% and 219.2\% with an average of 50.9\%.

For the policy obtained under the exogenous shelf-life uncertainty, we again observe close to zero optimality gaps for positive values of order size coefficients; see Figure \ref{fig:ImpEnd} (E1)-(E2). However, the gaps first increase and then decrease in the absolute values of negative order size coefficients, i.e., in Figure \ref{fig:ImpEnd} (E3)-(E8). Intuitively, this is because with small absolute values of coefficients in (E3), the dependence between order size and shelf-life is not strong and hence the impact of ignoring endogeneity is less significant. With large absolute values of coefficients in (E8), the shelf-life distribution converges quickly to the case where the probability of receiving units with remaining shelf-life of one is almost one (regardless of the order size), which can be captured using an exogenous shelf-life uncertainty model, hence reducing the impact. 

From (E4)-(E7), we observe a larger increase in the gap for larger values of fixed ordering cost. The increase is slightly greater with the exogenous uncertainty estimated under the Myopic policy, except for cases with higher unit cost of wastage. Intuitively, the larger gap is due to ordering larger amounts and the resulting increase in the chance of receiving units with remaining shelf-life of one, which is ignored under exogenous uncertainty. The Myopic policy that tends to order larger amounts results in a  conservative estimate of the exogenous uncertainty (with a higher probability of receiving units with remaining shelf-life of one), which could perform better when the unit cost of wastage is higher. In general, a higher unit cost of wastage that is still below the fixed ordering cost reduces the order size and hence the gap due to ignoring endogeneity. However, if the unit cost of wastage dominates the fixed ordering costs, the gap continues to grow as observed for cost combinations with $\kappa=10$.

In summary, the estimate of the expected optimality gap for the policy with exogenous uncertainty is on average 5.7\% among all cases when using data generated from the Myopic policy, and 5.5\% when using data from the optimal policy. The optimality gaps could be considerably large (respectively up to 31.6\% and 30.5\%) in cases with high dependence between order size and shelf-life and with high fixed ordering cost. 

\subsubsection{The Case with $m=5$.} \label{subsub:impact5} We next examine the impact of ignoring shelf-life uncertainty and endogeneity for larger problem instances using the ADP policy. As we cannot compute the exact policy with uncertain shelf-life, we instead calculate the ADP policies under the empirical exogenous probabilities as well as those estimated from the simulated shelf-life data using the ADP and Myopic policy obtained under the true endogenous uncertainty. Note that the empirical probabilities are estimated based on data generated under the unknown ordering policy of the hospital, and hence we cannot estimate them for alternative endogenous uncertainty scenarios. 

\begin{figure}
    \centering
    \includegraphics[width= 1.0 \textwidth]{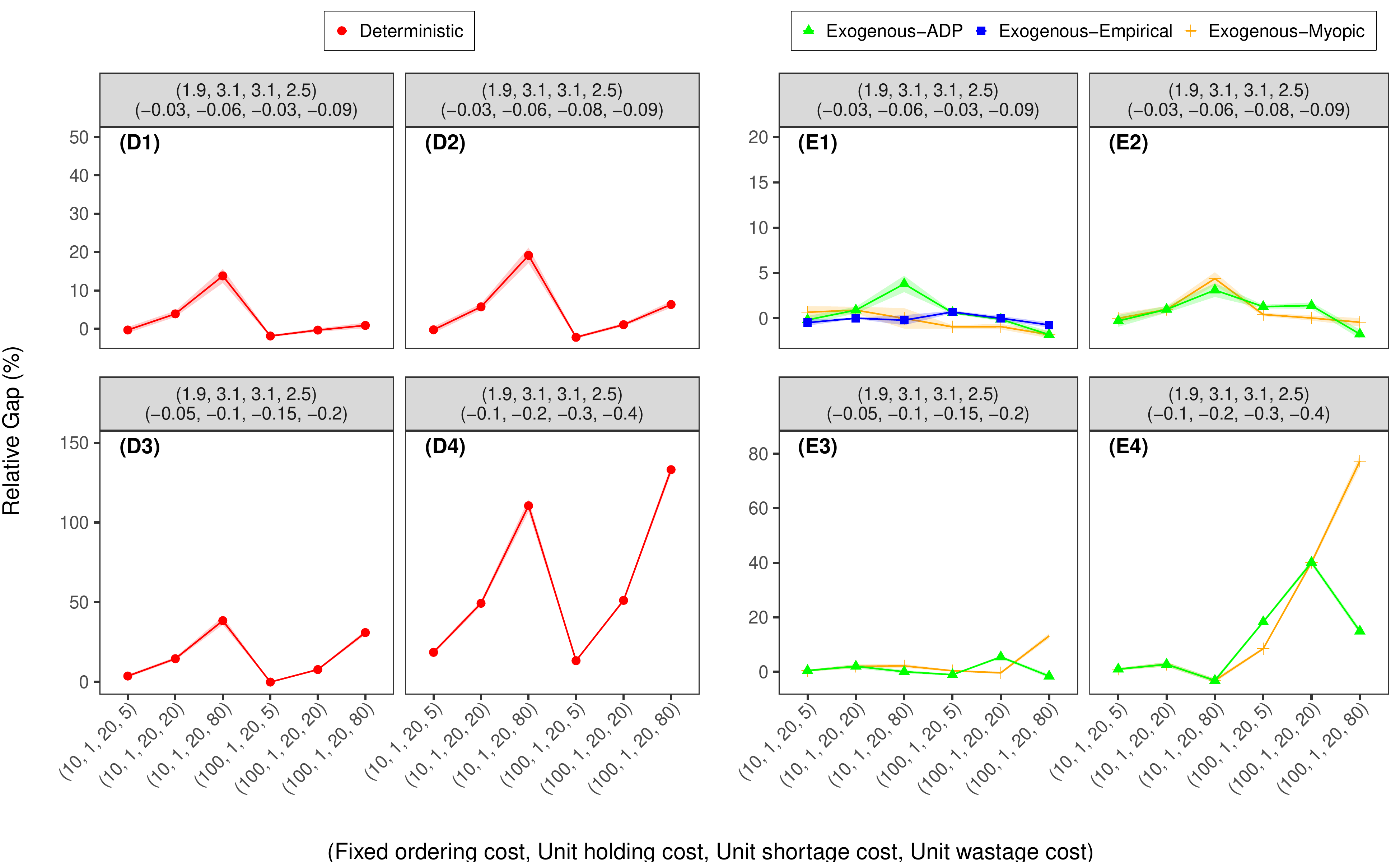}
    \caption{Impact of ignoring endogenous shelf-life uncertainty when $m=5$. On average, the estimate of the expected relative gap in cost compared with the ADP policy obtained under the true setting is 21.5\%, 6.1\%, and 3.6\% for the deterministic shelf-life, Myopic, and ADP exogenous shelf-life uncertainty, respectively.}
    \label{fig:ImpEndm5}
\end{figure}

In Figure \ref{fig:ImpEndm5}, we plot the performance of the calculated policies in comparison to the ADP policy obtained under the true endogenous setting with respect to estimated expected relative increase in cost. Each plot shows the relative gap for different cost and endogenous uncertainty settings (see Section \ref{subsub:shelf-life} for details). In (D1)-(D4) we illustrate the gap for the exact policies obtained under the assumption of deterministic shelf-life, and in (E1)-(E4) for the ADP policies obtained under exogenous uncertainties estimated based on the given endogenous uncertainty. For example, the blue line presents the increase for the policy obtained under the empirical exogenous probabilities estimated based on the historical data. Since the ordering policy of the hospital is unknown, we cannot generate the shelf-life data and estimate the empirical probabilities for alternative endogenous scenarios, and hence we do not provide the blue line in (E2)-(E4). 

Similar to the case with $m=3$, we observe that under the deterministic assumption the gap increases in the absolute values of negative order size coefficients and the unit cost of wastage; see Figure \ref{fig:ImpEndm5} (D1)-(D4). However, the larger fixed ordering cost only increases the gap in (D4) when the unit cost of wastage is $\theta=80$. Intuitively, with a longer maximum shelf-life, the chance of expiration is less and hence larger orders could reduce the inventory costs when $h<\kappa$ and the absolute values of negative order size coefficients \minorrevision{are not ``too large"}. In summary, the estimated gap among all cases with the deterministic shelf-life assumption is on average 21.5\% but could be much larger in certain cases. Also, as the impact is quantified based on the ADP policy, which is a sub-optimal policy, the real impact could be larger. Therefore, ignoring the shelf-life uncertainty could lead to considerably costlier ordering decisions. 


We next examine the impact of ignoring endogeneity. We observe that the impact is small for the empirical endogenous uncertainty estimated based on historical data; see Figure \ref{fig:ImpEndm5} (E1). This is due to the small absolute values of the order size coefficients in the fitted \revision{multinomial} logistic model, resulting in a relatively weak dependence between the order size and shelf-life of ordered units. Similar to the case with $m=3$, the gap could be large for alternative endogenous scenarios with greater absolute values of negative coefficients, e.g., E(4), where the dependence between the order size and shelf-life is stronger. In summary among all tested scenarios, the impact of ignoring endogeneity is on average 6.1\% and 3.6\% for the policy obtained under the exogenous probabilities estimated based on the simulated shelf-life data using the ADP and Myopic policy, respectively. The gap and impact of ignoring endogeneity could be considerably large in certain cases, e.g., when there is a relatively high probability of receiving units with remaining shelf-life of one and the fixed ordering and unit wastage costs are large. 

\section{Case Study: Hamilton General Hospital (HGH)} \label{sec:casestudy}
In this section, we conduct a case study using data from HGH and evaluate the out-of-sample performance of the ADP approach and compare it with other benchmarks. Our data comes from the TRUST (Transfusion Registry for Utilization, Surveillance and Tracking) dataset (see \citealt{abouee2022data} for details) which includes inventory information for blood products utilized in a network of hospitals in Hamilton, Ontario.  We use data from 2015-2016 to estimate the parameters of our model and tune the coefficients of the basis function. We then evaluate the out-of-sample performance of the policy using data from 2017 and compare to benchmarks.

\begin{figure}
    \centering
    \includegraphics[width=0.7\textwidth]{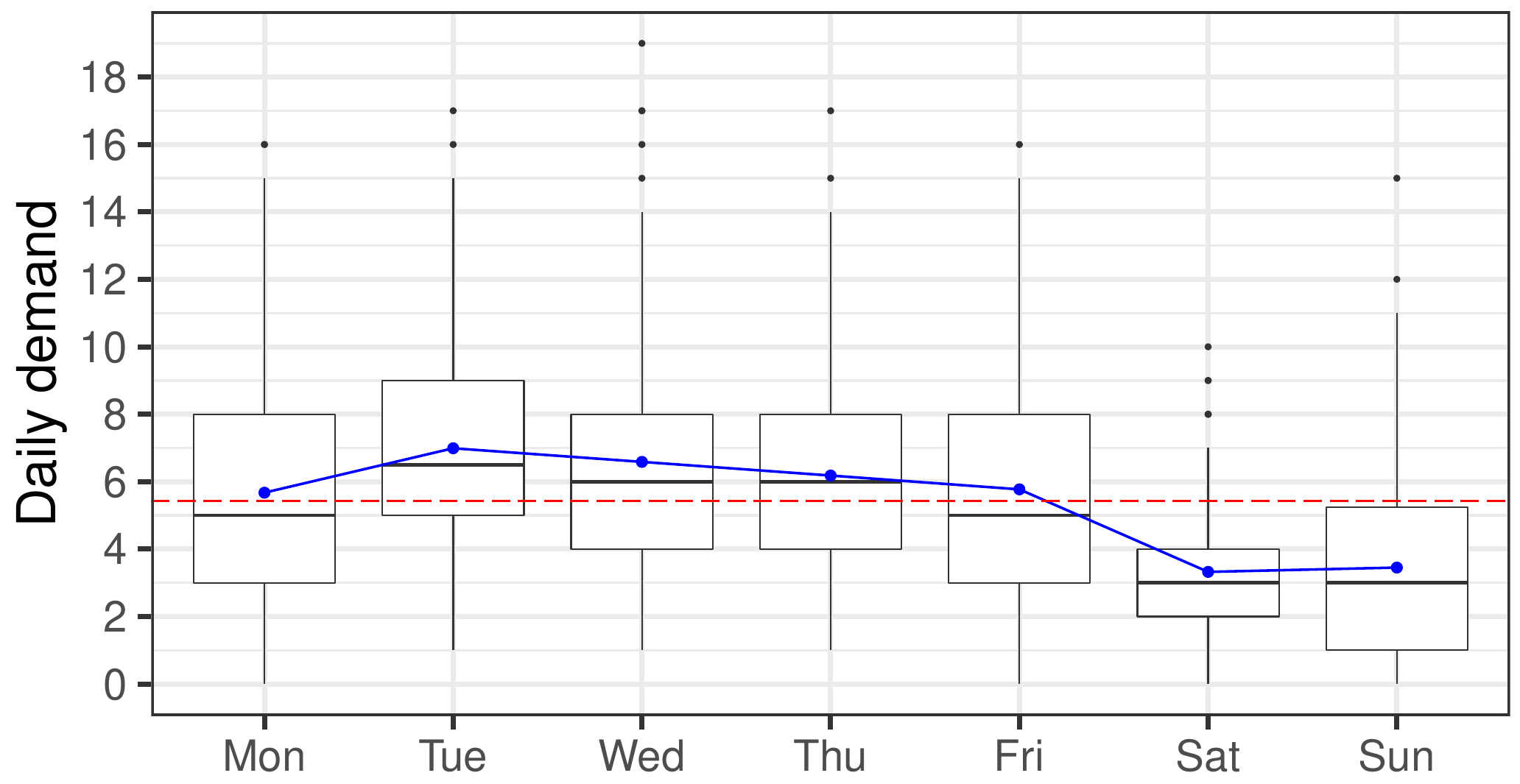}
    \caption{Daily platelet demand at HGH. The red dashed line is the estimated average daily demand within years 2015-2016. The blue line is the estimated average daily demand varying across days of the week.}
    \label{fig:DOW}
\end{figure}

\textbf{Model Inputs and Estimation}: There were 3965 transfusion records for platelets at HGH during  2015-2016. We assume that the hospital satisfied all demand, i.e., each patient who needed platelet transfusion eventually received it (possibly through emergency orders) and hence the observed demand is not censored. Therefore, we consider the total number of transfusion records for each day as the daily demand for platelets. Figure \ref{fig:DOW} illustrates how daily platelet demand varies across days of the week with overall average of 5.4 and variance of 12.3. Since demand data is overdispersed, we fit a \revision{negative binomial} distribution to each day of the week by estimating the parameters using \revision{maximum likelihood estimation} as presented in Section \ref{subsub:demand}. We then truncate the fitted distributions at $M=20$, the largest historical demand observed in the data. We use \revision{multinomial} regression to estimate the distribution of the shelf-life of deliveries and its dependence on order size as reported in Section \ref{subsub:shelf-life}. Since shortage is considered costlier than wastage for platelets, we consider three unit cost ratios of shortage to wastage $l/\theta \in\{10, 4, 1\}$ together with two fixed ordering costs $\kappa \in\{10, 20\}$. We fix the holding cost at $h=1$. 

\textbf{Benchmarks and Performance Metrics}: We consider ADP policies obtained under the endogenous (Endog.) and exogenous (Exoge.) shelf-life uncertainty, as well as three benchmarks. \minorrevision{The first benchmark is the exact policy under the deterministic maximum shelf-life of five days (Deter.). The second is the $(s, S)$ policy with parameters optimized based on the empirical in-sample cost. More specifically, we evaluate all possible $(s,S)$ policies using \eqref{eq:mincost} and pick the one resulting in the minimum sample average estimate of the expected long-run discounted cost. The last benchmark (referred to as LB) is obtained by using the information relaxation lower-bound of Section \ref{subsec:lb}. Specifically, we use the lower-bound as an approximation of the value function and compute the policy by solving the right-hand-side of \eqref{eq:mincost}.} For each policy, we compute the out-of-sample performance including the fraction of days with a non-zero regular order (Prob.), average order size (Orders), average daily number of units held in the inventory (Holding), shortage rate (Shortage), expiration rate (expiration), and the discounted cost function (Cost). Given the endogenous uncertainty in shelf-life, we simulate shelf-life values from the fitted \revision{multinomial} model under each policy and report the expected out-of-sample performance measures (with 95\% confidence intervals). 

In the data, we directly observe the final disposition of each delivered unit and hence can directly calculate historical expiration rates. However, we do not observe order types (i.e., routine versus emergency). Hence, we define a routine ordering window between 8 AM and 4 PM and count units received inside and outside this interval respectively as regular and emergency. We then estimate the historical shortage rate with emergency order rate as shortages due to insufficient inventory are satisfied using emergency orders.

 \begin{table}
    \centering
    \small
    \caption{Summary of the out-of-sample performance with $h=1$ and $l=20$. The values in brackets correspond to the half-with of a \%95 confidence interval for the corresponding estimates.}
    \begin{tabular}{ccccccccc}
        \hline 
         $\revision{\kappa}$ & $\theta$ & Policy & Prob. & Orders & Holding & \% Shortage & \% expiration & Cost  \\
        \hline
        10 & 2  & Endog. &  51.00 & 2005.31 & 6.98 & 3.97 & 3.94 & 303.77 \\[-3pt]
        & & & ($\pm$ 0.11) & ($\pm$ 1.13) & ($\pm$ 0.01) & ($\pm$ 0.06) & ($\pm$ 0.06) & ($\pm$ 1.26)\\
        & & Exoge. & 51.55 & 2015.05 & 7.10 & 3.70 & 4.40 & 308.36  \\[-3pt]
        & & & ($\pm$ 0.10) & ($\pm$ 1.33) & ($\pm$ 0.01) & ($\pm$ 0.05) & ($\pm$ 0.07) & ($\pm$ 1.49) \\
        & & Deter. & 45.58 & 2042.74 & 8.00 & 3.49 & 5.71 & 299.53\\
        [-3pt]
        & & & ($\pm$ 0.08) & ($\pm$ 1.33) & ($\pm$ 0.01) & ($\pm$ 0.05) & ($\pm$ 0.06) & ($\pm$ 1.28)\\
        & & \revision{$(s, S)$} & \revision{48.32} & \revision{2041.71} & \revision{7.66} & \revision{3.32} & \revision{5.68} & \revision{304.76}\\
        [-3pt]
        & & & \revision{($\pm$ 0.10)} & \revision{($\pm$ 1.33)} & \revision{($\pm$ 0.01)} & \revision{($\pm$ 0.06)} & \revision{($\pm$ 0.06)} & \revision{($\pm$ 1.30)}\\
        & & \minorrevision{LB} & \minorrevision{47.72} & \minorrevision{2028.51} & \minorrevision{7.57} & \minorrevision{3.08} & \minorrevision{5.04} & \minorrevision{309.16}\\
        [-3pt]
        & & & \minorrevision{($\pm$ 0.10)} & \minorrevision{($\pm$ 1.25)} & \minorrevision{($\pm$ 0.01)} & \minorrevision{($\pm$ 0.06)} & \minorrevision{($\pm$ 0.06)} & \minorrevision{($\pm$ 1.34)}\\
        \hline
        10 & 5  & Endog. & 53.92 & 2000.87 & 6.62 & 4.68 & 3.73 & 314.70 \\[-3pt]
        & & & ($\pm$ 0.12) & ($\pm$ 1.08) & ($\pm$ 0.01) & ($\pm$ 0.05) & ($\pm$ 0.05) & ($\pm$ 1.50)\\
        & & Exoge. & 52.92 & 2005.33 & 6.75 & 4.54 & 3.94 & 312.98 \\[-3pt]
        & & & ($\pm$ 0.12) & ($\pm$ 1.13) & ($\pm$ 0.01) & ($\pm$ 0.06) & ($\pm$ 0.06) & ($\pm$ 1.42)\\
        & & Deter. & 46.38 & 2040.02 & 7.88 & 3.62 & 5.57 & 315.50\\[-3pt]
        & & & ($\pm$ 0.08) & ($\pm$ 1.34) & ($\pm$ 0.01) & ($\pm$ 0.05) & ($\pm$ 0.06) & ($\pm$ 1.57)\\
        & & \revision{$(s, S)$} & \revision{52.93} & \revision{2021.96} & \revision{7.16} & \revision{3.62} & \revision{4.82} & \revision{324.19}\\
        [-3pt]
        & & & \revision{($\pm$ 0.08)} & \revision{($\pm$ 1.18)} & \revision{($\pm$ 0.01)} & \revision{($\pm$ 0.04)} & \revision{($\pm$ 0.06)} & \revision{($\pm$ 1.31)}\\
        & & \minorrevision{LB} & \minorrevision{49.81} & \minorrevision{2008.47} & \minorrevision{6.86} & \minorrevision{4.69} & \minorrevision{4.09} & \minorrevision{338.02}\\
        [-3pt]
        & & & \minorrevision{($\pm$ 0.09)} & \minorrevision{($\pm$ 1.23)} & \minorrevision{($\pm$ 0.01)} & \minorrevision{($\pm$ 0.06)} & \minorrevision{($\pm$ 0.06)} & \minorrevision{($\pm$ 1.63)}\\
        \hline
        10 & 20 & Endog. & 60.48 & 1970.34 & 5.56 & 6.45 & 2.44 & 345.19  \\[-3pt]
        & & & ($\pm$ 0.10) & ($\pm$ 0.99) & ($\pm$ 0.00) & ($\pm$ 0.03) & ($\pm$ 0.05) & ($\pm$ 2.52)\\
        & & Exoge. & 60.39 & 1970.33 & 5.56 & 6.46 & 2.43 & 345.19 \\[-3pt]
        & & & ($\pm$ 0.11) & ($\pm$ 0.99) & ($\pm$ 0.00) & ($\pm$ 0.03) & ($\pm$ 0.05) & ($\pm$ 2.46)\\
        & & Deter. & 49.80 & 2022.86 & 7.23 & 4.74 & 4.78 & 355.34 \\[-3pt]
        & & & ($\pm$ 0.09) & ($\pm$ 1.21) & ($\pm$ 0.01) & ($\pm$ 0.06) & ($\pm$ 0.06) & ($\pm$ 3.22)\\
        & & \revision{$(s, S)$} & \revision{56.60} & \revision{1979.95} & \revision{5.67} & \revision{5.99} & \revision{2.80} & \revision{329.25}\\
        [-3pt]
        & & & \revision{($\pm$ 0.07)} & \revision{($\pm$ 0.82)} & \revision{($\pm$ 0.00)} & \revision{($\pm$ 0.03)} & \revision{($\pm$ 0.04)} & \revision{($\pm$ 2.83)}\\
        & & \minorrevision{LB} & \minorrevision{54.94} & \minorrevision{1983.53} & \minorrevision{5.98} & \minorrevision{6.85} & \minorrevision{2.93} & \minorrevision{382.55}\\
        [-3pt]
        & & & \minorrevision{($\pm$ 0.14)} & \minorrevision{($\pm$ 1.16)} & \minorrevision{($\pm$ 0.01)} & \minorrevision{($\pm$ 0.09)} & \minorrevision{($\pm$ 0.06)} & \minorrevision{($\pm$ 2.42)}\\
        \hline
        20 & 2  & Endog. &  41.22 & 2045.12 & 8.14 & 3.56 & 5.77 & 378.16\\[-3pt]
        & & & ($\pm$ 0.09) & ($\pm$ 1.58) & ($\pm$ 0.01) & ($\pm$ 0.09) & ($\pm$ 0.07) & ($\pm$ 2.75)\\
        & & Exoge. & 40.61 & 2045.97 & 8.17 & 3.56 & 5.82 & 374.20 \\[-3pt]
        & & & ($\pm$ 0.10) & ($\pm$ 1.41) & ($\pm$ 0.01) & ($\pm$ 0.08) & ($\pm$ 0.07) & ($\pm$ 1.81)\\
        & & Deter. & 35.54 & 2099.39 & 9.05 & 3.30 & 7.96 & 375.45 \\[-3pt]
        & & & ($\pm$ 0.07) & ($\pm$ 1.61) & ($\pm$ 0.01) & ($\pm$ 0.07) & ($\pm$ 0.07) & ($\pm$ 1.56)\\
        & & \revision{$(s, S)$} & \revision{37.23} & \revision{2096.39} & \revision{8.86} & \revision{3.32} & \revision{8.12} & \revision{403.31}\\
        [-3pt]
        & & & \revision{($\pm$ 0.08)} & \revision{($\pm$ 1.70)} & \revision{($\pm$ 0.01)} & \revision{($\pm$ 0.08)} & \revision{($\pm$ 0.08)} & \revision{($\pm$ 2.40)}\\
        & & \minorrevision{LB} & \minorrevision{37.10} & \minorrevision{2083.20} & \minorrevision{8.70} & \minorrevision{3.34} & \minorrevision{7.52} & \minorrevision{373.34}\\
        [-3pt]
        & & & \minorrevision{($\pm$ 0.09)} & \minorrevision{($\pm$ 1.73)} & \minorrevision{($\pm$ 0.01)} & \minorrevision{($\pm$ 0.08)} & \minorrevision{($\pm$ 0.08)} & \minorrevision{($\pm$ 1.42)}\\
        \hline
        20 & 5  & Endog. &  44.52 & 2029.60 & 7.61 & 3.23 & 5.10 & 399.20 \\[-3pt]
        & & & ($\pm$ 0.10) & ($\pm$ 1.45) & ($\pm$ 0.01) & ($\pm$ 0.08) & ($\pm$ 0.07) & ($\pm$ 2.39)\\
        & & Exoge. & 43.53 & 2035.24 & 7.80 & 3.31 & 5.30 & 406.00 \\[-3pt]
        & & & ($\pm$ 0.09) & ($\pm$ 1.52) & ($\pm$ 0.01) & ($\pm$ 0.07) & ($\pm$ 0.07) & ($\pm$ 2.56)\\
        & & Deter. & 36.44 & 2080.41 & 8.67 & 3.69 & 7.59 & 398.01\\[-3pt]
        & & & ($\pm$ 0.06) & ($\pm$ 1.70) & ($\pm$ 0.01) & ($\pm$ 0.07) & ($\pm$ 0.07) & ($\pm$ 1.97)\\
        & & \revision{$(s, S)$} & \revision{37.23} & \revision{2096.39} & \revision{8.86} & \revision{3.32} & \revision{8.12} & \revision{421.68}\\
        [-3pt]
        & & & \revision{($\pm$ 0.08)} & \revision{($\pm$ 1.70)} & \revision{($\pm$ 0.01)} & \revision{($\pm$ 0.08)} & \revision{($\pm$ 0.08)} & \revision{($\pm$ 2.50)}\\
        & & \minorrevision{LB} & \minorrevision{39.96} & \minorrevision{2052.21} & \minorrevision{7.99} & \minorrevision{3.46} & \minorrevision{6.04} & \minorrevision{394.47}\\
        [-3pt]
        & & & \minorrevision{($\pm$ 0.10)} & \minorrevision{($\pm$ 1.60)} & \minorrevision{($\pm$ 0.01)} & \minorrevision{($\pm$ 0.07)} & \minorrevision{($\pm$ 0.08)} & \minorrevision{($\pm$ 2.05)}\\
        \hline
        20 & 20  & Endog. &  47.12 & 1994.12 & 6.38 & 7.86 & 3.42 & 441.33 \\[-3pt]
        & & & ($\pm$ 0.10) & ($\pm$ 1.25) & ($\pm$ 0.01) & ($\pm$ 0.09) & ($\pm$ 0.06) & ($\pm$ 5.15) \\
        & & Exoge. & 45.96 & 2010.96 & 6.71 & 6.85 & 4.20 & 451.25 \\[-3pt]
        & & & ($\pm$ 0.09) & ($\pm$ 1.30) & ($\pm$ 0.01) & ($\pm$ 0.07) & ($\pm$ 0.06) & ($\pm$ 2.86)\\
        & & Deter. & 41.53 & 2055.37 & 7.87 & 3.67 & 6.25 & 466.84  \\[-3pt]
        & & & ($\pm$ 0.07) & ($\pm$ 1.59) & ($\pm$ 0.01) & ($\pm$ 0.07) & ($\pm$ 0.07) & ($\pm$ 5.83)\\
        & & \revision{$(s, S)$} & \revision{44.13} & \revision{2003.48} & \revision{6.36} & \revision{6.10} & \revision{3.86} & \revision{494.07}\\
        [-3pt]
        & & & \revision{($\pm$ 0.08)} & \revision{($\pm$ 1.19)} & \revision{($\pm$ 0.01)} & \revision{($\pm$ 0.08)} & \revision{($\pm$ 0.06)} & \revision{($\pm$ 3.94)}\\
        & & \minorrevision{LB} & \minorrevision{45.48} & \minorrevision{2001.05} & \minorrevision{6.54} & \minorrevision{6.06} & \minorrevision{3.71} & \minorrevision{478.60}\\
        [-3pt]
        & & & \minorrevision{($\pm$ 0.09)} & \minorrevision{($\pm$ 1.22)} & \minorrevision{($\pm$ 0.01)} & \minorrevision{($\pm$ 0.08)} & \minorrevision{($\pm$ 0.06)} & \minorrevision{($\pm$ 3.01)}\\
        \hline 
    \end{tabular}
    \label{perf.ML}
\end{table}

\textbf{Results:} In 2017, the historical fraction of days with a non-zero regular order was 82.7\%, with a total number of units ordered, average daily holding units, (estimated) shortage rate, and expiration rate standing at 2099, 9.9, 13.9\%, and 8.7\%, respectively. Regardless of the cost setups, all approximate policies demonstrate improvements across all performance metrics when compared to historical performance. For instance, implementing the Endog. policy with parameters $\kappa=10, \, \theta=5, \, l=20$, and $h=1$ results in a significant 34.8\% reduction in the occurrence of days with non-zero order size. Furthermore, this policy leads to reductions of 4.7\% in order size, 33.1\% in average daily holding units, 66.3\% in shortage rate, and 57.1\% in expiration rate.

In comparison to the Deter. policy, ADP policies that account for shelf-life uncertainty tend to place smaller but more frequent orders. This helps to maintain lower inventory levels and mitigate the impact of shelf-life uncertainty on expiration rates. For instance, when the expiration cost equals the shortage cost ($\theta = l = 20$) and $\kappa=20$, implementing the Endog. policy yields a substantial 45.3\% reduction in expiration rate and a 5.5\% decrease in costs compared to Deter. policy. We also observe improvements under Exoge. policy, i.e., a 32.8\% reduction in the expiration rate and a 3.3\% decrease in costs. \revision{Both ADP policies outperform the $(s,S)$ policy. In particular, when $\theta\in \{5,20\}$ or $\kappa=20$, ADP policies result in substantial decreases in both expiration and shortage rates compared to the simple $(s,S)$ policy.} \minorrevision{The LB policy (which relies on a lower-bound approximation of the value function) consistently yields higher expiration costs compared to the ADP policies. When $\kappa=10$, the ADP policies substantially outperform the LB policy. When $\kappa=20$ and the ratio $l/\theta$ ratio is high, the LB policy performs similar or slightly better than the ADP policies with respect to the total discounted cost, but perform poorly when $l/\theta=1$.}

The Endog. policy achieves similar or better performance compared to the Exoge. policy. Given the small coefficients of order size in the fitted \revision{multinomial} model, the disparity between the two policies is more pronounced when both policies incline to order larger quantities under the larger fixed ordering cost of $\kappa=20$. This intensifies the impact of order size on the shelf-life of ordered units. For instance, when $\theta=5$, i.e., shortage is four times costlier than expiration, the Endog. policy achieves smaller shortage and expiration rates while keeping lower average inventory levels and placing less frequent orders. 

\begin{figure}
    \centering
    \includegraphics[width= 0.95\textwidth]{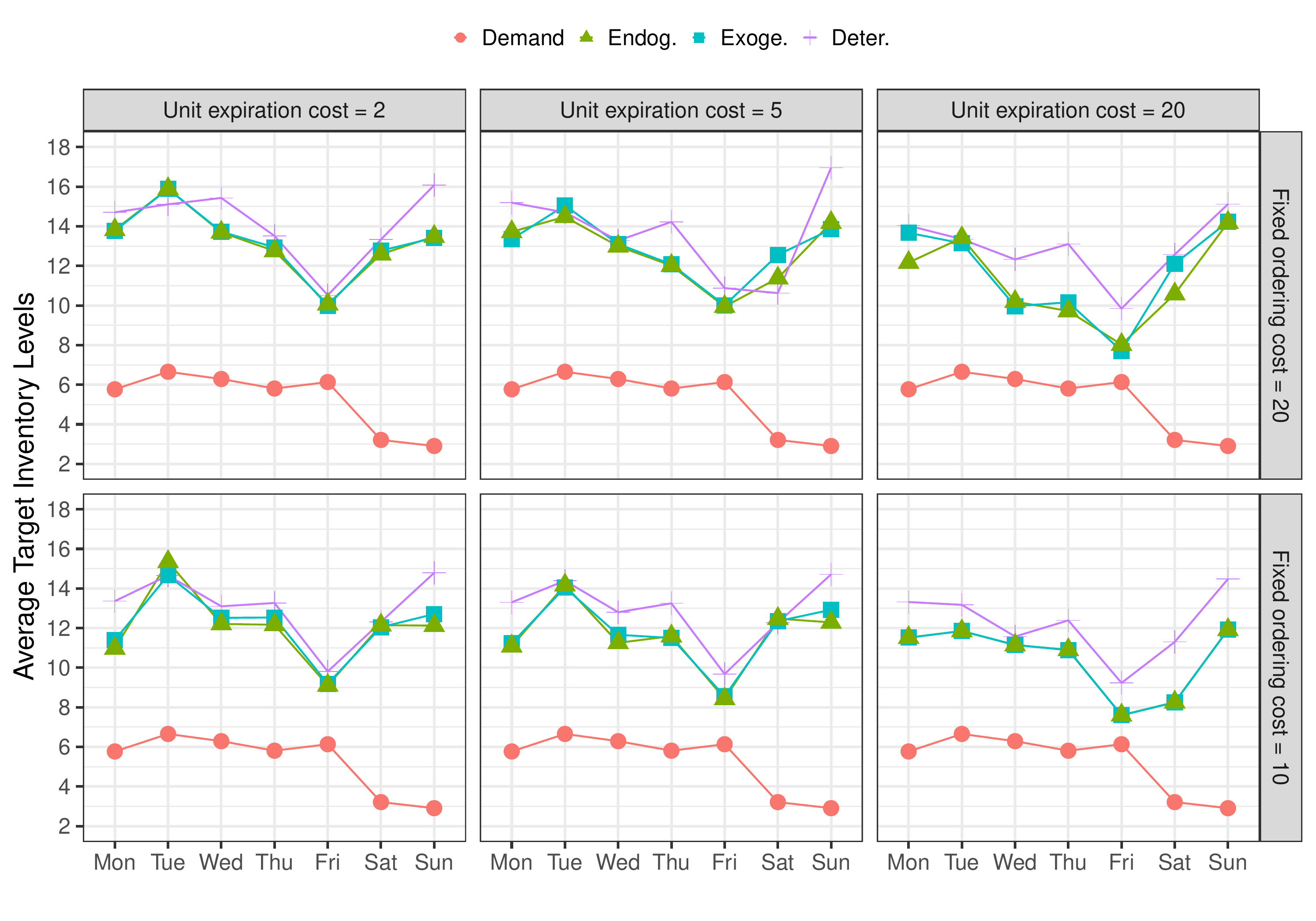}
    \caption{Comparing average daily demand in 2017 with average inventory levels after receiving orders placed using Endog., Exoge., and Deter. policy obtained under different cost settings in Table \ref{perf.ML}.}
    \label{fig:casestudy}
\end{figure}

In Figure \ref{fig:casestudy}, we plot the average daily demand over each day of the week in 2017 together with average inventory levels at the beginning of each day after receiving orders placed using Endog., Exoge., and Deter. policy obtained under different cost settings in Table \ref{perf.ML}. We observe that average inventory levels with Endog. and Exoge. tend to be lower compared to Deter. policy that ignores shelf-life uncertainty. We also observe the average inventory levels are lower with a larger unit cost of expiration ($\theta = 20$) and a smaller fixed ordering cost ($\kappa=10$). 

In summary, the ADP policies that account for shelf-life uncertainty consistently achieve similar or improved out-of-sample costs compared to an exact solution that ignores shelf-life uncertainty. The ADP algorithm can be solved online and in a few seconds, making it superior to the exact solution that needs to be solved and stored offline and could be hard to solve for larger problem instances.

 \section{Conclusion} \label{sec:conclusion}
We study a stochastic perishable inventory problem under endogenous shelf-life uncertainty and fixed ordering costs. Under these assumptions, the value function of the DP formulation of the problem is no longer convex, making it more challenging to find the optimal policy or high-quality heuristics.  We propose an online ADP algorithm that relies on a simple approximation of the value function. We illustrate the effectiveness of the proposed algorithm and the value of accounting for shelf-life uncertainty using extensive numerical studies particularly in settings relevant to blood platelet inventory management.   We further illustrate the out-of-sample quality of the ADP policy in a case study using platelet inventory data from a Canadian hospital. The ADP algorithm achieves significant improvements in all performance measures including the frequency of regular orders, order size, holding, expiration, and shortage rates compared to the historical performance at HGH. In addition, the ADP outperforms or performs as well as an optimal policy obtained under the deterministic shelf-life assumption and is computationally tractable for larger problem instances. Compared to recent data-driven approaches in the literature (e.g., \citealt{guan2017big, abouee2022data, motamedi2022}) the proposed approach relies only on time-related features (day of the week in our case) and can account for fixed ordering costs (i.e., lower frequency of orders) making it practically appealing. 

Our model accounts for key features observed in real data and typically ignored in the literature. However, it has some limitations. First, in practice, some patients may require fresher (i.e., with longer remaining shelf-life) platelet transfusions. In our data, we did not observe whether the demand was for fresh or regular units. As such, our estimates of historical shortage and expiration might be biased. Nevertheless, based on discussions with our collaborates we believe that patients requiring fresh transfusions constitute a small fraction in our hospital and hence we expect the bias to be small. Our ADP framework can easily be extended to account for different demand types (for different units) and include a mismatch (substitution) cost as in \cite{civelek2015blood} where data is available. \revision{Second, we assume a zero lead-time in our model. While this is consistent with current practice where the ordered platelet units are delivered the next day, our model can be extended to account for positive lead-time through expanding the state variable of the DP formulation. Third, we assume that demand and shelf-life uncertainty remain unchanged in time.} In practice, certain events (e.g., a pandemic or change in the policy of the supplier) can result in significant changes in distribution of demand or shelf-life uncertainty, requiring re-estimation of the input models to the ADP. Developing online algorithms that can learn and adapt to changes in demand and shelf-life distributions could be an interesting future direction.


%


\bibliographystyle{informs2014} 
\bibliography{manuscript} 


\ECSwitch
\section{Diminishing Value of Including Higher-Order Basis Functions}\label{app:basisfunctions}

In this section, we provide an analytical justification for excluding higher-order basis functions for a special case of the problem.  
\begin{proposition} \label{prop1}
   Assume $m=3$, uncertainty in shelf-life is exogenous, demand is uniformly distributed, and consider the class of policies that are linear in the state variables $x_1,x_2$. Denote the value function in iteration $n$ of the value iteration algorithm by $v^n(\tau,\textbf{x})$ with  $v^0(\tau, \textbf{x}) = 0$. We have,
   \begin{align}
       v^1(\tau, \textbf{x}) 
      =& \beta^1_{\tau,0} + \beta^1_{\tau,1} x_1+ \beta^1_{\tau,2} x_2+\beta^1_{\tau,3} x_1^2 + \beta^1_{\tau,4} x_1x_2 + \beta^1_{\tau,5} x_2^2, \hspace{2.2cm} \forall \tau, \textbf{x}, n = 1,\\ 
    v^n(\tau, \textbf{x}) 
      =& \beta^n_{\tau,0} + \beta^n_{\tau,1} x_1+ \beta^n_{\tau,2} x_2+\beta^n_{\tau,3} x_1^2 + \beta^n_{\tau,4} x_1x_2 + \beta^n_{\tau,5} x_2^2 +  \hspace{2cm} \forall \tau, \textbf{x},\, n>1,\label{appnz>=0}\\
      &\alpha \big[\beta^n_{\tau,6} x_1^3 + \beta^n_{\tau,7} x_1^2x_2 + \beta^n_{\tau,8} x_1x_2^2 + \beta^n_{\tau,9} x_2^3\big] + \nonumber\\
      &\alpha^2 \big[\beta^n_{\tau,10} x_1^4 + \beta^n_{\tau,11} x_1^3x_2 + \beta^n_{\tau,12} x_1^2x_2^2 + \beta^n_{\tau,13} x_1x_2^3 + \beta^n_{\tau,14} x_2^4\big]+ \nonumber\\
      & \vdots \nonumber\\
      &\alpha^{n-1} \big[\beta^{n}_{\tau,\frac{(n+1)(n+2)}{2}} x_1^{n+1} + \beta^{n}_{\tau,\frac{(n+1)(n+2)}{2}+1} x_1^{n}x_2 + \dots + \beta^{n}_{\tau,\frac{(n+1)(n+2)}{2}+n} x_1x_2^{n} + \beta^{n}_{\tau,\frac{(n+1)(n+4)}{2}} x_2^{n+1}\big]. \nonumber 
\end{align}
\end{proposition} 
\proof{Proof.} The proof is based on an induction argument. We set $n=1$ and initialize the value function $v^0(\tau, \textbf{x}) = 0$ for all $\tau \in \{0, \dots, \gamma-1\}$ and   $\textbf{x} \in \mathcal{X}$. Starting from an arbitrary time index within the demand period, we do the first update for each $\tau \in \{0, \dots, \gamma-1\}$ according to:
\begin{align}
v^1(\tau, \textbf{x}) = \min_{z\in \mathcal{Y}} \Exp \bigg[f\big(1_{z>0}\big) +h\Big(x_{2} + x_{1} + z -D_{\tau}\Big)^+ &+l\Big(D_{\tau}- x_{2} - x_{1} - z\Big)^+ 
    +\theta(x_{1}+Y_{1}-D_{\tau})^+ \label{app:optequind}\\ \nonumber
    &+ \alpha v^{0}\big(s(\tau, \textbf{x}, Y_3, Y_2, Y_1, D_\tau)\big)\bigg], \quad \forall \textbf{x} \in \mathcal{X}. 
\end{align}
Assuming $z^*$ minimizes the expected cost, we can rewrite \eqref{app:optequind} as follows:
\begin{align}
    v^1(\tau, \textbf{x}) = \sum_{y_1=0}^{z^*}P(Y_1 = y_1) \sum_{i=0}^{\infty} \prob(D_{\tau}=i) \bigg[f\big(1_{z^*>0}\big)  &+ h(x_2 +x_1 +z^*-i)^+ \\\nonumber 
    & + l(i - x_2 - x_1 -z^*)^+ +\theta(x_1+y_1-i)^+ \bigg],
\end{align}
and then due to the piece-wise linear structure of the cost function, we have
\begin{align}
     v^1(\tau, \textbf{x})  = \sum_{y_1=0}^{z^*}P(Y_1 = y_1) \Bigg\{\sum_{i=0}^{\infty} \prob(D_{\tau}=i) f\big(1_{z^*>0}\big)  + &\sum_{i=0}^{x_1+y_1} \prob(D_{\tau}=i) \bigg[\theta(x_1+y_1-i) \bigg] \\ \nonumber
    + &\sum_{i=0}^{x_1+ x_2+z^*} \prob(D_{\tau}=i) \bigg[h(x_2 +x_1+z^*-i) \bigg]\\ \nonumber
    + &\sum_{i=x_1 + x_2 +z^*+1}^{\infty} \prob(D_{\tau}=i) \bigg[l(i - x_2 - x_1-z^*)\bigg]\Bigg\}.
\end{align}
Since demand is uniformly distributed over $\{0, 1, \dots, M_\tau\}$, i.e., $\prob(D_{\tau}=i) = 1/(M_\tau+1)$ for every $i \in \{0, 1, \dots, M_\tau\}$, we have
\begin{align}
    v^1(\tau, \textbf{x}) &= \sum_{y_1=0}^{z^*}P(Y_1 = y_1)\times f \big(1_{z^*>0}\big) \\\nonumber
    &\hspace{4pt}+ \sum_{y_1=0}^{z^*}P(Y_1 = y_1) \times \frac{\theta}{M_\tau+1} \bigg[\frac{(x_1+y_1)(x_1+y_1+1)}{2} \bigg] \\ \nonumber
    &\hspace{4pt}+ \sum_{y_1=0}^{z^*}P(Y_1 = y_1)\times \frac{h}{M_\tau+1}\bigg[\frac{(x_1+x_2+z^*)(x_1+x_2+z^*+1)}{2}\bigg] \\\nonumber
    &\hspace{4pt}+ \sum_{y_1=0}^{z^*}P(Y_1 = y_1)\times \frac{l}{M_\tau+1} \bigg[\frac{M_\tau(M_\tau+1)}{2} - \frac{(x_1+x_2+z^*)(x_1+x_2+z^*+1)}{2} \\\nonumber
    &\hspace{6.2cm}- (M_\tau-(x_1+x_2+z^*)) (x_1+x_2+z^*)\bigg],
\end{align}

\begin{align}
     v^1(\tau, \textbf{x}) =  f\big(1_{z^*>0}\big) &+ \frac{\theta}{M_\tau+1} \bigg[ \frac{x_1(x_1+1)}{2} + \frac{(2x_1+1)\sum_{y_1=0}^{z^*}P(Y_1 = y_1)\times y_1}{2} + \frac{\sum_{y_1=0}^{z^*}P(Y_1 = y_1)\times y_1^2}{2}\bigg] \label{inductionstep11}\nonumber\\
     &+ \frac{h}{M_\tau+1}\bigg[\frac{(x_1+x_2+z^*)(x_1+x_2+z^*+1)}{2}\bigg] \\ \nonumber
    &+ \frac{l}{M_\tau+1} \bigg[\frac{M_\tau(M_\tau+1)}{2} - \frac{(x_1+x_2+z^*)(x_1+x_2+z^*+1)}{2} \\\nonumber
    &\hspace{6.2cm}- (M_\tau-(x_1+x_2+z^*)) (x_1+x_2+z^*)\bigg], \\
    v^1(\tau, \textbf{x}) =  f\big(1_{z^*>0}\big)  &+ \frac{\theta}{M_\tau+1} \bigg[\frac{x_1(x_1+1)}{2} + \frac{(2x_1+1)z^*p_1}{2} + \frac{z^*p_1(1-p_1)+{z^*}^2p_1^2}{2} \bigg] \label{inductionstep12}\\\nonumber &+ \frac{h}{M_\tau+1}\bigg[\frac{(x_1+x_2+z^*)(x_1+x_2+z^*+1)}{2}\bigg]\\ \nonumber
    &+ \frac{l}{M_\tau+1} \bigg[\frac{M_\tau(M_\tau+1)}{2} - \frac{(x_1+x_2+z^*)(x_1+x_2+z^*+1)}{2} \\\nonumber
    &\hspace{6.2cm}- (M_\tau-(x_1+x_2+z^*)) (x_1+x_2+z^*)\bigg].
\end{align}
Since $z^*$ is a linear function of $x_1$ and $x_2$ and uncertainty in  shelf-life is exogenous, we can summarize the first update for the value function at each $\tau \in \{0, \dots, \gamma-1\}$ as follows:
\begin{align}
    v^1(\tau, \textbf{x}) &= \beta^1_{\tau,0} + \beta^1_{\tau,1} x_1+ \beta^1_{\tau,2} x_2+\beta^1_{\tau,3} x_1^2 + \beta^1_{\tau,4} x_1x_2 + \beta^1_{\tau,5} x_2^2, \quad \forall \textbf{x} \in \mathcal{X}. 
\end{align}

In step 2 of induction, we assume the argument is true for any stage $n = k$ of value iteration. We have 
\begin{align}
    v^{k}(\tau, \textbf{x}) &= \sum_{y_1=0}^{z^*}\sum_{y_2=0}^{(z^*-y_1)}\sum_{y_3=0}^{(z^*-y_1-y_2)}P(Y_3 = y_3, Y_2= y_2, Y_1 =y_1) \sum_{i=0}^{\infty} \prob(D_{\tau}=i) \bigg[f\big(1_{z^*>0}\big) + h(x_2 +x_1 + z^*-i)^+  \nonumber\\
    &+ l(i - x_2 - x_1-z^*)^+ +\theta(x_1+y_1-i)^+  + \alpha v^{k-1}\big(s(\tau, \textbf{x}, y_3, y_2, y_1, D_\tau) \big) \bigg],
\end{align}
where $s(\tau, \textbf{x}, y_3, y_2, y_1, D_\tau)= \Big(\tau+1, \, \big((y_3 - (i-x_2-x_1-y_1-y_2)^+)^+, (x_2+y_2 - (i-x_1-y_1)^+)^+\big)\Big)$ using the convention that $\tau+1 = \big(\tau+1\big) \mod \gamma$. 
We can rewrite 
\small
\begin{align}
     v^k(\tau, \textbf{x})  &=  \sum_{y_1=0}^{z^*}\sum_{y_2=0}^{(z^*-y_1)}\sum_{y_3=0}^{(z^*-y_1-y_2)}P(Y_3 = y_3, Y_2= y_2, Y_1 =y_1) \sum_{i=0}^{x_1+y_1} \prob(D_{\tau}=i) \bigg[f\big(1_{z^*>0}\big)+ h(x_2 +x_1+z^*-i) \nonumber\\ 
     &+\theta(x_1+y_1-i) + \alpha v^{k-1}\big(\tau+1, (y_3,x_2+y_2) \big) \bigg] \\ \nonumber
    &+ \sum_{y_1=0}^{z^*}\sum_{y_2=0}^{(z^*-y_1)}\sum_{y_3=0}^{(z^*-y_1-y_2)}P(Y_3 = y_3, Y_2= y_2, Y_1 =y_1)\sum_{i=x_1+y_1+1}^{x_1 +y_1 + x_2 +y_2} \prob(D_{\tau}=i) \bigg[f\big(1_{z^*>0}\big)+h(x_2 +x_1+z^*-i) \\\nonumber
    &+ \alpha v^{k-1}\big(\tau+1, (y_3,x_2+y_2+x_1+y_1-i) \big) \bigg]\\ \nonumber
    &+ \sum_{y_1=0}^{z^*}\sum_{y_2=0}^{(z^*-y_1)}\sum_{y_3=0}^{(z^*-y_1-y_2)}P(Y_3 = y_3, Y_2= y_2, Y_1 =y_1)\sum_{i=x_1 +y_1 + x_2 +y_2+1}^{x_1+x_2+z^*} \prob(D_{\tau}=i) \bigg[f\big(1_{z^*>0}\big) + h(x_2 +x_1+z^*-i) \\\nonumber
    &+ \alpha v^{k-1}\big(\tau+1, (x_1+x_2+z^*-i,0) \big) \bigg]\\ \nonumber
    &+ \sum_{y_1=0}^{z^*}\sum_{y_2=0}^{(z^*-y_1)}\sum_{y_3=0}^{(z^*-y_1-y_2)}P(Y_3 = y_3, Y_2= y_2, Y_1 =y_1)\sum_{i=x_1 + x_2 +z^*+1}^{\infty} \prob(D_{\tau}=i) \bigg[f\big(1_{z^*>0}\big)+l(i - x_2 - x_1 - z^*)  \\\nonumber
    &+ \alpha v^{k-1}\big(\tau+1, (0,0) \big)\bigg].
\end{align}
With step 2 of induction, we also have
   \begin{align}
    v^k(\tau, \textbf{x}) 
      =& \beta^k_{\tau,0} + \beta^k_{\tau,1} x_1+ \beta^k_{\tau,2} x_2+\beta^k_{\tau,3} x_1^2 + \beta^k_{\tau,4} x_1x_2 + \beta^k_{\tau,5} x_2^2 + \\
      &\alpha \big[\beta^k_{\tau,6} x_1^3 + \beta^k_{\tau,7} x_1^2x_2 + \beta^k_{\tau,8} x_1x_2^2 + \beta^k_{\tau,9} x_2^3\big] + \nonumber\\
      &\alpha^2 \big[\beta^k_{\tau,10} x_1^4 + \beta^k_{\tau,11} x_1^3x_2 + \beta^k_{\tau,12} x_1^2x_2^2 + \beta^k_{\tau,13} x_1x_2^3 + \beta^k_{\tau,14} x_2^4\big]+ \nonumber\\
      & \vdots \nonumber\\
      &\alpha^{k-1} \big[\beta^{k}_{\tau,\frac{(k+1)(k+2)}{2}} x_1^{k+1} + \beta^{k}_{\tau,\frac{(k+1)(k+2)}{2}+1} x_1^{k}x_2 + \dots + \beta^{k}_{\tau,\frac{(k+1)(k+2)}{2}+k} x_1x_2^{k} + \beta^{k}_{\tau,\frac{(k+1)(k+4)}{2}} x_2^{k+1}\big], \quad \forall \tau, \textbf{x}.\nonumber 
\end{align}
We complete the proof by showing that the argument is true for $n=k+1$. Similarly, we have
\begin{align}
    v^{k+1}(\tau, \textbf{x}) &= \sum_{y_1=0}^{z^*}\sum_{y_2=0}^{(z^*-y_1)}\sum_{y_3=0}^{(z^*-y_1-y_2)}P(Y_3 = y_3, Y_2= y_2, Y_1 =y_1) \sum_{i=0}^{\infty} \prob(D_{\tau}=i) \bigg[f\big(1_{z^*>0}\big) + h(x_2 +x_1 + z^*-i)^+  \nonumber\\
    &+ l(i - x_2 - x_1-z^*)^+ +\theta(x_1+y_1-i)^+  + \alpha v^{k}\big(s(\tau, \textbf{x}, y_3, y_2, y_1, D_\tau) \big) \bigg],
\end{align}
where $s(\tau, \textbf{x}, y_3, y_2, y_1, D_\tau)= \Big(\tau+1, \, \big((y_3 - (i-x_2-x_1-y_1-y_2)^+)^+, (x_2+y_2 - (i-x_1-y_1)^+)^+\big)\Big)$ using the convention that $\tau+1 = \big(\tau+1\big) \mod \gamma$. 
We can rewrite 
\begin{align}
     v^{k+1}(\tau, \textbf{x})  &=  \sum_{y_1=0}^{z^*}\sum_{y_2=0}^{(z^*-y_1)}\sum_{y_3=0}^{(z^*-y_1-y_2)}P(Y_3 = y_3, Y_2= y_2, Y_1 =y_1) \sum_{i=0}^{x_1+y_1} \prob(D_{\tau}=i) \bigg[f\big(1_{z^*>0}\big) + h(x_2 +x_1+z^*-i) \nonumber\\ 
     &+\theta(x_1+y_1-i) + \alpha v^{k}\big(\tau+1, (y_3,x_2+y_2) \big) \bigg] \\ \nonumber
    &+ \sum_{y_1=0}^{z^*}\sum_{y_2=0}^{(z^*-y_1)}\sum_{y_3=0}^{(z^*-y_1-y_2)}P(Y_3 = y_3, Y_2= y_2, Y_1 =y_1)\sum_{i=x_1+y_1+1}^{x_1 +y_1 + x_2 +y_2} \prob(D_{\tau}=i) \bigg[f\big(1_{z^*>0}\big)+h(x_2 +x_1+z^*-i) \\\nonumber
    &+ \alpha v^{k}\big(\tau+1, (y_3,x_2+y_2+x_1+y_1-i) \big) \bigg]\\ \nonumber
    &+ \sum_{y_1=0}^{z^*}\sum_{y_2=0}^{(z^*-y_1)}\sum_{y_3=0}^{(z^*-y_1-y_2)}P(Y_3 = y_3, Y_2= y_2, Y_1 =y_1)\sum_{i=x_1 +y_1 + x_2 +y_2+1}^{x_1+x_2+z^*} \prob(D_{\tau}=i) \bigg[f\big(1_{z^*>0}\big) + h(x_2 +x_1+z^*-i) \\\nonumber
    &+ \alpha v^{k}\big(\tau+1, (x_1+x_2+z^*-i,0) \big) \bigg]\\ \nonumber
    &+ \sum_{y_1=0}^{z^*}\sum_{y_2=0}^{(z^*-y_1)}\sum_{y_3=0}^{(z^*-y_1-y_2)}P(Y_3 = y_3, Y_2= y_2, Y_1 =y_1)\sum_{i=x_1 + x_2 +z^*+1}^{\infty} \prob(D_{\tau}=i) \bigg[f\big(1_{z^*>0}\big)+l(i - x_2 - x_1 - z^*)  \\\nonumber
    &+ \alpha v^{k}\big(\tau+1, (0,0) \big)\bigg].
\end{align}
Using step 2 of induction we have
\begin{align}
     v^{k+1}(\tau, \textbf{x})  =& \beta^{k+1}_{\tau,0} + \beta^{k+1}_{\tau,1} x_1+ \beta^{k+1}_{\tau,2} x_2+\beta^{k+1}_{\tau,3} x_1^2 + \beta^{k+1}_{\tau,4} x_1x_2 + \beta^{k+1}_{\tau,5} x_2^2 \label{induction}\\
      &+\alpha \big[\beta^{k+1}_{\tau,6} x_1^3 + \beta^{k+1}_{\tau,7} x_1^2x_2 + \beta^{k+1}_{\tau,8} x_1x_2^2 + \beta^{k+1}_{\tau,9} x_2^3\big]  \nonumber\\
      &+\alpha^2 \big[\beta^{k+1}_{\tau,10} x_1^4 + \beta^{k+1}_{\tau,11} x_1^3x_2 + \beta^{k+1}_{\tau,12} x_1^2x_2^2 + \beta^{k+1}_{\tau,13} x_1x_2^3 + \beta^{k+1}_{\tau,14} x_2^4\big] \nonumber\\
      & \vdots \nonumber\\
      &+\alpha^{k-1} \big[\beta^{k}_{\tau,\frac{(k+1)(k+2)}{2}} x_1^{k+1} + \beta^{k}_{\tau,\frac{(k+1)(k+2)}{2}+1} x_1^{k}x_2 + \dots + \beta^{k}_{\tau,\frac{(k+1)(k+2)}{2}+k} x_1x_2^{k} + \beta^{k}_{\tau,\frac{(k+1)(k+4)}{2}} x_2^{k+1}\big] \nonumber\\
     &+ \sum_{y_1=0}^{z^*}\sum_{y_2=0}^{(z^*-y_1)}\sum_{y_3=0}^{(z^*-y_1-y_2)}P(Y_3 = y_3, Y_2= y_2, Y_1 =y_1) \sum_{i=0}^{x_1+y_1} \frac{\alpha^k}{M_\tau +1} 
     \big[\beta^{k}_{\tau+1,\frac{(k+1)(k+2)}{2}} (x_2+y_2)^{k+1} \nonumber\\
     &\hspace{9cm}+ \dots + \beta^{k}_{\tau+1,\frac{(k+1)(k+4)}{2}} y_3^{k+1}\big]\nonumber\\
     &+ \sum_{y_1=0}^{z^*}\sum_{y_2=0}^{(z^*-y_1)}\sum_{y_3=0}^{(z^*-y_1-y_2)}P(Y_3 = y_3, Y_2= y_2, Y_1 =y_1)\sum_{i=x_1+y_1+1}^{x_1 +y_1 + x_2 +y_2} \frac{\alpha^k}{M_\tau +1} \nonumber\\
     &\hspace{5cm}\big[\beta^{k}_{\tau+1,\frac{(k+1)(k+2)}{2}} (x_2+y_2+x_1+y_1-i)^{k+1} 
     + \dots + \beta^{k}_{\tau+1,\frac{(k+1)(k+4)}{2}} y_3^{k+1}\big]\nonumber\\
    &+ \sum_{y_1=0}^{z^*}\sum_{y_2=0}^{(z^*-y_1)}\sum_{y_3=0}^{(z^*-y_1-y_2)}P(Y_3 = y_3, Y_2= y_2, Y_1 =y_1)\sum_{i=x_1 +y_1 + x_2 +y_2+1}^{x_1+x_2+z^*} \frac{\alpha^k}{M_\tau +1} \nonumber\\
     &\hspace{9cm}\big[\dots + \beta^{k}_{\tau+1,\frac{(k+1)(k+4)}{2}} {(x_1+x_2+z^*-i)}^{k+1}\big].\nonumber
\end{align}
The last three terms in \eqref{induction} represent basis functions of order up to $k+2$, each discounted by $\alpha^k$. The increase in the maximum order by one is due to the summation over $i$, which allows for the expansion of terms involving both $x_1$ and $x_2$. Since $$\sum_{y_1=0}^{z^*}\sum_{y_2=0}^{(z^*-y_1)}\sum_{y_3=0}^{(z^*-y_1-y_2)}P(Y_3 = y_3, Y_2= y_2, Y_1 =y_1) = 1,$$ and we can bring out terms involving $x_1$ and $x_2$ from this summation, the maximum order increase for $x_1$ and $x_2$ remains one. Moreover, the summation over $i$ for terms involving $y_1$, $y_2$, and $y_3$ can also increase their maximum order by one and so the summation over $(y_3, y_2, y_1)$ can generate $z^*$ with the maximum order of $k+2$ similar to equations \eqref{inductionstep11}-\eqref{inductionstep12}. Considering that $z^*$ is a linear function of $x_1$ and $x_2$, and uncertainty in shelf-life is exogenous, the maximum order of $x_1$ and $x_2$ generated by ${z^*}^{k+2}$ is $k+2$. Therefore, we can write
\begin{align}
    v^{k+1}(\tau, \textbf{x}) 
      =& \beta^{k+1}_{\tau,0} + \beta^{k+1}_{\tau,1} x_1+ \beta^{k+1}_{\tau,2} x_2+\beta^{k+1}_{\tau,3} x_1^2 + \beta^{k+1}_{\tau,4} x_1x_2 + \beta^{k+1}_{\tau,5} x_2^2 + \\
      &\alpha \big[\beta^{k+1}_{\tau,6} x_1^3 + \beta^{k+1}_{\tau,7} x_1^2x_2 + \beta^{k+1}_{\tau,8} x_1x_2^2 + \beta^{k+1}_{\tau,9} x_2^3\big] + \nonumber\\
      &\alpha^2 \big[\beta^{k+1}_{\tau,10} x_1^4 + \beta^{k+1}_{\tau,11} x_1^3x_2 + \beta^{k+1}_{\tau,12} x_1^2x_2^2 + \beta^{k+1}_{\tau,13} x_1x_2^3 + \beta^{k+1}_{\tau,14} x_2^4\big]+ \nonumber\\
      & \vdots \nonumber\\
      &\alpha^{k-1} \big[\beta^{k}_{\tau,\frac{(k+1)(k+2)}{2}} x_1^{k+1} + \beta^{k}_{\tau,\frac{(k+1)(k+2)}{2}+1} x_1^{k}x_2 + \dots + \beta^{k}_{\tau,\frac{(k+1)(k+2)}{2}+k} x_1x_2^{k} + \beta^{k}_{\tau,\frac{(k+1)(k+4)}{2}} x_2^{k+1}\big]+ \nonumber\\ 
      &\alpha^{k} \big[\beta^{{k+1}}_{\tau,\frac{(k+2)(k+3)}{2}} x_1^{k+2} + \beta^{{k+1}}_{\tau,\frac{(k+2)(k+3)}{2}+1} x_1^{k+1}x_2 +\dots + \beta^{{k+1}}_{\tau,\frac{(k+2)(k+3)}{2}+k+1} x_1x_2^{k+1} + \beta^{k+1}_{\tau,\frac{(k+2)(k+5)}{2}} x_2^{k+2}\big].\nonumber
\end{align}
for $\forall \tau, \textbf{x}$. The proof is complete. \halmos

Proposition \ref{prop1} highlights the diminishing impact of employing higher-order basis functions for approximating the value function as $n$ increases. 

The proposition includes a restrictive assumption on the structure of the optimal policy, namely it being a linear function in state variables. Consider the first iteration of the value iteration algorithm with no assumption on the structure of the optimal policy in \eqref{inductionstep12}. If we assume a piecewise linear optimal policy instead, the structure of the first iteration of the value iteration algorithm becomes piecewise quadratic. Furthermore, if we approximate the piecewise linear policy with a quadratic function (similar to \citealt{dai2019inpatient}), the structure of the first iteration involves polynomial terms with orders up to four. While cubic and fourth-degree basis functions may hold significance, especially in scenarios involving piece-wise linear ordering policies, we decide to utilize the value function of the corresponding non-perishable inventory problem as a basis function to represent all ignored higher-order basis functions. Through numerical analysis, we also demonstrate that cubic terms fail to contribute as significantly as quadratic terms. 

\newpage


\revision{\section{Details of the Approximate Policy Iteration Algorithm}\label{app:algorithm}

\minorrevision{This section provides details of the ADP algorithm as well as a discussion of its implementation.}

\begin{center}\label{al:API}
\OneAndAHalfSpacedXII
\scalebox{0.9}{
\begin{minipage}{\linewidth}
\begin{algorithm}[H]
\SetAlgoLined
\textbf{Input:} Total number of iterations $N$, initial coefficient vectors $\boldsymbol{\beta}^0_\tau=0 \: \forall \tau$, number of replications $Q$, learning rate $\lambda_n = 1/n, \, \forall n \in \{1,2,\dots, N\}.$

\textbf{Output:} Updated coefficient vectors $\boldsymbol{\beta}^*_\tau \in \{\boldsymbol{\beta}^0_\tau, \boldsymbol{\beta}^1_\tau, \dots, \boldsymbol{\beta}^N_\tau\}$ for all $\tau$ which minimize the sample average estimate of expected long-run discounted cost.

Set $n \leftarrow 1$

\While{$n \leq N$}{
        \textbf{Step 1.} \emph{Policy improvement}: Simulate $Q$ replications of shelf-life and demand sample paths to generate inventory states $\big \{\textbf{x}_{\tau,\psi}^n:\psi=1,\dots,\Psi_\tau(Q)\big\}$ under policy $\mu^n$ \eqref{alg:policy} and real-time decisions made at each visited state.
        \begin{align}
            \mu^n(\tau, \textbf{x})  = \argmin_{z \in \mathcal{Y}} \Exp \bigg[C\big(\textbf{x}, \, z\big) + \alpha \sum_{i=1}^k\beta^{n-1}_{\tau+1,i}\phi_{i}\big(s(\tau, \textbf{x}, Y_m^z, \dots, Y_1^z, D_\tau)\big)\bigg] \label{alg:policy}
        \end{align}  
        If the exact calculation of expectation in \eqref{alg:policy} is computationally expensive, then minimize the sample average estimate using Monte Carlo simulation and by leveraging CRN to reduce the variance of estimates.  
        
        \textbf{Step 2.} \emph{Policy evaluation}: estimate expected long-run discounted cost $c_{\tau,\psi}^n$ for each state $\textbf{x}_{\tau,\psi}^n$ using Monte Carlo simulation. Solve the least squares problem  \eqref{alg:ls}; Update coefficient vectors using \eqref{alg:upd}:
        \begin{align}
            \boldsymbol{\beta}_\tau^{n^*} &= \argmin_{\boldsymbol{\beta}_\tau} \sum_{\psi=1}^{\Psi_\tau(Q)}\Big[c_{\tau,\psi}^n-\boldsymbol{\beta}_\tau \cdot \boldsymbol{\phi}(\tau, \textbf{x}_{\tau,\psi}^n) \Big]^2 \label{alg:ls}\\ 
            \boldsymbol{\beta}_\tau^{n} &= (1-\lambda_n) \times \boldsymbol{\beta}_\tau^{n-1} + \lambda_n \times 
        \boldsymbol{\beta}_\tau^{n^*} \label{alg:upd}
        \end{align}
    
    \textbf{Step 3.} Set $n \leftarrow n + 1$
}         
\caption{Approximate Policy Iteration Algorithm}
\label{algo:approx_policy_iteration}
\end{algorithm}
\end{minipage}
}
\end{center}
\vspace{5mm}
Once the coefficients $\boldsymbol{\beta}^*_\tau$ are obtained, one can make real-time decisions at each visited state using $\mu^*(\tau, \textbf{x})$ obtained by solving, 
\begin{align}
    \mu^*(\tau, \textbf{x})  = \argmin_{z \in \mathcal{Y}} \Exp \bigg[C\big(\textbf{x}, \, z\big) + \alpha \sum_{i=1}^k\beta^{*}_{\tau+1,i}\phi_{i}\big(s(\tau, \textbf{x}, Y_m^z, \dots, Y_1^z, D_\tau)\big)\bigg]. \label{ADPolicy}
\end{align} 
The online decisions at each visited state typically require only a few seconds in the case study. The algorithm's longest runtime for instances with $m=5$ is approximately 1.6 hours per iteration, while for instances with $m=8$, it extends to about 4.2 hours per iteration with a 2.4GH CPU and 4GB of RAM.

\minorrevision{\section{The Case with $m=8$} \label{sec:adperfm8}
To examine the scalability of the ADP algorithm, we also investigate larger problem instances with a maximum shelf-life of eight days (see Table \ref{tab:scen_m58} for the shelf-life settings). For problem instances with $m=8$, we set $(c_0^2, c_0^3, c_0^4, c_0^5, c_0^6, c_0^7, c_0^8) = (0.8, 1.4, 1.9, 2.3, 1.7, 1.2, 0.8)$ resulting in the base exogenous distribution: $p_8 = 0.06$, $p_7 = 0.09$, $p_6 = 0.16$, $p_5 = 0.29$, $p_4 = 0.19$, $p_3=0.12$, $p_2 = 0.06$, $p_1 =0.03$, and consider five different scenarios for $(c_1^2, c_1^3, c_1^4, c_1^5, c_1^6, c_1^7, c_1^8)$ such that the chance of receiving units with remaining shelf-life of one increases in the order size as well as absolute values of coefficients, except for the exogenous case for which we set the coefficients to zero.

Figure \ref{fig:ADPerfm8} illustrates the performance with respect to the estimate of the expected relative reduction in cost compared to the initial Myopic policy. Similar to the case with $m=5$, we observe  small relative reductions in cost for problems with the fixed ordering cost of \revision{$\kappa=10$} as the Myopic policy performs close to optimal. However, for problems with the fixed ordering cost of \revision{$\kappa=100$} the ADP policy can significantly improve the initial upper-bound by up to 31.4\% and achieves an average of 18.7\% reduction in cost among all tested scenarios. }

\begin{figure}
    \centering
    \includegraphics[width= 1.0 \textwidth]{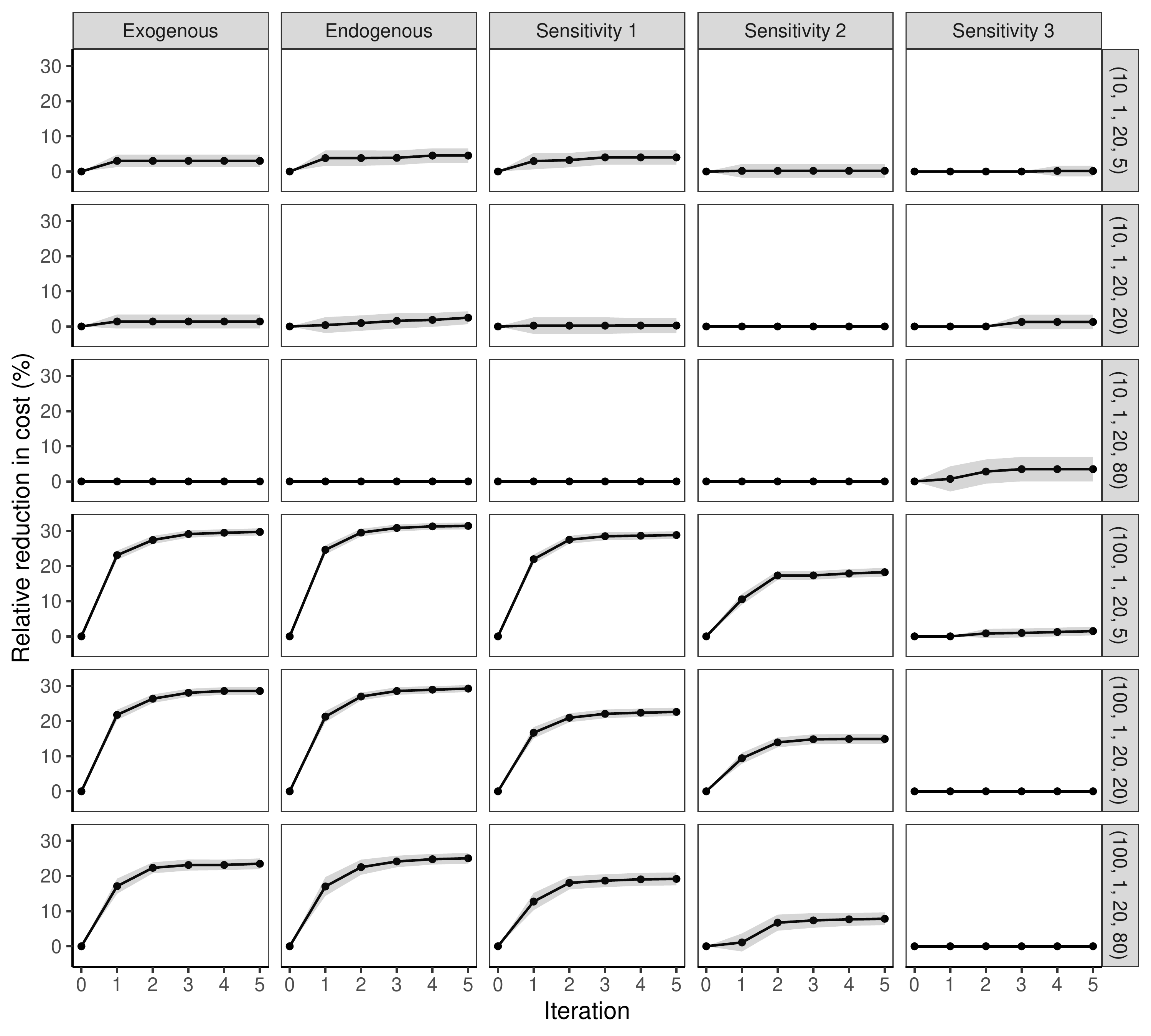}
    \caption{Performance of the algorithm. The black line is the estimate of the expected relative reduction in cost compared to the initial Myopic policy at iteration zero. The gray area is the 95\% confidence interval. On average, the estimate of the expected relative reduction in cost among cases with $\kappa=10$ and $\kappa=100$ is 1.4\% and 18.7\%, respectively.}
    \label{fig:ADPerfm8}
\end{figure}

\section{Demand Sensitivity Analysis}\label{app:sensdem}
We increase the mean and variance of the \revision{negative binomial} distribution by a factor of two for each day of the week and investigate the performance of the ADP approach for cases with $m \in \{3, 5\}$. To ensure that the effective Coefficient of Variation (CV) after truncation remains unchanged, we also increase the truncation level from $M=20$ to $M=30$. Table \ref{tab:demCV} presents the true and effective mean, variance, and CV before and after truncation for different days of the week and demand settings.

The higher truncation level of $M=30$ results in a problem with a larger state-space for which computing the optimal policy or lower-bound becomes computationally challenging. Therefore, we evaluate the performance of the ADP compared to the initial Myopic policy in terms of the expected relative improvement in the upper-bound. 
Figure \ref{fig:DemSens3} illustrates the estimate of the expected relative reduction in cost compared to the initial (exact) Myopic policy at iteration zero together with its 95\% confidence interval for cases with $m=3$ and the baseline demand ($M=20$) as well as larger demand ($M=30$). We observe for problems with the fixed ordering cost of $\kappa=10$ and positive coefficients of the order size that the Myopic policy with larger demand performs better than that with baseline demand as the unit cost of expiration goes from 5 to 80. Intuitively, larger demand indicates a greater utilization of the inventory and hence a lower likelihood of future expiration. However, when the fixed ordering cost is $\kappa=100$, the Myopic policy with baseline demand performs better than that with larger demand. This is because the larger fixed ordering cost together with the larger demand results in placing larger orders less frequently and hence a higher likelihood of future expiration. For negative coefficients of order size that increases the chance of receiving units with remaining shelf-life of one, the Myopic policy with larger demand again performs better than that with baseline demand. In summary, the estimate of the expected relative reduction in cost on average among all tested scenarios is 4.9\% and 4.3\% with baseline and larger demand, respectively.

\begin{table}[]
\small
\centering
\caption{The true and effective mean, variance, and CV before and after truncation for each day of the week.}
\resizebox{\textwidth}{!}{\begin{tabular}{cccclccclccclccc}
\hline
\multicolumn{1}{c}{} & \multicolumn{3}{c}{Data} & \multicolumn{1}{l}{} & \multicolumn{3}{c}{$M=20$} & \multicolumn{1}{l}{} & \multicolumn{3}{c}{Counterfactual} & \multicolumn{1}{l}{} & \multicolumn{3}{c}{$M=30$} \\ \cline{2-4} \cline{6-8} \cline{10-12} \cline{14-16} 
$\tau$ & Mean & Variance & CV &  & Mean & Variance & CV &  & Mean & Variance & CV &  & Mean & Variance & CV \\
\hline
0 & 5.66 & 14.82 & 2.62 &  & 5.65 & 14.48 & 2.56 &  & 11.32 & 29.64 & 2.62 &  & 11.31 & 29.13 & 2.58 \\
1 & 6.92 & 11.28 & 1.63 &  & 6.92 & 11.22 & 1.62 &  & 13.85 & 22.57 & 1.63 &  & 13.84 & 22.40 & 1.62 \\
2 & 6.50 & 12.39 & 1.91 &  & 6.50 & 12.28 & 1.89 &  & 13.01 & 24.79 & 1.91 &  & 13.00 & 24.54 & 1.89 \\
3 & 6.17 & 9.60 & 1.56 &  & 6.16 & 9.58 & 1.55 &  & 12.33 & 19.20 & 1.56 &  & 12.33 & 19.16 & 1.55 \\
4 & 5.82 & 11.52 & 1.98 &  & 5.81 & 11.44 & 1.97 &  & 11.63 & 23.04 & 1.98 &  & 11.63 & 22.91 & 1.97 \\
5 & 3.33 & 5.35 & 1.61 &  & 3.33 & 5.35 & 1.61 &  & 6.65 & 10.70 & 1.61 &  & 6.65 & 10.70 & 1.61 \\
6 & 3.43 & 8.78 & 2.56 &  & 3.43 & 8.74 & 2.55 &  & 6.85 & 17.56 & 2.56 &  & 6.85 & 17.54 & 2.56 \\
\hline
\end{tabular}}
\label{tab:demCV}
\end{table}

\begin{figure}
    \centering
    \includegraphics[width= 1.0 \textwidth]{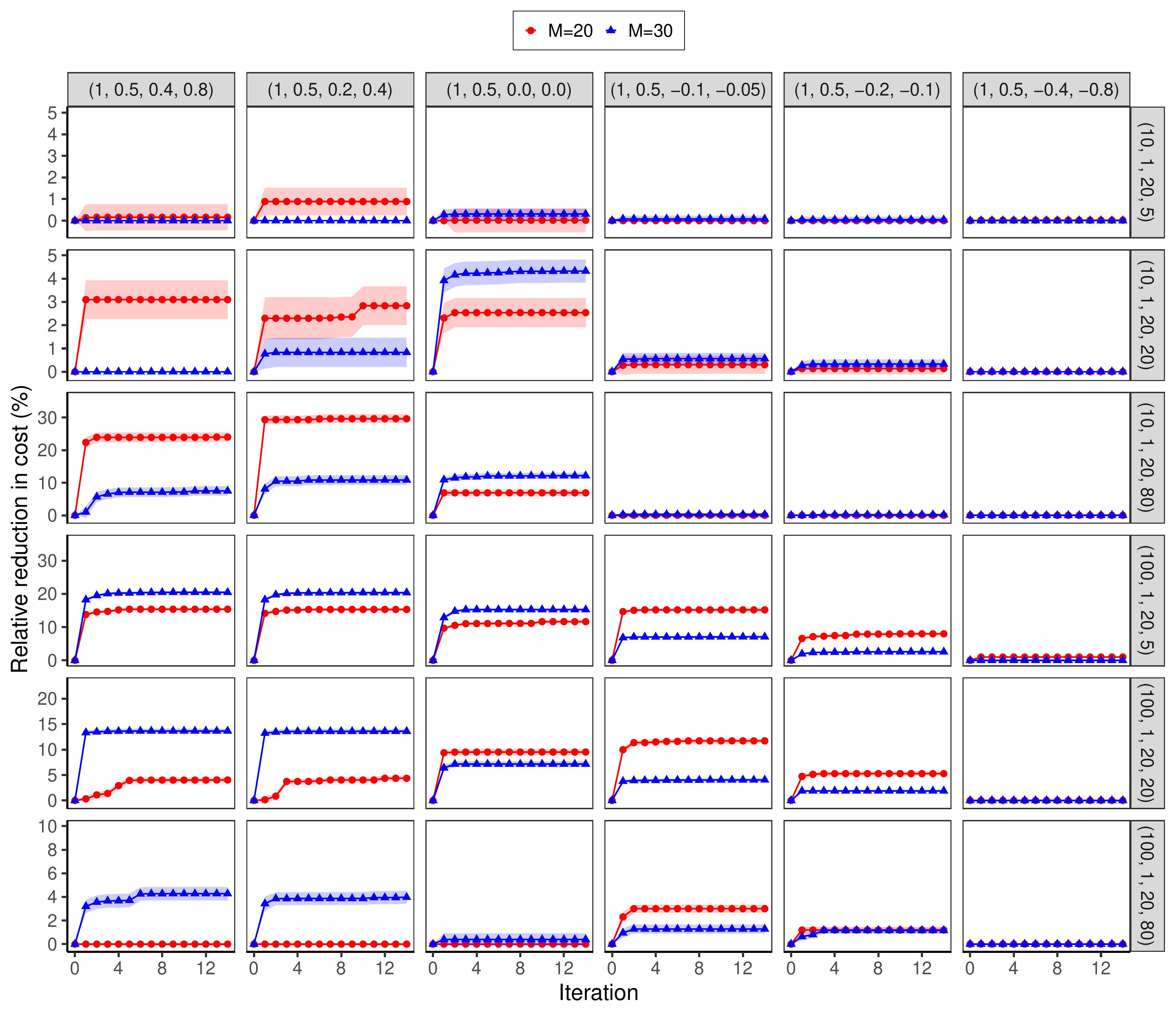}
    \caption{Sensitivity to demand for cases with $m=3$. On average among all tested scenarios, the estimate of the expected relative improvement in the upper-bound, i.e., Myopic policy at iteration zero, after using the ADP approach is 4.9\% and 4.3\% for the real (red) and larger (blue) demand, respectively.}
    \label{fig:DemSens3}
\end{figure}

For cases with $m=5$, because computing the exact Myopic policy is also challenging, we use Monte Carlo simulation to determine ordering decisions based on the least sample average estimate of the expected cost in  \eqref{greedy}. In Figure \ref{fig:DemSens5}, we provide the estimate of the expected relative reduction in cost compared to the initial (approximate) Myopic policy at iteration zero together with its 95\% confidence interval for cases with $m=5$ and the baseline demand ($M=20$) as well as larger demand  ($M=30$). Similarly, we observe for the fixed ordering cost of $\kappa=10$ the Myopic policy would perform well and close to optimal for larger demand. However, for the fixed ordering cost of $\kappa=100$ the ADP policy with the larger demand results in a higher relative reduction in cost particularly when the unit expiration cost is larger and the chance of receiving units with remaining shelf-life of one is smaller; see, e.g., exogenous case with $(f,h,l,\theta)=(100, 1, 20, 80)$.  

\begin{figure}
    \centering
    \includegraphics[width= 1.0 \textwidth]{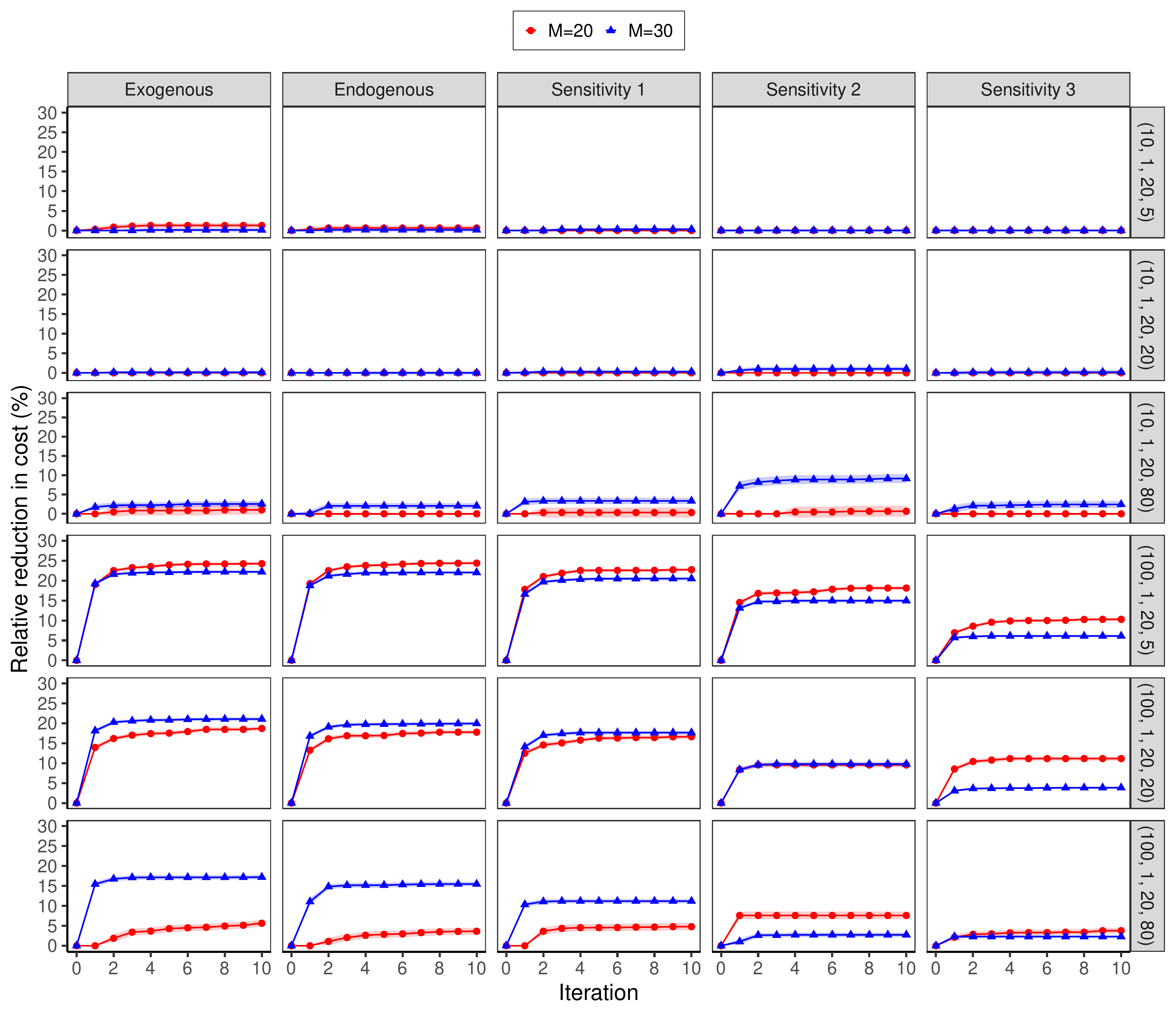}
    \caption{Sensitivity to demand for cases with $m=5$. On average among all tested scenarios, the estimate of the expected relative improvement in the upper-bound, i.e., Myopic policy at iteration zero, after using the ADP approach is 6.8\% and 7.6\% for the real (red) and larger (blue) demand, respectively.}
    \label{fig:DemSens5}
\end{figure}

\section{Validation of the Parametric Choices of Demand and Shelf-life Distributions using Data} \label{ap:data}

Figures \ref{fig:GOF-Mon} and \ref{fig:GOF-Sun} present a comparison between \revision{negative binomial} and Poisson distributions fitted to historical daily demand data for Mondays ($\tau=0$) and Sundays ($\tau=6$) throughout the years 2015 and 2016. The chi-square goodness of fit test returns a $p$-value between 0.396 for Sundays and  0.841 for Mondays suggesting that negative binomial fits the demand data well. In contrast, the chi-square test rejects the hypothesis that the data is generated from the Poisson distribution.

\begin{figure}[!htbp]
    \centering
    \includegraphics[width= 1.0 \textwidth]{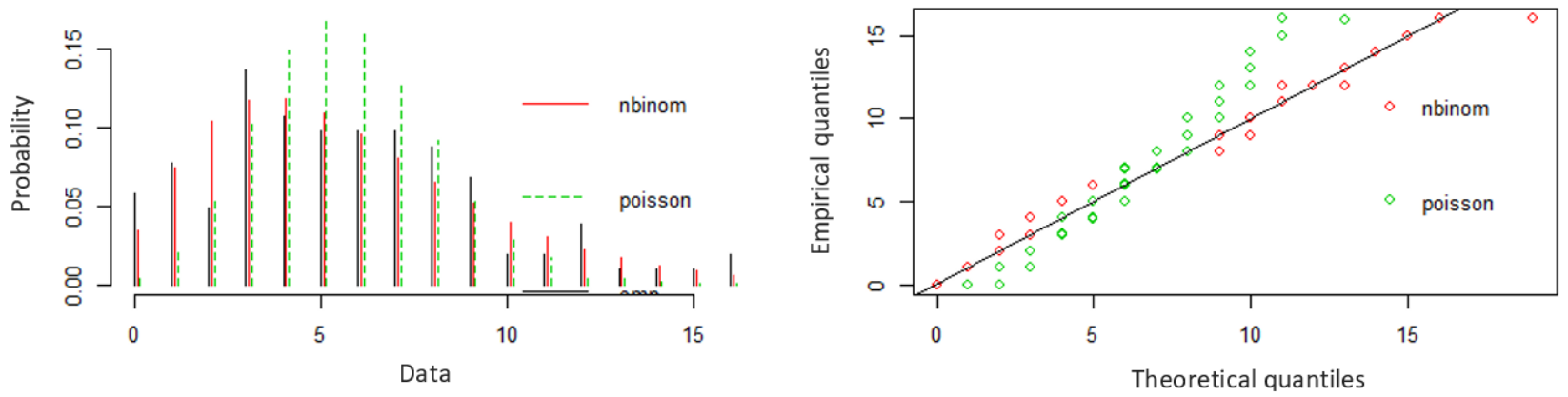}
    \caption{\revision{Comparison of fitted negative binomial and Poisson distributions with historical demand data for Mondays ($\tau = 0$) during the years 2015 and 2016.}}
    \label{fig:GOF-Mon}
\end{figure}

\begin{figure}[!htbp]
    \centering
    \includegraphics[width= 1.0 \textwidth]{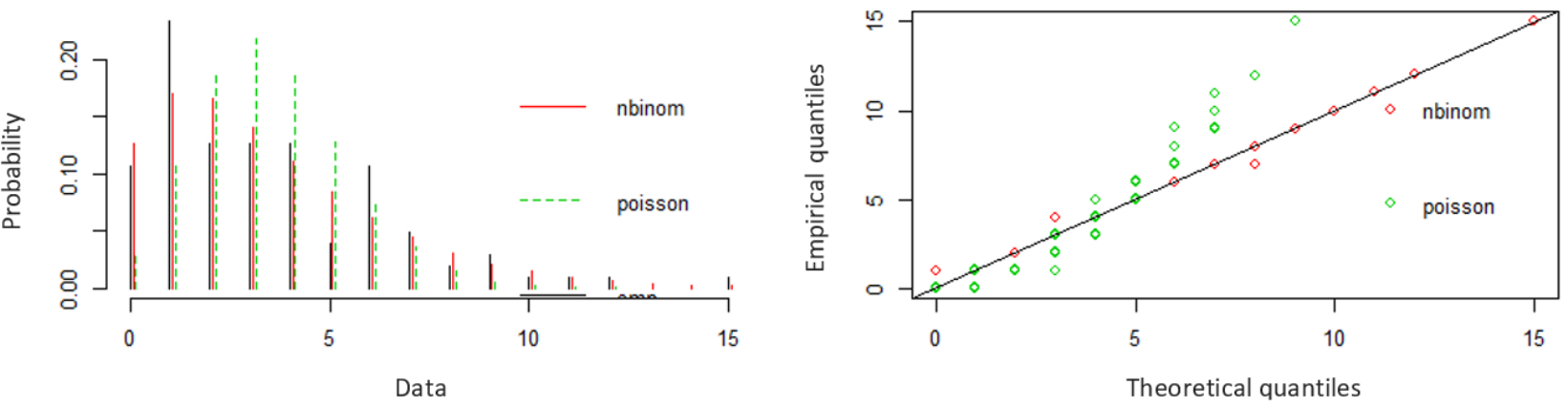}
    \caption{\revision{Comparison of fitted negative binomial and Poisson distributions with historical demand data for Sundays ($\tau = 6$) during the years 2015 and 2016.}}
    \label{fig:GOF-Sun}
\end{figure}

Table \ref{app:Reg.} presents the results of a \revision{negative binomial} regression model for daily demand at HGH with covariates for days of the week and months of the year. The coefficients for days of the week are statistically significant. 

\begin{table}[!htbp] \centering 
    \OneAndAHalfSpacedXII
    \small
    \caption{\revision{Negative binomial regression results for daily demand at HGH}.}
    \label{app:Reg.} 
    \begin{tabular}{@{\extracolsep{5pt}}lcc} 
    \\[-1.8ex]\hline 
    \hline \\[-1.8ex] 
     & \multicolumn{2}{c}{\textit{Dependent variable:}} \\ 
    \cline{2-3} 
    \\[-1.8ex] & demand & Std. Error\\ 
    \hline \\[-1.8ex] 
     Intercept & 1.263$^{***}$ & 0.098 \\ [5pt]
      Mon & 0.501$^{***}$ & 0.088 \\[5pt] 
      Tue & 0.710$^{***}$ & 0.086 \\[5pt]
      Wed & 0.653$^{***}$ & 0.086 \\ [5pt]
     Thu & 0.584$^{***}$  & 0.087 \\ [5pt]
     Fri & 0.514$^{***}$  & 0.087 \\ [5pt]
     Sat & $-$0.037 & 0.094 \\ [5pt]
     Jan & 0.006 & 0.107 \\ [5pt]
     Feb &  $-$0.058 & 0.110  \\ [5pt]
     Mar & $-$0.113 & 0.108 \\ [5pt]
     Apr & $-$0.031 & 0.108  \\ [5pt]
     May & 0.049 & 0.106\\ [5pt]
     Jun & 0.060 & 0.106\\ [5pt]
     Jul &  0.071 & 0.106 \\ [5pt]
     Aug & $-$0.174 & 0.109 \\ [5pt]
     Sep & $-$0.107 & 0.109 \\ [5pt]
     Oct & 0.061  & 0.106 \\ [5pt]
     Nov & $-$0.129 & 0.109 \\ 
    \hline \\[-1.8ex] 
    Observations & 731 \\ 
    Log Likelihood & $-$1,825.634 \\ 
    $\theta$ & 5.984$^{***}$  (0.702) \\ 
    Akaike Inf. Crit. & 3,687.269 \\ 
    \hline 
    \hline \\[-1.8ex] 
    \textit{Note:}  & \multicolumn{1}{r}{$^{*}$p$<$0.1; $^{**}$p$<$0.05; $^{***}$p$<$0.01} \\ 
    \end{tabular} 
\end{table} 

Table \ref{app:multilogit} presents the estimated \revision{multinomial} \revision{logistic} regression model, demonstrating that the remaining shelf-life of received units are associated with the order size. Finally, Figure \ref{fig:shelf-lifedata} compares the empirical probabilities for the remaining shelf-life of delivered units under two different order sizes and compares them to the multinomial distribution with the estimated parameters through the logistic regression.

\begin{table}[] \centering 
    \centering 
    \caption{\revision{Multinomial logistic regression results for the remaining shelf-life of orders.}}
\begin{tabular}{@{\extracolsep{5pt}}lcccc} 
\\[-1.8ex]\hline 
 & \multicolumn{4}{c}{\textit{Remaining shelf-life}} \\ 
\cline{2-5} 
\\[-1.8ex] & 2 & 3 & 4 & 5 \\ 
\hline \\[-2ex] 
 Constant & 1.895$^{***}$ & 3.127$^{***}$ & 3.071$^{***}$ & 2.471$^{***}$ \\ 
  & (0.376) & (0.356) & (0.353) & (0.375) \\ [1ex]
 OrderSize & $-$0.032 & $-$0.064$^{*}$ & $-$0.025 & $-$0.095$^{**}$ \\ 
  & (0.037) & (0.035) & (0.034) & (0.037) \\ 
\hline \\[-1.8ex] 
Akaike Inf. Crit. & 5,602.048 & 5,602.048 & 5,602.048 & 5,602.048 \\ 
\hline 
\hline \\[-1.8ex] 
\textit{Note:}  & \multicolumn{4}{r}{$^{*}$p$<$0.1; $^{**}$p$<$0.05; $^{***}$p$<$0.01} \\ 
\end{tabular} 
    \label{app:multilogit}
\end{table} 

\begin{figure}%
    \centering
    \captionsetup[subfigure]{labelformat=empty}
    \subfloat[(A) Order size of $z=6$.]{\includegraphics[width=0.49\textwidth]{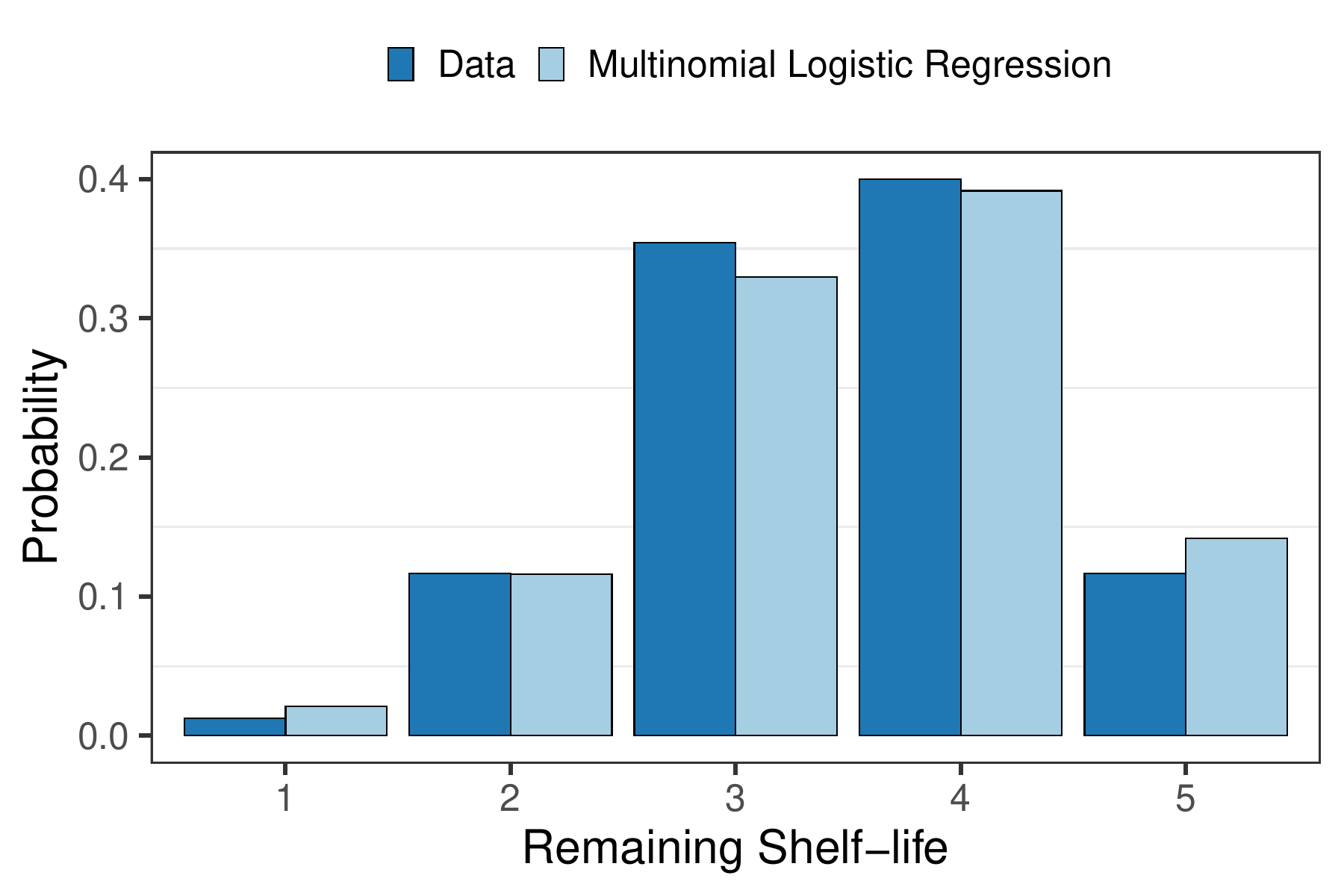}}%
    \subfloat[(B) Order size of $z=8$.]{\includegraphics[width=0.49\textwidth]{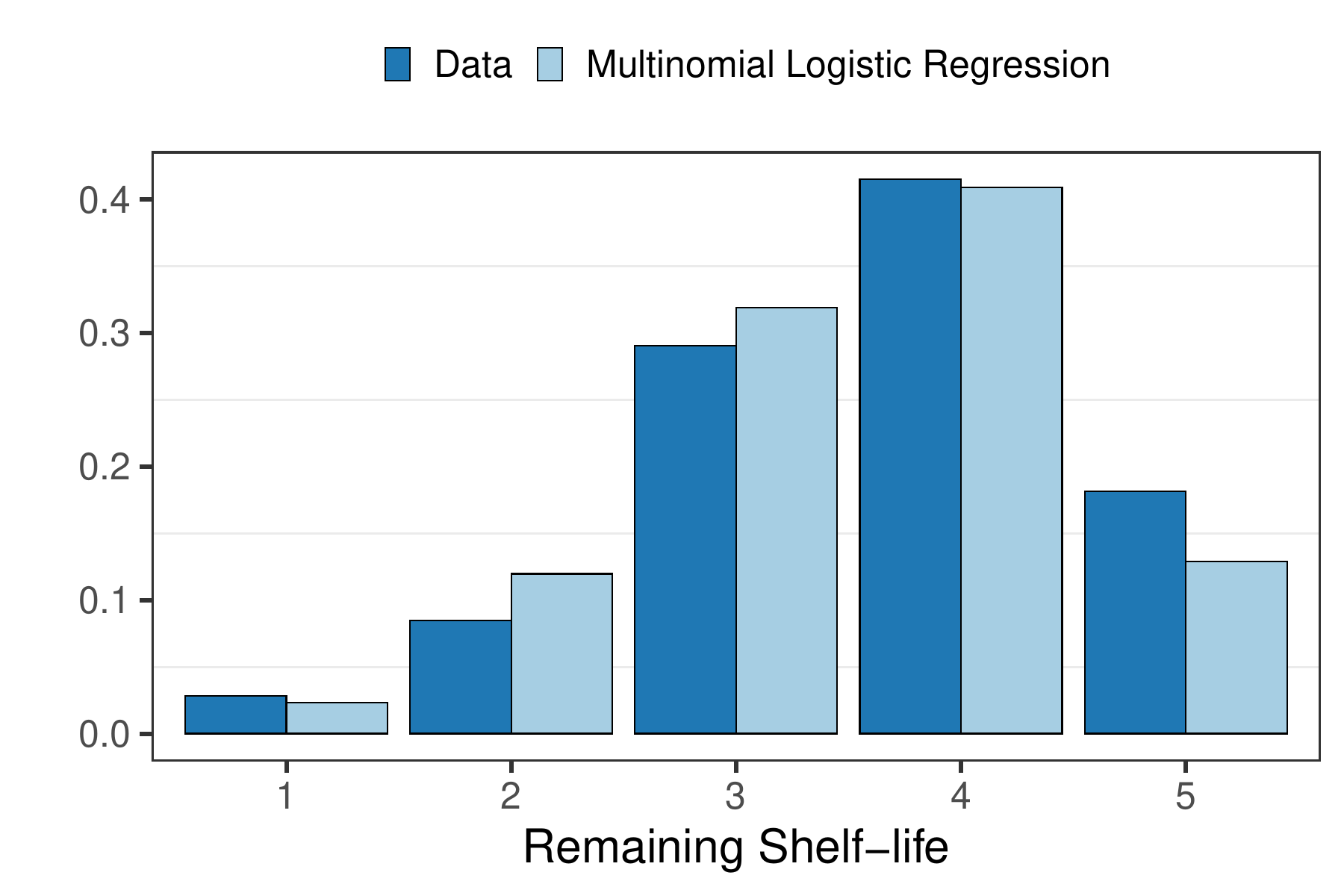} }
    \caption{\revision{\revision{Comparing the historical distribution of remaining shelf-life with the fitted multinomial distribution for two common order sizes, 6 and 8, at HGH in 2017}.}}%
    \label{fig:shelf-lifedata}%
\end{figure}
}

\end{document}